\documentclass{article}
\usepackage[margin=1in]{geometry}  

\usepackage{listings}
\usepackage{amsmath,mathtools}
\usepackage{amsthm}
\usepackage{tikz}
\usetikzlibrary{positioning}
\usepackage{caption,subcaption}
\usepackage{array}
\usepackage{mdwmath}
\usepackage{multirow}
\usepackage{mdwtab}
\usepackage{eqparbox}
\usepackage{multicol}
\usepackage{amsfonts}
\usepackage{multirow,bigstrut,threeparttable}
\usepackage{amsthm}
\usepackage{array}
\usepackage{bbm}
\usepackage{subcaption}
\usepackage{epstopdf}
\usepackage{mdwmath}
\usepackage{mdwtab}
\usepackage{eqparbox}
\usepackage{latexsym}
\usepackage{amssymb}
\usepackage{bm}
\usepackage{amssymb}
\usepackage{graphicx}
\usepackage{mathrsfs}
\usepackage{epsfig}
\usepackage{psfrag}
\usepackage{setspace}
\usepackage{tikz-cd}

\usepackage{wrapfig}
\usepackage{caption}
\usepackage{floatflt}

\usepackage[
CJKbookmarks=true,
bookmarksnumbered=true,
bookmarksopen=true,
colorlinks=true,
citecolor=red,
linkcolor=blue,
anchorcolor=red,
urlcolor=blue
]{hyperref}
\usepackage{cleveref}
\usepackage{algorithm}
\usepackage{algorithmic}
\usepackage{stfloats}
\usepackage[
autocite    = superscript,
backend     = bibtex, 
sortcites   = true,
style       = authoryear 
]{biblatex} 
\usepackage{nameref}
\usepackage[normalem]{ulem}

\usepackage{soul}

\input xy
\xyoption{all}

\theoremstyle{plain}
\newtheorem{theorem}{Theorem}[section]
\newtheorem{proposition}{Proposition}
\newtheorem{lemma}{Lemma}[section]
\newtheorem{corollary}{Corollary}[section]

\Crefname{theorem}{Theorem}{Theorems}
\Crefname{proposition}{Proposition}{Propositions}
\Crefname{corollary}{Corollary}{Corollaries}

\theoremstyle{definition}

\newtheorem{example}{Example}

\Crefname{definition}{Definition}{Definitions}
\Crefname{remark}{Remark}{Remarks}

\def\EE{\mathbb{E}}

\def\II{\mathbb{I}}

\def\PP{\mathbb{P}}

\def\RR{\mathbb{R}}

\def\calI{\mathcal{I}}

\def\calN{\mathcal{N}}

\newcommand\diag{\textup{diag}}

\def\1{\mathbbm{1}}
\def\var{\mathsf{Var}}

\newcommand{\argmax}{\mathop{\mathrm{argmax}}}

\usepackage{booktabs}
\usepackage{adjustbox}
\usepackage{multirow}
\usepackage{listings}

\theoremstyle{plain}

\def \var {\mathsf{Var}}

\usepackage{xspace}

\definecolor{myblue}{rgb}{.8, .8, 1}
\definecolor{mathblue}{rgb}{0.2472, 0.24, 0.6} 
\definecolor{mathred}{rgb}{0.6, 0.24, 0.442893}
\definecolor{mathyellow}{rgb}{0.6, 0.547014, 0.24}

\usepackage{enumitem}

\usepackage{pgfplots}
\pgfplotsset{compat=1.17}

\newcommand{\debiased}{\text{db}}

\title{MUSE: Multi-Treatment Experiment Design for Winner Selection and Effect Estimation
}
\author{
Jiachen Xu\textsuperscript{1} \qquad 
Jian Qian\textsuperscript{2}\qquad 
Zijun Gao\textsuperscript{3}\\
{\small \textsuperscript{1}Department of Industrial Engineering and Decision Analytics, HKUST, China} \\
{\small \textsuperscript{2}Courant Institute of Mathematical Sciences, New York University, USA} \\
{\small \textsuperscript{3}Marshall School of Business, University of Southern California, USA}
}

\begin{document}

\maketitle

\begin{abstract}

We study the design of experiments with multiple treatment levels, a setting common in clinical trials and online A/B/n testing.
Unlike single-treatment studies, practical analyses of multi-treatment experiments typically first select a winning treatment, and then only estimate the effect therein.
Motivated by this analysis paradigm, we propose a design for MUlti-treatment experiments that jointly maximizes the accuracy of  winner Selection and effect Estimation (MUSE).
Explicitly, we introduce a single objective that balances selection and estimation, and determine the unit allocation to treatments and control by optimizing this objective.
Theoretically, we establish finite-sample guarantees and asymptotic equivalence between our proposal and the Neyman allocation for the true optimal treatment and control.
Across simulations and a real data application, our method performs favorably in both selection and estimation compared to various standard alternatives.

\end{abstract}



\section{Introduction}\label{sec:introduction}

Experiments with multi-level treatments have become increasingly prominent across domains, including dose-finding trials in medicine \parencite{oquigley1990continual, iasonos2014adaptive}, educational programs of varying intensity \parencite{Krueger1999star, heckman2007productivity, durlak2010meta}, and large-scale A/B/n experiments on technology platforms \parencite{kohavi2020trustworthy, bakshy2014designing}.
The design of experiment, which determines how experimental resources are distributed across arms, is crucial and significantly influences the downstream analysis, conclusions, and decisions. 
In this paper, we study the design problem with multiple treatment levels under a fixed budget.

There is an extensive and rapidly growing literature on experiment design; nevertheless, existing approaches have not fully addressed all of the considerations below.
\begin{enumerate}
    \item {Asymptotic v.s. \textbf{Non-asymptotic}}. 
    A large body of research has been devoted to asymptotically optimal experiment design \parencite{van2000asymptotic, tsiatis2006semiparametric, hahn2011adaptive, kato2020efficient, cook2023semiparametric, li2024optimal}.
    However, in cases where time or financial constraints limit the number of experimental units, non-asymptotic analysis is more relevant.

    \item {Regret v.s. \textbf{Estimation accuracy}}.
    The non-asymptotic literature typically examines regret rates accumulated in the experiment (looking backward) \parencite{cook2023semiparametric, wager2024causal, noarov2025stronger, neopane2025optimistic, neopane2025logarithmic}, whereas our focus is on the bias and the variance of the estimated treatment effect supporting downstream analysis (looking forward).

    \item {Without selection v.s. \textbf{With selection}}.
    Design focusing on non-asymptotic estimator accuracy typically involves no selection: either there is only one treatment \parencite{neyman1934representative}, or the variance of a pre-specified contrast is considered \parencite{blackwell2022batch}.
    However, in practice, as only one treatment out of multiple is ultimately selected to deploy, it is arguably more natural to minimize the variance of the data-dependent contrast between the selected treatment and the control.
\end{enumerate}

To fill this gap, we study the design of multi-treatment experiments aimed at non-asymptotic estimation accuracy with selection.
A key difficulty is that selection and estimation ``pull'' in opposite directions, favoring different unit allocations. 
To address this inherent conflict, we propose to use the natural mean-squared error (MSE) comparing the estimated effect of the selected treatment to the true optimal treatment's effect as our design objective.
This MSE balances the selection and the estimation, as it admits the decomposition (rigorously stated in \eqref{eq:decomposition}),
\begin{align*}
    \text{Proposed MSE} = \text{Treatment effect gap} \cdot \text{Selection error} + \text{Estimation error},
\end{align*}
and it penalizes the selection error more heavily when the treatment effect gap is large (so an incorrect selection is more consequential).
We justify our design theoretically by establishing its finite-sample guarantee (\Cref{prop:adp.alloc.convergence}) and asymptotic efficiency (\Cref{prop:adp.clt}), and empirically on simulated and real datasets compared to various common alternatives (\Cref{sec:simulation,sec:real.data}). 

\vspace{0.5cm}
\noindent\textbf{Contributions}. We summarize our main contributions below:
\begin{itemize}
    \item Our methodological contribution (\Cref{sec:method}) is a single objective that balances the selection and the estimation in multi-treatment experiments, and the derivation of the associated optimal unit allocation.
   The proposed objective naturally extends to other designs facing the conflict of selection and estimation, e.g., sequential multi-stage experiments.

    \item 
    Our theoretical contributions (\Cref{sec:theory}) showcase that the proposed procedure is decision-theoretically optimal in selection and statistically efficient in estimation.
    \begin{itemize}
        \item We analyze the statistical efficiency of our two-stage adaptive allocation procedure. We prove that it achieves the same asymptotic variance as the strategy that knows the optimal treatment in advance and applies the Neyman allocation (\Cref{prop:adp.clt}). Furthermore, we complement the asymptotic result with finite-sample guarantees, establishing nonasymptotic rates at which the data-driven allocation converges to the  Neyman allocation for the true optimal treatment and control (\Cref{prop:adp.alloc.convergence}). 
        \item We establish the minimax optimality of the winner selection rule among decision rules based on the normalized treatment-effect gap. While selecting the treatment with the largest treatment effect is minimax optimal under fixed allocations, this may fail with allocations based on the treatment-effect gap (\Cref{exam:counter.selected.winner}). Normalizing by the variance of the treatment-effect gap restores minimax optimality under a natural monotonicity condition on the signal-to-noise ratio, thereby justifying our adaptive design (\Cref{theo:optimal.rule}).
    \end{itemize}
    \end{itemize}

\vspace{0.5cm}
\noindent\textbf{Organization}.
In \Cref{sec:formulation}, we state the formulation of experiments with multiple treatments and specify the goals of selection and estimation therein.
In \Cref{sec:method}, 
we introduce the MSE that balances selection and estimation, and derive the unit allocation that optimizes it.
In \Cref{sec:theory}, we derive finite-sample and asymptotic bounds on the gap between our design and the Neyman allocation for the true optimal and control, as well as the minimax-optimality of the selection rule adopted.
In \Cref{sec:simulation,sec:real.data}, we compare our proposal with alternative designs on simulated and the STAR dataset.


\section{Problem formulation}\label{sec:formulation}

\subsection{Model}\label{sec:model}

We adopt the potential outcome framework \parencite{rubin1974estimating}.
We focus on the case with $K=3$ arms, i.e.,  two treatment levels and one control level, and our method, theoretical results directly extend to a finite number of treatment levels.
For unit~$i$, let the treatment assignment be $W_i \in \{0,1,2\}$,
where $W_i=0$ indicates assignment to control and $W_i=1$ or $2$ corresponds to treatment level~$1$ or~$2$, respectively. 
We impose the stable unit treatment value assumption (SUTVA) that the observed outcome of a unit depends only on its own treatment assignment. 
For the treatment described above, each unit~$i$ is associated with three potential outcomes $Y_i(w)$, $w \in \{0,1,2\}$,
but only the outcome corresponding to the received treatment $Y_i = Y_i(W_i)$ is observed.

To enable the definition of the optimal treatment level, we further assume that units are drawn from a super-population model, \begin{align}\label{eq:super.population}
    (Y_i(0), Y_i(1), Y_i(2)) \stackrel{\text{i.i.d.}}{\sim} \mathbb{P}, 
\end{align}
for some unknown distribution~$\mathbb{P}$. 
Without the super-population, the optimal treatment may depend on the particular realization of the finite sample, which is difficult to interpret and does not generalize to future data.
We denote by $\mu_w := \EE[Y(w)]$ the mean outcome under level~$w$, $w \in \{0,1,2\}$, where $\mu_0$ corresponds to the control group, and $\mu_1$, $\mu_2$ correspond to the two treatment levels.
We define the treatment effect as $\tau_w := \mu_w - \mu_0$, $w \in \{1,2\}$.
We regard the treatment with a larger ATE as better, and the optimal treatment is 
\begin{align}\label{eq:winner}
    w_{\max} := \argmax_{w \in \{1,2\}} \tau_w.     
\end{align}
We use $\Delta := \tau_2 - \tau_1 = \mu_2-\mu_1$ to denote the gap between the two treatment effects.

To facilitate the removal of selection bias, we further suppose that $\PP$ in \eqref{eq:super.population} is multivariate normal,
\begin{align}\label{eq:normal}
    \mathcal{N}\!\left((\mu_0, \mu_1, \mu_2)^\top,\; \mathrm{Diag}(\sigma_0^2, \sigma_1^2, \sigma_2^2)\right),
\end{align}
with unknown hyperparameters $\mu_w \in \mathbb{R}$ and $\sigma_w^2 > 0$ for $0 \leq w \leq 2$. 
Without specifying the class of super-population $\PP$, debiasing methods like the jackknife are needed, but they cannot be readily incorporated into the experiment design that requires explicit characterization of the debiased estimator's variance.

\subsection{Experiment design}

Suppose we are provided with a total budget of $T$ units. 
We use $n_0$, $n_1$, $n_2$ to denote the number of units assigned to the control group and in treatment 1 and 2, respectively.
Here $T = n_0 + n_1 + n_2$.

\subsubsection{Objective of treatment selection}\label{sec:optimal.treatment}

We use $\hat{w}_{\max} \in \{1,2\}$, a function of the observed data 
$\{(W_i, Y_i)\}_{i=1}^T$, to denote the selected treatment.
To quantitatively compare selection rules, we employ the error of selection
\begin{align}\label{eq:selection.error}
    \PP\left(\hat{w}_{\max} \neq w_{\max}\right)
    = \PP\left(\hat{w}_{\max} =1~\text{and}~\Delta>0 ~\text{or}~\hat{w}_{\max}=2~\text{and}~\Delta<0\right).
\end{align}

Given $\sigma_w^2$, no single selection rule can minimize error uniformly across all values of $\Delta$. 
In fact, the rule $\hat{w}_{\max}\equiv 1$ achieves zero error whenever $\Delta<0$, while $\hat{w}_{\max}\equiv 2$ achieves zero error whenever $\Delta>0$, and a uniformly optimal rule would thus require zero error for every $\Delta$. 
Alternatively, we adopt a minimax perspective and use the worst-case error
\begin{align}\label{eq:selection.error.minimax}
     \max_{\Delta \in \calI} 
     \PP\!\left(\hat{w}_{\max}=1 \ \text{and}\ \Delta>0 \ \text{or}\ \hat{w}_{\max}=2 \ \text{and}\ \Delta<0\right), ~\text{for some}~\calI\subseteq \mathbb{R}.
\end{align}

\subsubsection{Objective of winner's effect estimation}\label{sec:estimation}

We use $\hat{\tau}_{\hat{w}_{\max}}$ to denote the estimated effect of the selected treatment $\hat{w}_{\max}$,
which again can depend on all observed data $\{(W_i, Y_i)\}_{i=1}^T$. 
Biased estimators are misleading in downstream decision making: 
an upward-biased estimator may motivate the deployment of a treatment whose true benefit is smaller than its cost, 
while a downward-biased estimator may cause useful treatments to be overlooked.
Therefore, it is desirable for the estimator to satisfy the following unbiasedness property,
\begin{align}\label{eq:conditional.unbiasedness}        
    \mathbb{E}\left[\hat{\tau}_{\hat{w}_{\max}} - 
    \tau_{\hat{w}_{\max}} \mid \hat{w}_{\max} = w \right] = 0, \quad w \in \{1, 2\}.
\end{align}
We require the unbiasedness in the \emph{conditional} sense: given that the selected treatment level is $w$, the estimator is still guaranteed to be unbiased, as all decisions after the trial are inherently conditional on the selected treatment.
The conditional unbiasedness implies the marginal unbiasedness 
$\EE\!\left[\hat{\tau}_{\hat{w}_{\max}} - \tau_{\hat{w}_{\max}}\right] = 0$, where the latter may hold even if the effect estimator is conditionally biased for any configuration of $\hat{w}_{\max}$.

Given a conditionally unbiased estimator, the next goal is to maximize its accuracy.
We quantify the estimator precision by the expected conditional variance
\begin{align}\label{eq:variance}
\EE[\var(\hat{\tau}_{\hat{w}_{\max}} \mid \hat{w}_{\max})],
\end{align}


\subsection{Literature}\label{sec:literature}

Our work is closely related to the literature on efficient estimation of the Average Treatment Effect (ATE), defined as the expected difference in outcomes between treatments. The classical idea of optimal allocation originates from \textcite{neyman1934representative}, who proposed minimizing the variance of a difference-in-means estimator by allocating samples proportional to the standard deviations of outcomes. While Neyman allocation offers statistical efficiency, it requires prior knowledge of outcome variances, which are typically unavailable in practice.
To address this limitation, adaptive procedures have been developed to estimate variances on the fly \parencite{robbins1952some}.
\textcite{hahn2011adaptive} proposed a two-stage design for improving the efficiency of ATE estimation using an inverse probability weighting (IPW) estimator: the first stage uses uniform allocation to estimate variances, and the second stage applies an empirical approximation of Neyman allocation. They established asymptotic normality and showed the corresponding estimator achieves the semiparametric efficiency bound.
Extending this line of work, \textcite{kato2020efficient}
 introduced the adaptive AIPW estimator (A2IPW) in a complete adaptive setting, which incorporates machine learning-based estimation for non-parametric components \parencite{chernozhukov2018double}.
Their method also attains semiparametric efficiency asymptotically. \textcite{cook2023semiparametric}
 proposed the Clipped Standard-Deviation Tracking (ClipSDT) algorithm, which clips variance estimates during early exploration to improve stability. Under weaker assumptions, they replicate the efficiency and normality results of A2IPW and leverage the anytime-valid inference framework of \textcite{waudby2022anytime}. \textcite{li2024optimal} proposed a low-switching algorithm that updates the allocation in a batch setting and showed that it inherits the asymptotic efficiency and normality properties of the A2IPW estimator.

While these approaches emphasize asymptotic guarantees, their finite-sample performance remains less understood. Addressing this, \textcite{dai2023clipogd} analyzed the problem in a design-based framework \parencite{wager2024causal}, introducing the notion of Neyman regret—the variance gap between an adaptive design and the (infeasible) optimal Bernoulli design with fixed treatment probability using IPW estimator. Leveraging the convexity of the objective variance function, they proposed Clip-OGD, a clipped variant of online stochastic projected gradient descent, and proved a $\tilde{O}( \sqrt{T})$ bound on expected Neyman regret.
Subsequent work has improved upon these results. \textcite{noarov2025stronger}
 established a $\tilde{O}(\log T)$ regret bound by assuming stronger moment conditions that ensure strongly convexity of the variance function. \textcite{neopane2025logarithmic} introduced the Clipped Second Moment Tracking (ClipSMT) algorithm, which is similar in spirit to \textcite{cook2023semiparametric}, but achieves $\tilde{O}(\log T)$ Neyman regret and reduces the exponential dependence on problem parameters to a polynomial one in a superpopulation setting.
Most recently,  \textcite{neopane2025optimistic} refined the notion of Neyman regret to consider the minimum attainable variance over all estimators and allocations—not just IPW. They introduced the Optimistic Policy Tracking (OPT) algorithm, which projects a confidence set for the Neyman allocation onto the uniform design and achieves $\tilde{O}(\log T)$ regret with improved constants over prior work.
Despite this substantial progress, most adaptive Neyman strategies in multi-arm settings focus on minimizing the average variance across all pairwise treatment contrasts \parencite{blackwell2022batch}, treating each arm symmetrically. However, in many real-world applications, there is a natural asymmetry: one control arm and multiple candidate treatment arms. In such cases, the primary goal is often to estimate the effect of the best treatment relative to control, rather than assigning equal weight to all pairwise comparisons. This asymmetry is not well-addressed by existing Neyman-inspired methods, motivating new adaptive allocation strategies tailored to this setting.

In contrast, The Multi-Armed Bandit (MAB) framework~\parencite{lattimore2020bandit, auer2002finite} provides a principled approach to sequential decision-making under uncertainty, balancing exploration and exploitation to optimize a cumulative reward criterion. Classical regret minimization algorithms, such as UCB~\parencite{lai1985asymptotically, agrawal1995sample, garivier2011upper} and Thompson Sampling~\parencite{thompson1933likelihood, russo2017tutorial, agrawal2012analysis}, have been widely studied for their theoretical guarantees and empirical effectiveness. In contrast to regret minimization, the best-arm identification (BAI) setting~\parencite{kaufmann2016complexity, garivier2016optimal} seeks to efficiently identify the arm with the highest mean reward, often under fixed-budget or fixed-confidence constraints. While some recent efforts~\parencite{hadad2021confidence, zhang2020inference, zhang2021statistical} have focused on developing valid inference methods for data collected via bandit algorithms, the primary objective of these algorithms remains minimizing regret or error probability in arm selection, rather than optimizing the statistical precision of treatment effect estimation.

Another relevant line of research is about selection bias, particularly the phenomenon known as the Winner’s Curse—a selection-induced bias where estimated effects chosen for being large tend to overestimate the true effect \parencite{zollner2007overcoming}. The bias arises because extreme observed effects are more likely to be selected, even when driven by sampling noise.
Several works have proposed methods to reduce this bias. \textcite{xiao2009quantifying, xu2011bayesian} addressed the Winner’s Curse by developing bias-reduced or shrinkage-based estimators for selected effects. More broadly, the Winner’s Curse is closely linked to selective inference and post-model-selection bias, as discussed by \textcite{leeb2006conditional, leeb2008sparse}. Shrinkage and empirical Bayes approaches
 \parencite{efron2011tweedie} offer partial corrections.
Selective inference frameworks
 \parencite{lee2016exact} develop valid post-selection confidence intervals conditional on the selection event.
Beyond individual estimates, \textcite{lee2018winner} 
 study unconditional bias in large-scale A/B testing. They examine how aggregating effects from only significant experiments can inflate overall impact estimates and propose debiased estimators that adjust for this selection-induced upward bias.


\section{Experiment design}\label{sec:method}


Accurate optimal treatment selection and precise unbiased treatment effect estimation typically favor different unit allocations. 
Intuitively, accurate selection requires sufficient exploration of all treatment levels, including the suboptimal one, whereas precise estimation of the winner's treatment effect favors early exploitation of the more promising treatment (see \Cref{exam:conflict} for a concrete illustration).

To balance the winner selection and the post-selection estimation, we choose the allocation by minimizing the MSE between the estimated effect of the selected treatment and the true effect of the better treatment, that is 
$\EE\left[(\hat{\tau}_{\hat{w}_{\max}}-\tau_{w_{\max}})^2\right]$.
Notably, for a conditionally unbiased $\hat{\tau}_{\hat{w}_{\max}}$, the proposed MSE admits the decomposition
\begin{align}\label{eq:decomposition}
\underbrace{\EE\!\left[(\hat{\tau}_{\hat{w}_{\max}}-\tau_{w_{\max}})^2\right]}_{\text{Proposed MSE}}
= \Delta^2 \cdot \underbrace{\PP(\hat{w}_{\max}\neq w_{\max})}_{\text{(a) Selection error}} 
+ \underbrace{\EE[\var(\hat{\tau}_{\hat{w}_{\max}} \mid \hat{w}_{\max})]}_{\text{(b) Estimation error}}.
\end{align}
Term (a) in Eq.~\eqref{eq:decomposition} corresponds to the mis-selection rate and further weighted by the signal size $\Delta^2$; term (b) is the expected variance~\eqref{eq:variance} used to quantify the precision of the estimator. 
Thus, minimizing the proposed MSE naturally trades off selection accuracy and estimation precision (illustrated in \Cref{exam:conflict}).

\begin{wrapfigure}{r}{0.4\textwidth} 
    \centering
    \includegraphics[width=\linewidth]{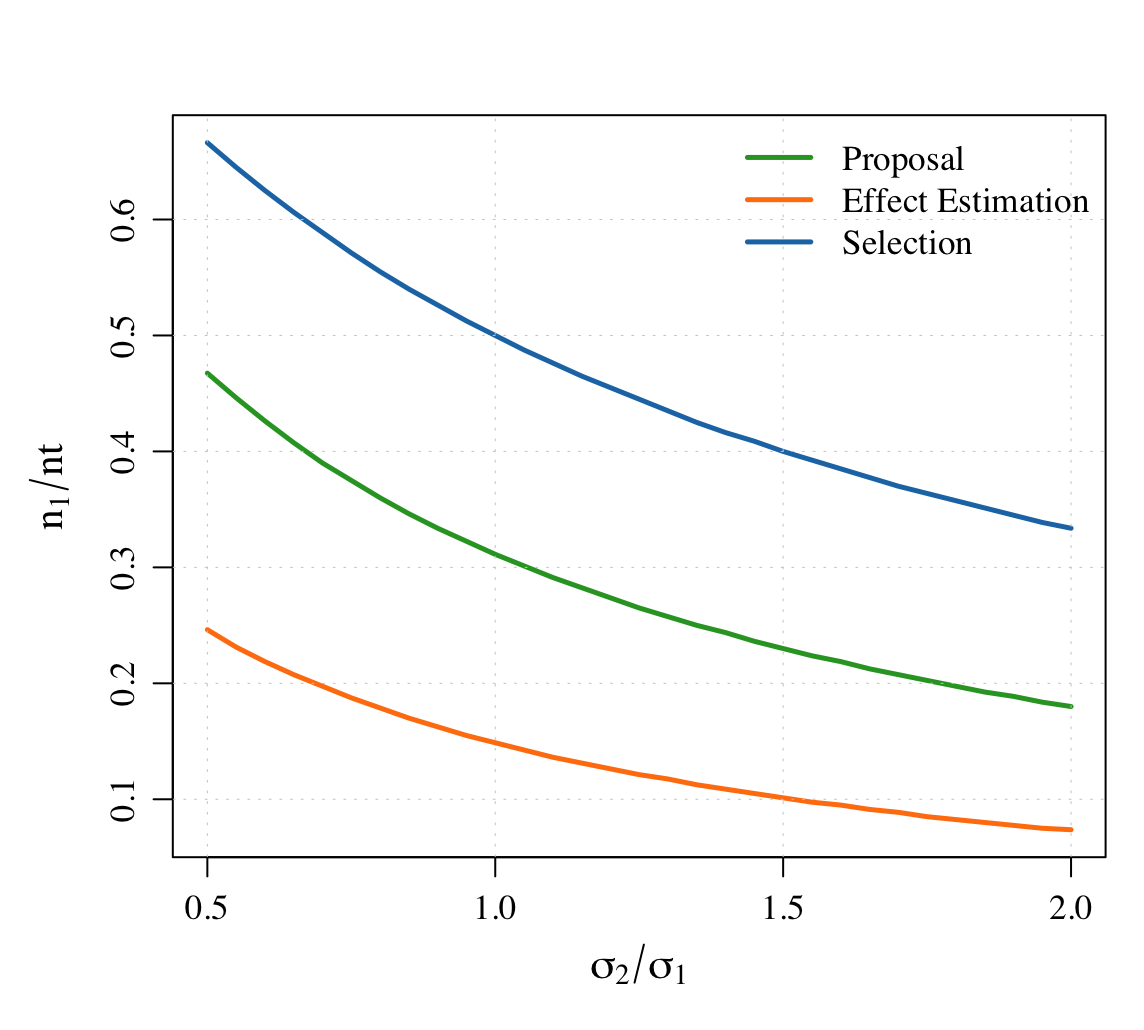} 
    \captionof{figure}{Optimal allocation fraction for treatment level 1 ($n_1/n_t$) with different objectives for $\Delta = 0.1$ (treatment 2 better) and various standard deviation ratios $\sigma_2/\sigma_1$ ($T=800$).}
    \label{fig:fig:proposal.decompose}
\end{wrapfigure}

\begin{example}[Trade-off selection and estimation]\label{exam:conflict}
As shown in \Cref{fig:fig:proposal.decompose},
when $\Delta>0$ (treatment 2 is better),
for $\hat{w}_{\max}$ in ~\eqref{eq:selected.winner} and $\hat{\tau}^{\debiased}_{\hat{w}_{\max}}$ in~\eqref{eq:tau.hat.debiased} below, the precision-optimal allocation (minimizing $\var(\hat{\tau}^{\debiased}_{\hat{w}_{\max}})$) assigns more units to the better treatment (larger $n_2$, smaller $n_1$) to increase the accuracy of the effect estimator $\hat{\tau}_{\hat{w}_{\max}}^\debiased$, the selection-optimal allocation (minimizing incorrect selection probability) is more balanced to ensure both treatments receive enough observations to estimate the sign of their difference reliably, and our MSE-based design lies between them.
\end{example}

To introduce our design optimizing the proposed MSE, we first lay out a few prerequisites.
Given a design $n_w$, $w \in \{0,1,2\}$, we define 
\begin{align*}
    \hat{\mu}_w := \frac{\sum_{W_i = w} Y_i}{n_w}, ~w \in \{0,1,2\},\quad 
    \hat{\tau}_w := \hat{\mu}_w -\hat{\mu}_0, ~w \in \{1,2\}.
\end{align*}

\noindent For winner selection, we adopt the natural rule
\begin{align}\label{eq:selected.winner}
    \hat{w}_{\max} = \argmax_{w \in \{1,2\}} \hat{\mu}_w,
\end{align}
which is proved later (in \Cref{sec:theory}) to achieve the minimax bound for the incorrect–selection probability when $n_1$, $n_2$ satisfy a natural monotonicity condition.
Under the selection rule~\eqref{eq:selected.winner}, the vanilla estimator $\hat{\tau}_{\hat{w}_{\max}}$ is upward biased, and we construct the conditionally unbiased estimator under~\eqref{eq:normal} as 
\begin{align}\label{eq:tau.hat.debiased}
  \begin{split}  \hat{\tau}^{\debiased}_{\hat{w}_{\max}}
  :=& \hat{\tau}_{\hat{w}_{\max}} - b_{\hat{w}_{\max}}, \\
  b_{\hat{w}_{\max}}
    :=& \EE\left[\hat{\tau}_{\hat{w}_{\max}}-\tau_{\hat{w}_{\max}} \mid \hat{w}_{\max}\right]
    =
    \begin{cases}
         \frac{\sigma^2_1}{n_1\sqrt{V}}\frac{\phi(\frac{\Delta}{\sqrt{V}})}{\Phi(\frac{-\Delta}{\sqrt{V}})}, & \hat{w}_{\max} = 1, \\
         \frac{\sigma^2_2}{n_2\sqrt{V}}\frac{\phi(\frac{\Delta}{\sqrt{V}})}{\Phi(\frac{\Delta}{\sqrt{V}})}, & \hat{w}_{\max}=2,
    \end{cases} 
\end{split}
\end{align}
where $\phi$, $\Phi$ denote the probability density function and cumulative distribution function of the standard normal distribution, respectively.
See \Cref{appe:derive:mse.A} for the derivation and comment on the conditional bias $b_{\hat{w}_{\max}}$.

\subsection{Oracle design}\label{sec:allocation}

We now solve the optimization problem based on our proposed MSE with $\hat w_{\max}$ in \eqref{eq:selected.winner}, $\hat\tau^{\mathrm{debiased}}_{\hat w_{\max}}$ in \eqref{eq:tau.hat.debiased} given the oracle information $\Delta$, $\sigma_w^2$.
Unlike standard multi-arm bandits, the treatment arms and the control arm play \emph{asymmetric} roles in our problem: selection is performed only among the treatment levels and does not involve the control. 
This inspires a second decomposition of the proposed MSE,
\begin{align}\label{eq:decomposition.2}
\underbrace{\EE\!\left[(\hat{\tau}^{\debiased}_{\hat{w}_{\max}}-\tau_{w_{\max}})^2\right]}_{\text{Proposed MSE}}
= \underbrace{\EE\!\left[(\hat{\mu}^{\debiased}_{\hat{w}_{\max}}-\mu_{w_{\max}})^2\right]}_{\text{(t) Treatment}}
\;+\;
\underbrace{\operatorname{Var}(\hat{\mu}_0)}_{\text{(c) Control}},
\end{align}
The treatment component (t) is more complex involving selection and estimation, while the control component (c) is simpler and only involves estimation.
The decomposition~\eqref{eq:decomposition.2} motivates us to handle treatment and control allocations separately and solve the optimization problem in two hierarchies.
\begin{enumerate}
    \item {Within-treatment allocation.} 
    Given a total treatment budget \(n_t\) to be split across the two treatment levels, choose \(n_1,n_2\) with \(n_1+n_2=n_t\) to minimize term (t) in Eq.~\eqref{eq:decomposition.2}. Denote the optimal solution by $(n_1(\Delta, n_t), n_2(\Delta, n_t))$, and the corresponding term (t) by $\text{MSE}_t(n_t)$.
    
    \item {Treatment vs.\ control split.}
    Given the total sample size \(T\), choose \(n_t\) and \(n_0\) with \(n_t+n_0=T\) to minimize the proposed $\text{MSE}=\text{MSE}_t(n_t)+{\sigma_0^2}/{n_0}$, where $\text{MSE}_t(n_t)$ is obtained from Step 1.
\end{enumerate}

\begin{figure}[tbp]
\centering
 \begin{minipage}{0.48\textwidth}
    \centering
    \includegraphics[width = 1\textwidth]{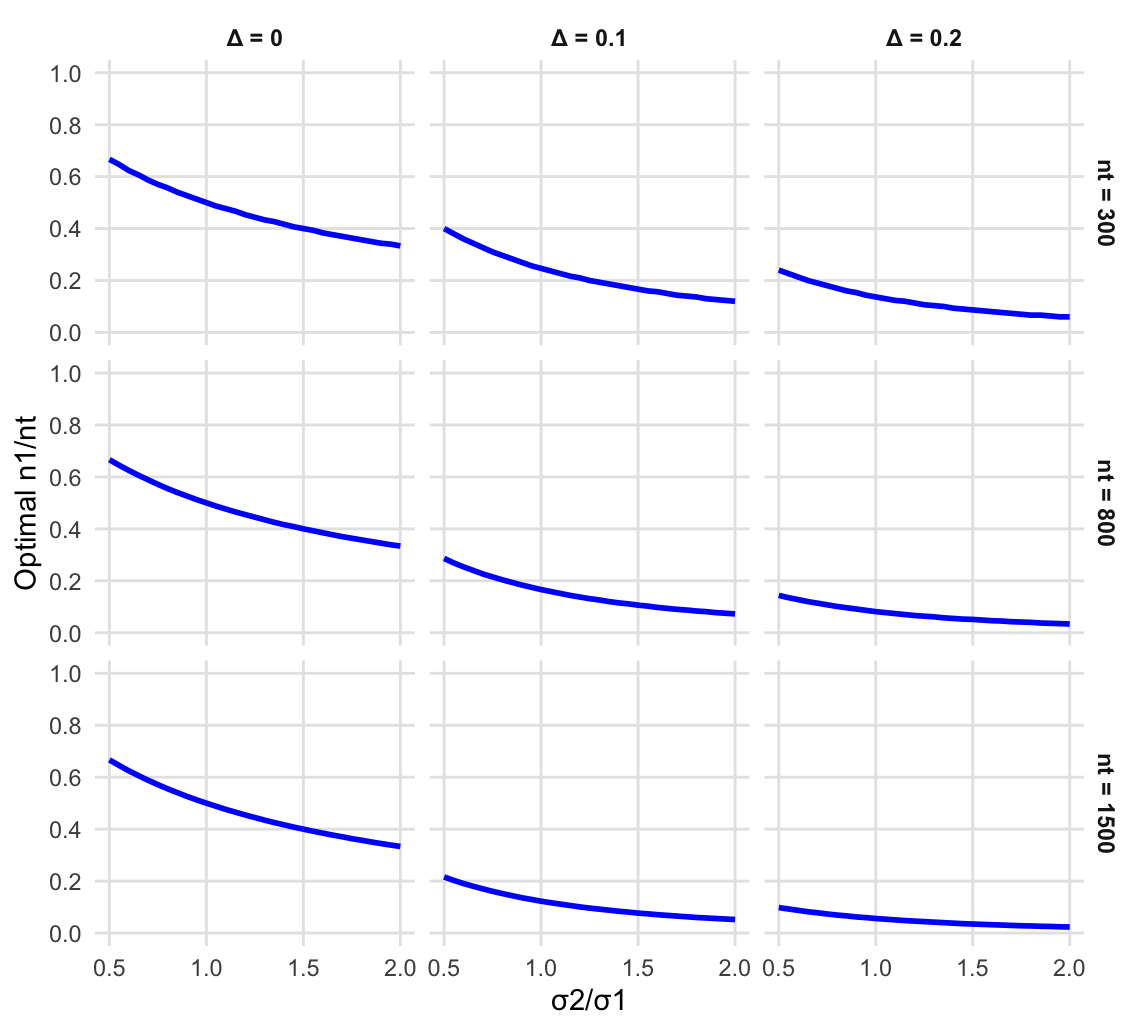}
  \end{minipage}
  \begin{minipage}{0.48\textwidth}
    \centering
    \includegraphics[width = 1\textwidth]{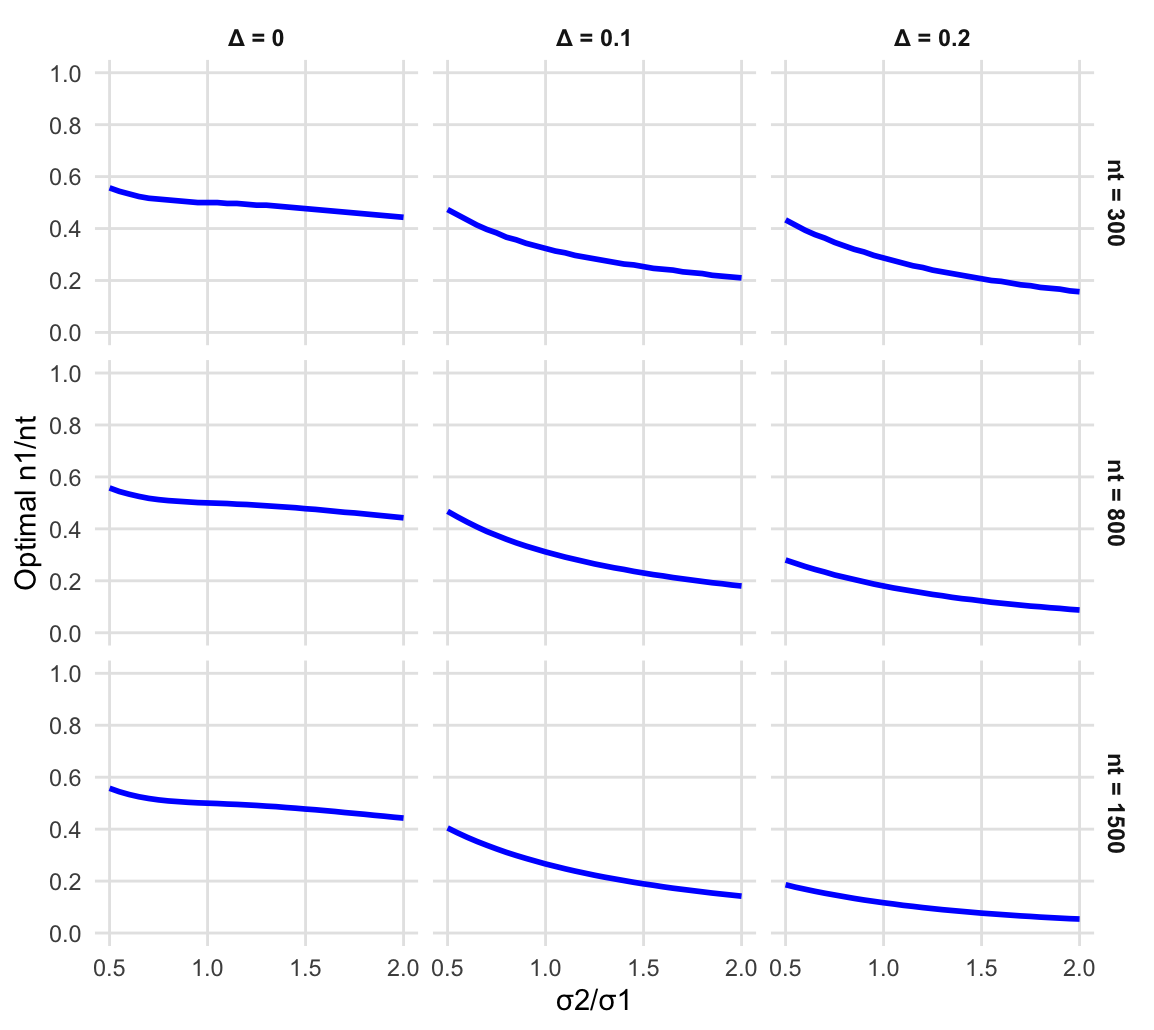}
  \end{minipage}
  \caption{Dependence of optimal allocation ${n_1}/{n_t}$ which minimize the MSE of the estimator without debias (left panel) and debiased estimator (right panel) on $\Delta$  and $\sigma_2/\sigma_1$. 
}\label{fig:op1}
\end{figure}

For Step 2, since $\mathrm{MSE}_t(n_t)$ typically has no closed-form, we optimize the treatment–control split via a grid search over $n_t$.
For Step~1, term (t) takes the form for  our debiased estimator~\eqref{eq:tau.hat.debiased}, 
\begin{align}\label{eq:n1.oracle}
    \begin{split}   
    \text{MSE}_t(n_t)
    &= \frac{\sigma^2_2}{n_2(\Delta, n_t)}\Phi(\frac{\Delta}{\sqrt{V}})- \frac{\sigma^4_2}{n_2^2(\Delta, n_t)\cdot V}\phi(\frac{\Delta}{\sqrt{V}})\cdot\frac{\Delta}{\sqrt{V}} - \frac{\sigma^4_2}{n_2^2(\Delta, n_t)\cdot V}\frac{\phi^2(\frac{\Delta}{\sqrt{V}})}{\Phi(\frac{\Delta}{\sqrt{V}})} \\
    &\quad~+ \frac{\sigma^2_1}{n_1(\Delta, n_t)}\Phi(\frac{-\Delta}{\sqrt{V}})+ \frac{\sigma^4_1}{n_1^2(\Delta, n_t)\cdot V}\phi(\frac{\Delta}{\sqrt{V}})\cdot\frac{\Delta}{\sqrt{V}} - \frac{\sigma^4_1}{n_1^2(\Delta, n_t)\cdot V}\frac{\phi^2(\frac{\Delta}{\sqrt{V}})}{\Phi(\frac{-\Delta}{\sqrt{V}})}\\
    &\quad~+
    \1\left\{\Delta>0\right\}\cdot\Delta^2\Phi(\frac{-\Delta}{\sqrt{V}}) + \1\left\{\Delta<0\right\}\cdot\Delta^2\Phi(\frac{\Delta}{\sqrt{V}}),
    \end{split}
\end{align}
where $V=\sigma_1^2/n_1(\Delta, n_t)+\sigma_2^2/n_2(\Delta, n_t)$, $n_1(\Delta, n_t)+n_2(\Delta, n_t)=n_t$.
The complete derivation of Eq.~\eqref{eq:n1.oracle} is given in \Cref{appe:derive:mse.A}. 
A closed-form expression for the minimizer $n_1(\Delta,n_t)$ is generally unavailable, so we examine its dependence on $\Delta$ and the standard-deviation ratio $\sigma_2/\sigma_1$ numerically. 
As shown in \Cref{fig:op1} right panel, the optimal share $n_1/n_t$ decreases as $\sigma_2/\sigma_1$ increases, reflecting a Neyman-type tendency to allocate more units to the higher-variance treatment. 
Compared with the optimal design based on the biased estimator $\hat{\tau}_{\hat{w}_{\max}}$, our debiased counterpart is less sensitive to $\Delta$ and $\sigma_2/\sigma_1$.


\subsection{Adaptive design}\label{sec:adaptive.method}

Both the debiasing step and the allocation depend on unknown parameters in~\eqref{eq:normal}, so we adopt an adaptive analogue of the oracle.
\begin{enumerate}
    \item {Pilot stage.} Collect a small pilot of size $T_0$ to estimate the nuisance parameters, then compute the oracle plan for the full budget $T$ by plugging in these estimates.
    \item {Post-pilot stage.} Implement the plan by assigning the remaining $T_1:=T-T_0$ units so that the realized totals match the $T$-budget targets.
\end{enumerate}

Particularly, in the pilot stage,  \( m_0 \) units are assigned to control group, \( m_1 \) units are assigned to treatment 1 and \( m_2  \) to treatment 2, with prefixed \( m_0 \), \( m_1 \), \( m_2 \) (we can set \(m_0 =  m_1 = m_2 \) when no prior information is available).
Let $\hat{\mu}_w^{(0)} := {\sum_{W_j=w, j \in \text{pilot}} Y_j}/{m_w}$, $w \in \{1,2\}$, and estimate $\Delta$, \( \sigma_0^2 \),\( \sigma_1^2 \), and \( \sigma_2^2 \) by
\begin{align*}
    \hat{\Delta} 
     = \hat{\mu}_2^{(0)} - \hat{\mu}_1^{(0)},  \quad \hat{\sigma}_w^2
    = \frac{\sum_{W_j=w} \left(Y_j - \hat{\mu}_w^{(0)}\right)^2}{m_w-1}, \quad w \in \{0,1,2\}. 
\end{align*}

After the pilot stage, we proceed to the post-pilot stage with a budget of \( T_1 := T - T_0 \) units. 
To determine the optimal \( n_1 \), since our two-stage procedure is equivalent to that applied to a single experiment aggregating both stages with estimated hyperparameters, we apply Eq.~\eqref{eq:n1.oracle} with estimated hyperparameters from the pilot stage to identify the optimal total allocation \( m_1 + n_1 \) to treatment level 1, and compute \( n_1 = n_1(\hat{\Delta}, \hat{\sigma}_1^2, \hat{\sigma}_2^2, T, T_0)\) by further subtracting \( m_1 \).
Specifically,  we find the optimal $(n_0, n_1, n_2)$ which minimizes
\begin{align}\label{eq:MSE.estimate}
\begin{split}
\hat{E} &= \frac{\hat{\sigma}^2_2}{n_2+m_2}\Phi\left(\frac{\hat{\Delta}}{\sqrt{\hat{V}}}\right)- \frac{\hat{\sigma}^4_2}{(n_2+m_2)^2\cdot \hat{V}}\phi\left(\frac{\hat{\Delta}}{\sqrt{\hat{V}}}\right)\cdot\frac{\hat{\Delta}}{\sqrt{\hat{V}}} - \frac{\hat{\sigma}^4_2}{(n_2+m_2)^2\cdot \hat{V}}\frac{\phi^2\left(\frac{\hat{\Delta}}{\sqrt{\hat{V}}}\right)}{\Phi(\frac{\hat{\Delta}}{\sqrt{\hat{V}}})} \\
&\quad~+ (\frac{\hat{\sigma}^2_1}{n_1+m_1})\Phi\left(\frac{-\hat{\Delta}}{\sqrt{\hat{V}}}\right)+ \frac{\hat{\sigma}^4_1}{(n_1+m_1)^2\cdot \hat{V}}\phi\left(\frac{\hat{\Delta}}{\sqrt{\hat{V}}}\right)\cdot\frac{\hat{\Delta}}{\sqrt{\hat{V}}} - \frac{\hat{\sigma}^4_1}{(n_1+m_1)^2\cdot \hat{V}}\frac{\phi^2\left(\frac{\hat{\Delta}}{\sqrt{\hat{V}}}\right)}{\Phi\left(\frac{-\hat{\Delta}}{\sqrt{\hat{V}}}\right)} \\
&\quad~+
    \1\left\{\hat{\Delta}>0\right\}\cdot\hat{\Delta}^2\Phi(\frac{-\hat{\Delta}}{\sqrt{\hat{V}}}) + \1\left\{\hat{\Delta}<0\right\}\cdot\hat{\Delta}^2\Phi(\frac{\hat{\Delta}}{\sqrt{\hat{V}}})+ \frac{\hat{\sigma}^2_0}{n_0+m_0}
\end{split}
\end{align}
on $\mathcal{N}_1 = \{\left(n_0, n_1, n_2\right) \in [ 1, T_1-2 ]^3:  n_0 + n_1 + n_2 = T_1 \}$.

After the post-pilot stage, we compute 
$\hat{\mu}^{(1)}_w := {\sum_{W_i = w, i \in \text{post-pilot}} Y_i}/{n_w}$, for $w \in \{1, 2\}$,
We combine the data from the pilot stage and the post-pilot stage to estimate $\mu_w$ as
\[
\hat{\mu}_w
:= \frac{\sum_{j \in \text{pilot}, W_j = w} Y_j + \sum_{i \in \text{post-pilot}, W_i = w} Y_i}{m_w + n_w}
= \omega \hat{\mu}^{(0)}_w + (1 - \omega) \hat{\mu}^{(1)}_w, \quad \omega = \frac{m_w}{m_w + n_w}, ~w \in \{1, 2\}.
\]
We select the winner by
\[
\hat{w}_{\max} := \arg\max_{w \in \{1,2\}} \left\{ \hat{\mu}_w\right\},
\]
and break ties randomly.
We estimate the bias in \eqref{eq:tau.hat.debiased} by replacing the hyperparameters with their estimates from the pilot stage, and construct the debiased estimator accordingly. This design is formally presented in \Cref{alg:two-stage}.

\begin{algorithm}[t]
\caption{Two-Stage Adaptive Design}
\label{alg:two-stage}
\begin{algorithmic}[1]
\STATE \textbf{Input:} Total budget $T$, pilot stage budget $T_0$, initial allocation $(m_0,m_1,m_2)$
\STATE \textbf{Pilot stage (Hyperparameter Estimation)}  
\STATE \quad Allocate $m_w$ units to each arm $w \in \{0,1,2\}$ and compute $\hat{\mu}_w^{(0)}$, $\hat{\sigma}_w^2$, $\hat{\Delta}$
\STATE \textbf{Post-pilot stage (Optimal Allocation)}  
\STATE \quad Set $T_1 = T - T_0$
\STATE \quad Find $(n_0^*,n_1^*,n_2^*) = \arg\min_{(n_0,n_1,n_2) \in \mathcal{N}_1} \hat{E}$ for the objective $\hat{E}$ in \eqref{eq:MSE.estimate}.
\STATE \quad Allocate $n_w^*$ units to each arm $w \in \{0,1,2\}$ and compute $\hat{\mu}_w^{(1)}$
\STATE \textbf{Combine Estimates and Select Winner}  
\STATE \quad $\hat{\mu}_w = \omega \hat{\mu}_w^{(0)} + (1-\omega)\hat{\mu}_w^{(1)}$, \quad $\omega = \tfrac{m_w}{m_w+n_w^*}$
\STATE \quad $\hat{w}_{\max} = \arg\max_{w \in \{1,2\}} \hat{\mu}_w$
\STATE \textbf{Output:} Debiased estimator $\hat{\tau}^{\text{debiased}}_{\hat{w}_{\max}} = \hat{\tau}_{\hat{w}_{\max}} - \hat{b}_{\hat{w}_{\max}}$

\end{algorithmic}
\end{algorithm}

The choice of $T_0$ is formalized in \Cref{sec:theory}, and we assess performance via numerical experiments (\Cref{fig:T_0.choose,fig:two_stage_50,fig:two_stage_100,fig:two_stage_300}).
We remark that we optimize for the full budget rather than only the post-pilot stage, keeping the adaptive design aligned with its oracle counterpart (see \Cref{exam:optimize.full.sample}).

\section{Theory}\label{sec:theory}

In \Cref{sec:asymptotic}, we show that our design is asymptotically equivalent to the Neyman allocation for the true optimal treatment and control.
In \Cref{sec:nonasymptotic}, we provide the associated non-asymptotic bound.
In \Cref{sec:selection.minimax}, we display the minimax-optimality of the selection rule~\eqref{eq:selected.winner}.

\subsection{Asymptotic result}\label{sec:asymptotic}
\begin{theorem}\label{prop:adp.clt}
For any fixed $T,T_0,T_1,m_0,m_1,m_2$ in the two batch design, let $n_0,n_1,n_2$ be the numbers of allocations for each arm in the second batch and $p_{0,T} = n_0/T_1$, $p_{1,T} = n_1/T_1$, $p_{2,T} = n_2/T_1$ be the allocation proportions for each arm in the second batch. $\hat{p}_{T} := \left(\hat{p}_{0,T}, \hat{p}_{1,T}, \hat{p}_{2,T}\right)$ be the optimal allocation proportions for each arm in the second batch that minimizes objective function~\eqref{eq:MSE.estimate} on the domain $\mathcal{P}_{T} = \{(p_0, p_1, p_2) \in [\delta_T, 1-2\cdot\delta_T]^3 : p_0 + p_1 + p_2 = 1\}$ for adaptive two-stage
algorithm, where $\delta_T= \frac{1}{2}t^{-\alpha}$ and $\alpha \in (0, \frac{1}{2})$. Suppose further $T_0 = O(\sqrt{T})$. Then as $T \to \infty$,
\begin{align}\label{prop:eq:adp.clt}
\begin{split}
\sqrt{T}\left(\hat{\tau}_{\hat{w}_{\max}} - \tau_{w_{\max}}\right) \xrightarrow{d} \mathcal{N}\left(0, (\sigma_0 +\sigma_{w_{\text{max}}})^2 \right)
\end{split}
\end{align}
\end{theorem}

\Cref{prop:adp.clt} shows that our adaptive design attains the same asymptotic variance for estimating the winner's effect as the optimal strategy that knows the optimal treatment a priori and applies the Neyman allocation (i.e., the optimal allocation once the winner is known).

The proof of \Cref{prop:adp.clt} can be decomposed into two parts: the convergence of the allocation proportion 
\[
\hat{p}_T \xrightarrow{P} p^* = (p_0^*, p^*_{w_{\text{sub}}}, p^*_{w_{\text{max}}}) 
= \left(\frac{\sigma_0}{\sigma_0 + \sigma_{w_{\text{max}}}}, \; 0, \; 
\frac{\sigma_{w_{\text{max}}}}{\sigma_0 + \sigma_{w_{\text{max}}}}\right),
\]  
and the asymptotic normality of the sample mean estimator. The first part builds upon the analysis in Theorem~5.7 of~\textcite{van2000asymptotic}, relying on a uniform convergence argument to justify interchanging the $\arg\min$ and the limit of the objective function.
To accommodate the case where the limit may lie at the boundary, we establish uniform convergence over expanding sets approaching the boundary as the sample size increases. To establish the $\sqrt{T_0}$-consistency of the stage-0 sample mean and variance, 
we impose the condition that the fourth moments of the potential outcomes are finite.
The finite sample guarantee is given in \Cref{sec:nonasymptotic}. For the asymptotic normality of the sample mean estimator, we apply the central limit theorem for martingale difference sequences. In particular, we invoke the central limit theorem as stated in \parencite{dvoretzky1972asymptotic, zhang2020inference, cook2023semiparametric}. \textcite{zhang2020inference} assumes a condition slightly stronger than the existence of finite second moments and uses it to show that the average of the terms converges to zero by bounding the largest term to verify a Lindeberg-type condition. In contrast, we directly establish that the average converges to zero while assuming only finite second moments.
Details are given in \Cref{sec:proof:asymptotic}.

\subsection{Nonasymptotic result}\label{sec:nonasymptotic}

\begin{proposition}\label{prop:adp.alloc.convergence}
For any two batch design with $T_0 = T^{\beta}$ with $\beta \in (0,1)$, let $n_0,n_1,n_2$ be the numbers of allocations for each arm in the second batch and $p_{0,T} = n_0/T_1$, $p_{1,T} = n_1/T_1$, $p_{2,T} = n_2/T_1$ be the allocation proportions for each arm in the second batch. Let $\hat{p}_{T} := \left(\hat{p}_{0,T}, \hat{p}_{1,T}, \hat{p}_{2,T}\right)$ be the optimal allocation proportions that minimize objective function~\eqref{eq:MSE.estimate} on the domain $\mathcal{P}_{T} = \{(p_0, p_1, p_2) \in [\delta_T, 1-2\cdot\delta_T]^3 : p_0 + p_1 + p_2 = 1\}$ for adaptive two-stage
algorithm, where $\delta_T= \frac{1}{2}t^{-\alpha}$ and $\alpha \in (0, \frac{1}{2})$. Then with probability at least $1-\varepsilon$, for all $T  \ge \tau =O((\frac{1}{\varepsilon})^{\frac{1}{\beta}})$, we have
\begin{align}
\|\hat{p}_T - p^*\|_2 = O(T^{\min \left\{-(1/2-\alpha)\beta, -\alpha\beta, (2\alpha+1)\beta-1\right\}/2})
\end{align}
where $p^* = (p_0^*, p^*_{w_{\text{sub}}},p^*_{w_{\text{max}}}) = (\frac{\sigma_0}{\sigma_0 + \sigma_{w_{\text{max}}}},0,\frac{\sigma_{w_{\text{max}}}}{\sigma_0 + \sigma_{w_{\text{max}}}})$
which is the Neyman allocation between the optimal treatment arm and the control arm, with zero allocation to the sub-optimal treatment arm, minimizes the variance of the average treatment effect associated with the optimal treatment.
Setting $\alpha = \frac{1}{4}$ and $\beta = \frac{4}{7}$, we have $(2\alpha+1)\beta-1 = -(1/2-\alpha)\beta= -\alpha\beta = -1/7$, namely
\begin{align*}
\|\hat{p}_T - p^*\|_2 = O(T^{-1/14})
\end{align*}
\end{proposition}
To obtain finite-sample guarantees, we impose the condition that the fourth moments of the potential outcomes are finite, which yields $\tau = O\!\left((1/\varepsilon)^{1/\beta}\right)$. Stronger assumptions, such as sub-Gaussian tails or bounded outcomes, further improve the rate to $\tau = O\!\left((\log(1/\varepsilon))^{1/\beta}\right)$.
To prove \Cref{prop:adp.alloc.convergence},
we multiply the objective function~\eqref{eq:MSE.estimate} by $T_1$ and denote it as $\mathcal{R}_{T}$ to emphasis its dependence on $T$. For any fixed $T$, the optimal allocation proportions that minimize objective function~\eqref{eq:MSE.estimate}  is same as the one minimizes $\mathcal{R}_{T}$. Hence it's sufficient to work on the function $\mathcal{R}_{T}$. $\mathcal{R}_{T}$ can be written as a function of $p_{0,T}=n_0/T, p_{1,T} = n1/T,p_{2,T} = n_2/T$. For abbreviation, we drop the subscript $T$ in $p$. 
 Define 
\begin{align*}
\mathcal{R}(p_0, p_1, p_2) = \frac{\sigma^2_{w_{\text{max}}}}{p_{w_{\text{max}}}}+ \frac{\sigma^2_0}{p_0} 
\end{align*}
Which is the Neyman variance for the ATE estimation of associated with the optimal-treatment arm. Let $p^*_T$ be the optimal solution for $\mathcal{R}(p)$ solved on $\mathcal{P}_T$,
then we have the following decomposition
\begin{align*}
\begin{split}
\mathcal{R}(\hat{p}_T) - \mathcal{R}(p^*)
&  =\mathcal{R}(\hat{p}_T)  - \mathcal{R}_T(p^*_T) +  \mathcal{R}_T(p^*_T) - R(p^*_T)+ R(p^*_T)- R(p^*) \\
& \leq  \mathcal{R}(\hat{p}_T)  - \mathcal{R}_T(\hat{p}_T) +  \mathcal{R}_T(p^*_T) - R(p^*_T)+ R(p^*_T)- R(p^*)\\
& \leq \sup_{p \in \mathcal{P}_{T}} \mid \mathcal{R}_{T}(p) - \mathcal{R}(p)\mid + \sup_{p \in \mathcal{P}_{T}} \mid \mathcal{R}_{T}(p) - \mathcal{R}(p)\mid + \big|\mathcal{R}(p^*_T) - \mathcal{R}(p^*)\big| \\ 
& = \underbrace{2\sup_{p \in \mathcal{P}_{T}} \mid \mathcal{R}_{T}(p) - \mathcal{R}(p)\mid}_{O(T^{\min \left\{-(1/2-\alpha)\beta,  (2\alpha+1)\beta-1\right\}})} + \underbrace{\big|\mathcal{R}(p^*_T) - \mathcal{R}(p^*)\big|}_{O(T^{-\alpha\beta})}
\end{split}
\end{align*}
To guarantee uniform convergence over $\mathcal{P}_{T}$, $\alpha$ should be chosen small so that the set does not expand too rapidly. Conversely, bounding $\big|\mathcal{R}(p^*_{T}) - \mathcal{R}(p^*)\big|$ requires that $p^*_{T}$ converges to $p^*$ at a sufficiently fast rate, which favors a larger value of $\alpha$. Thus, $\alpha$ must balance these competing considerations. Similarly, a large $T_0$ and, correspondingly, a large $\beta$ improve the $O_p(T_0^{-1/2})$ rate of the stage0 sample estimator. However, a large $\beta$ also increases the discrepancy between the overall mean squared error $R_T$ and the Neyman variance $R$, potentially leading to the unboundedness of $\mathcal{R}_{T}(p) - \mathcal{R}(p)$. Hence, the choice of $\beta$ also involves a trade-off.
Lastly, for any $\delta \in (0,1)$ such that 
$\mathcal{R}(\hat{p}_T) - \mathcal{R}(p^*) < \delta$, we have $\|\hat{p}_T - p^* \|_2 = O(\sqrt{\delta})$.
The rate might be sharpened by improving the uniform convergence result used in the proof. 
Details are provided in \Cref{sec:proof:nonasymptotic}.

\subsection{Minimax-optimality of selection rule}\label{sec:selection.minimax}

When $n_1$ and $n_2$ do not depend on $\Delta$, the selection rule~\eqref{eq:selected.winner} is minimax optimal among all rules based on $\hat{\Delta} := \hat{\mu}_2 - \hat{\mu}_1$ for any $\calI$ in~\eqref{eq:selection.error.minimax} symmetric around zero\footnote{
This follows from Theorem~1 of \cite{Karlin1956decision} and the fact that $\hat{\Delta} \sim \mathcal{N}(\Delta, {\sigma_1^2}/{n_1}+{\sigma_2^2}/{n_2})$ admits the monotone likelihood ratio property in $\Delta$.
}.
However, when $n_1(\Delta)$ and $n_2(\Delta)$ vary with $\Delta$, the variance of $\hat{\Delta}$, denoted by $V(\Delta):={\sigma_1^2}/{n_1(\Delta)}+{\sigma_2^2}/{n_2(\Delta)}$, changes with $\Delta$, and the optimality of~\eqref{eq:selected.winner} is no longer guaranteed (see \Cref{exam:counter.selected.winner} for a counter example).
To account for the varying variance of $\hat{\Delta}$, it is natural to base the selection on the normalized counterpart
$\hat{\Delta}^N := {\hat{\Delta}}/{\sqrt{V(\Delta)}}$.
Fortunately, selection rule~\eqref{eq:selected.winner} remains minimax-optimal among rules based on $\hat{\Delta}^N$ under a natural assumption on $n_1(\Delta)$, $n_2(\Delta)$.

\begin{theorem}[Optimality of the winner selection strategy]\label{theo:optimal.rule}
For any $V(\Delta)$ such that the normalized difference $r(\Delta):={\Delta}/{\sqrt{V(\Delta)}}$ increases in $\Delta$ for any $\Delta \in \RR$, and any $\calI$ such that $\{r(\Delta): \Delta\in\calI\}$ is symmetric around zero, the selection rule~\eqref{eq:selected.winner} is minimax optimal among all rules based on $\hat{\Delta}^N$.
\end{theorem}
The proof of \Cref{theo:optimal.rule} is provided in~\Cref{appe:proof:optimal.rule}. 
Note that $r(\Delta)$ represents the signal-to-noise ratio in the winner selection task, and reasonable choices of $n_1(\Delta)$ and $n_2(\Delta)$ should ensure that $r(\Delta)$ increases with $\Delta$, i.e., the monotone assumption in \Cref{theo:optimal.rule}. 
In particular, the monotonicity is satisfied when $n_1(\Delta)$ and $n_2(\Delta)$ do not depend on $\Delta$, and also by our proposal in \Cref{sec:method} below (validated numerically in \Cref{fig:deri0}).

\section{Simulation}\label{sec:simulation}

\subsection{Data generation mechanism}
We simulate the experiment from $(Y(0), Y(1), Y(2)) \sim \calN\left((\mu_0, \mu_1, \mu_2      )^\top, \text{Diag}(\sigma^2_0, \sigma_1^2, \sigma_2^2)\right)$. without loss of generality, we set $\mu_0 = \mu_1 = 0$ and $\mu_2 = \Delta+\mu_1 = \Delta$. We simulate 2 effect sizes: $\Delta = 0.1, 0.15$, fix $\sigma^2_1 + \sigma^2_2= 1$, $\sigma^2_0 = 1$ and consider two instances $\frac{\sigma_2}{\sigma_1} = 0.8, 1.25$. For  each of these problems, we run the experiment for $T:=T_0+T_1$ ranging from 300 to 1500 over 500,000 simulations and set $T_0 = T/3$.

\subsection{Method for comparison}
We compare five designs including our proposal and the oracle counterpart. 
All methods adopt the winner selection rule~\eqref{eq:selected.winner} and the treatment effect estimator $\hat{\tau}^{\debiased}_{\hat{w}_{\max}}$ or $\hat{\tau}_{\hat{w}_{\max}}$.
\begin{itemize}
       \item \textbf{Non-adaptive method}. This method adopts the equal allocation \( n_0 =n_1 = n_2 = T/3 \) without adapting to the data, and includes the bias correction in the treatment effect estimation.

       \item \textbf{Sample splitting for selection and estimation (SS.SE)}. 
    This method adopts a two-stage design, uses a pilot stage for  winner selection.
    In the post-pilot stage, the method only samples from the selected winner level and estimates the corresponding mean using the data in the post-pilot stage.
    Since the selection and the final estimation are done on different folds, no bias correction is needed.

   \item \textbf{Sample splitting for hyperparameter estimation (SS.Hyper)}. 
    This method adopts a two-stage design. The pilot stage is purely used for hyperparameter estimation. In the post-pilot stage, we select the winner and estimate its effect; unit allocation is chosen to minimize the MSE of the debiased estimator based on the post-pilot data only.

    \item \textbf{Proposal without bias correction}.
    This method is similar to our proposal but with no bias correction. 

    \item \textbf{Proposal}. We implement \Cref{alg:two-stage}. 
    
    \item \textbf{Oracle}. The proposed design with true hyperparameters.

\end{itemize}

\begin{figure}[htbp]
    \centering

    \setlength{\tabcolsep}{2pt}
    \renewcommand{\arraystretch}{0.8}
    \begin{tabular}{cccc}
        \begin{tabular}{c}
            \includegraphics[width=0.20\textwidth]{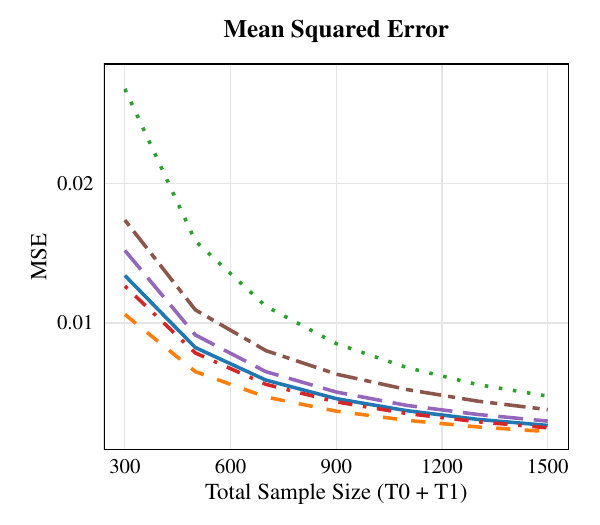} \\[-0.35em]
            \includegraphics[width=0.20\textwidth]{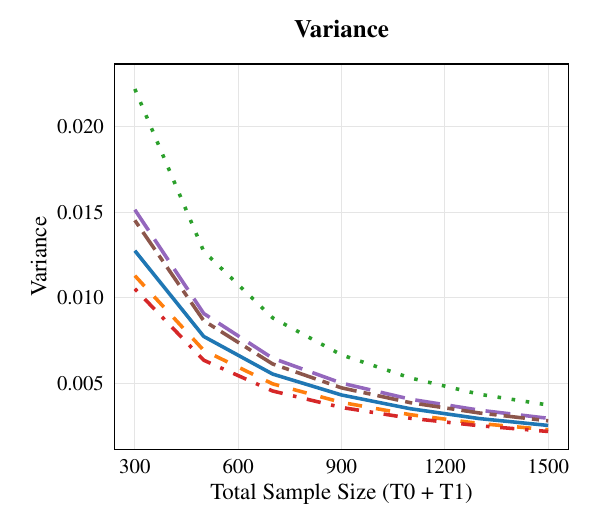} \\[-0.35em]
            \includegraphics[width=0.20\textwidth]{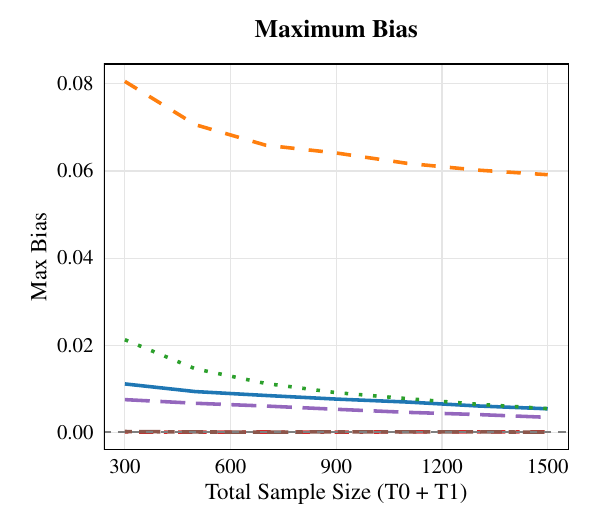} \\[-0.35em]
            \includegraphics[width=0.20\textwidth]{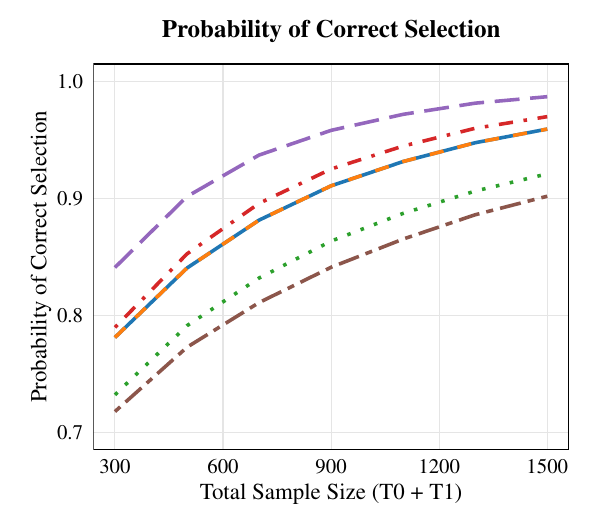} \\
            \subcaptionbox{$\Delta=0.1$, $\tfrac{\sigma_2}{\sigma_1}=0.8$}[0.21\textwidth]{}
        \end{tabular}
        &
        \begin{tabular}{c}
            \includegraphics[width=0.20\textwidth]{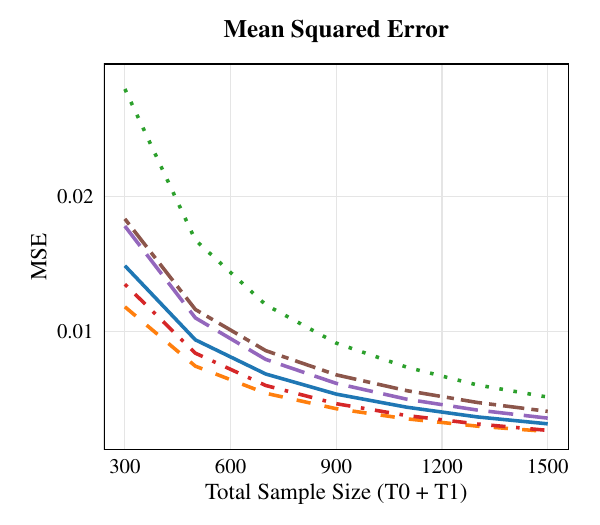} \\[-0.35em]
            \includegraphics[width=0.20\textwidth]{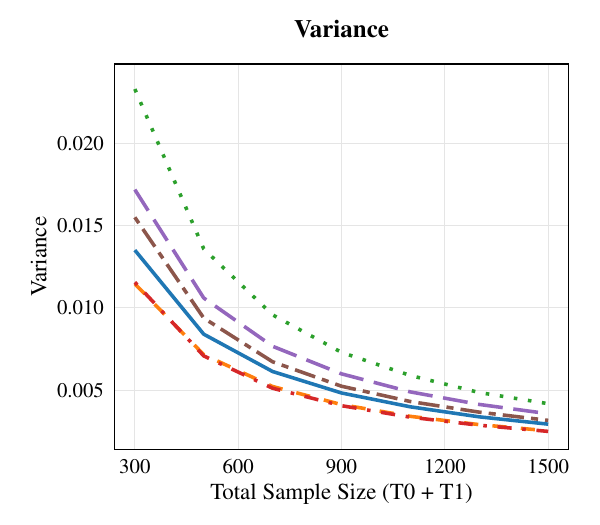} \\[-0.35em]
            \includegraphics[width=0.20\textwidth]{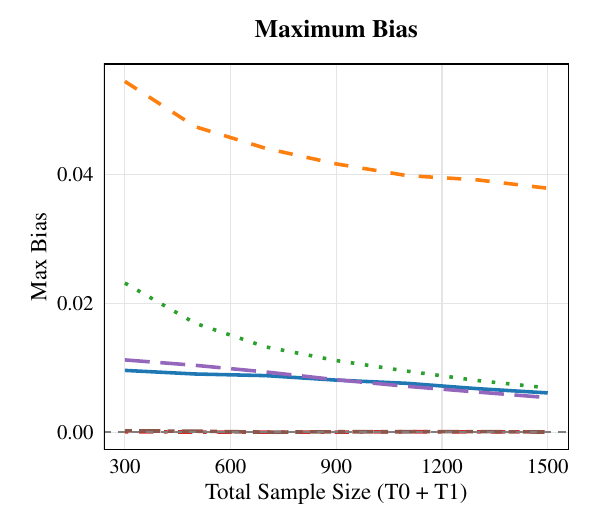} \\[-0.35em]
            \includegraphics[width=0.20\textwidth]{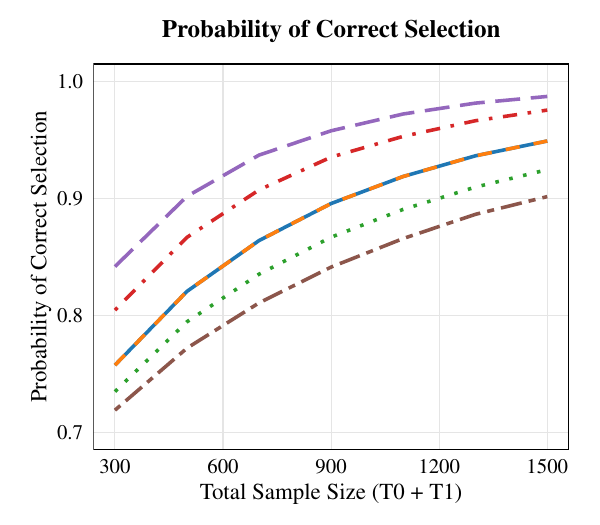} \\
            \subcaptionbox{$\Delta=0.1$, $\tfrac{\sigma_2}{\sigma_1}=1.25$}[0.21\textwidth]{}
        \end{tabular}
        &
        \begin{tabular}{c}
            \includegraphics[width=0.20\textwidth]{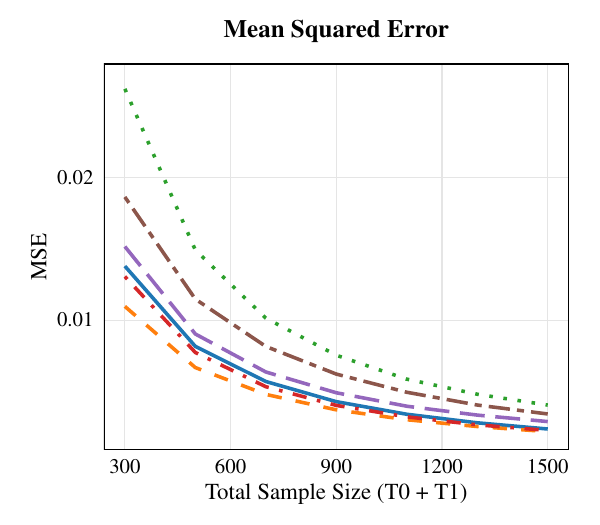} \\[-0.35em]
            \includegraphics[width=0.20\textwidth]{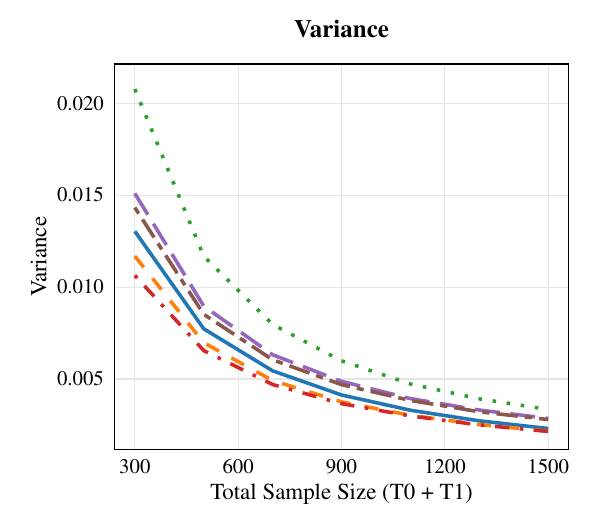} \\[-0.35em]
            \includegraphics[width=0.20\textwidth]{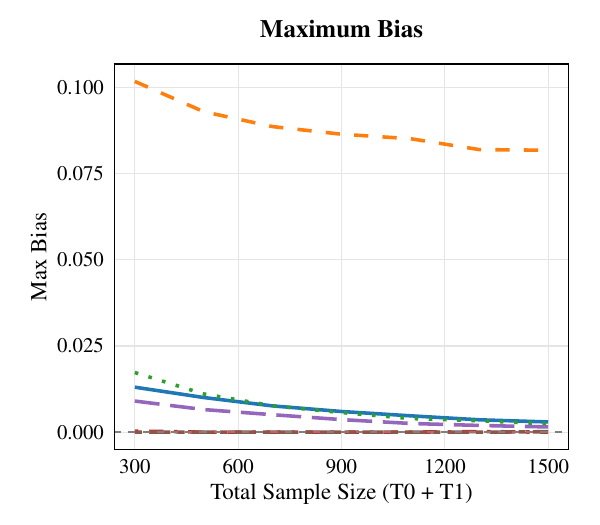} \\[-0.35em]
            \includegraphics[width=0.20\textwidth]{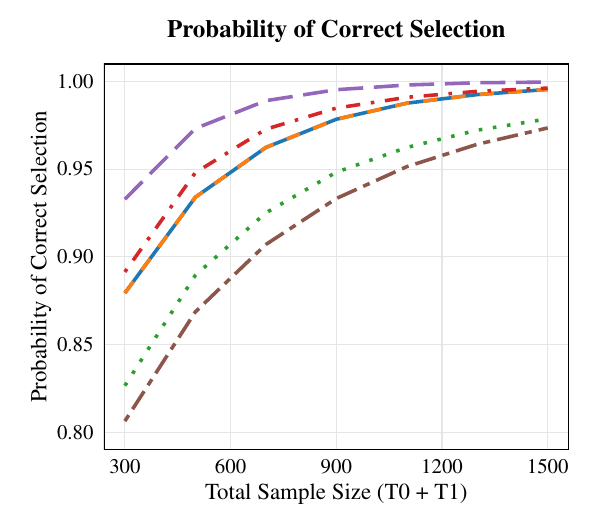} \\
            \subcaptionbox{$\Delta=0.15$, $\tfrac{\sigma_2}{\sigma_1}=0.8$}[0.21\textwidth]{}
        \end{tabular}
        &
        \begin{tabular}{c}
            \includegraphics[width=0.20\textwidth]{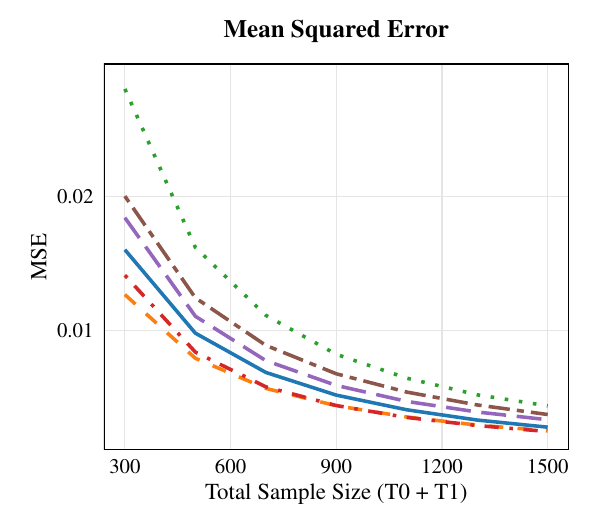} \\[-0.35em]
            \includegraphics[width=0.20\textwidth]{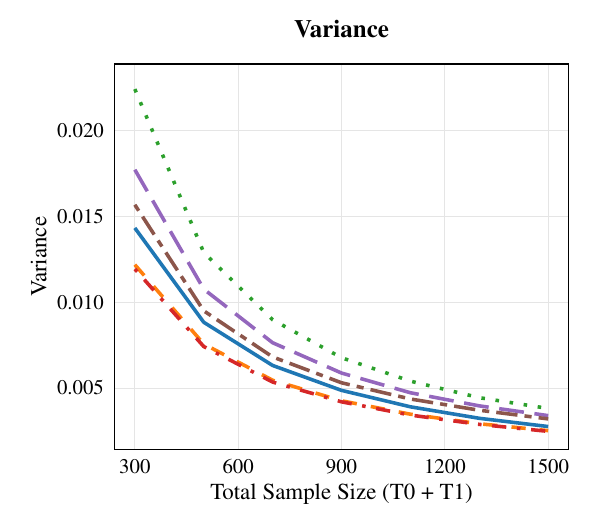} \\[-0.35em]
            \includegraphics[width=0.20\textwidth]{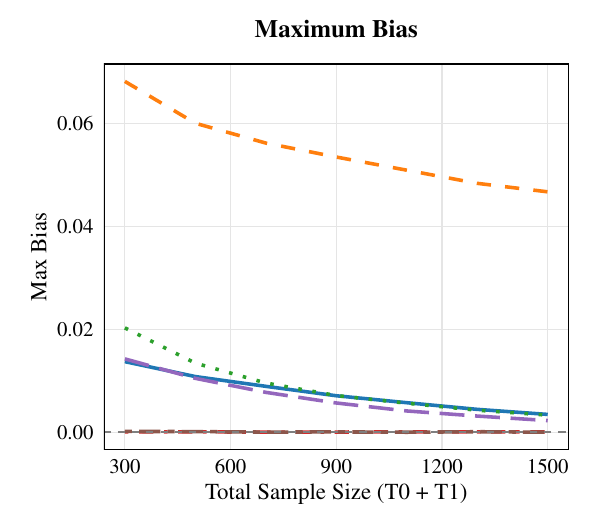} \\[-0.35em]
            \includegraphics[width=0.20\textwidth]{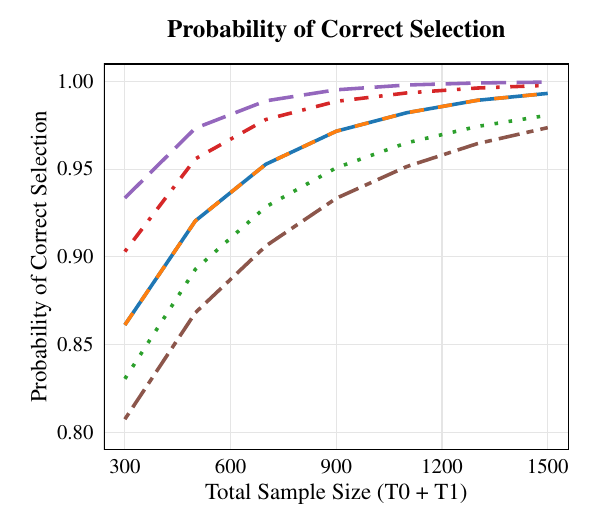} \\
            \subcaptionbox{$\Delta=0.15$, $\tfrac{\sigma_2}{\sigma_1}=1.25$}[0.21\textwidth]{}
        \end{tabular}
    \end{tabular}

    \vspace{0.15em}
\begin{tikzpicture}
    \begin{axis}[
        hide axis,
        xmin=0, xmax=1, ymin=0, ymax=1,
        legend style={
            at={(0.5,0)}, anchor=north,
            legend columns=6,  
            column sep=0.1cm,  
            row sep=0.1cm,
            font=\footnotesize,
            draw=none,
            /tikz/every even column/.append style={column sep=0.1cm}  
        },
        legend cell align=left
    ]
        \addlegendimage{no markers, solid, color={rgb,255:red,31;green,119;blue,180}, thick}
        \addlegendentry{Proposal}

        \addlegendimage{no markers, dashed, color={rgb,255:red,255;green,127;blue,14}, thick}
        \addlegendentry{Proposal w/o bias correction}

        \addlegendimage{no markers, dotted, color={rgb,255:red,44;green,160;blue,44}, thick}
        \addlegendentry{SS.Hyper}

        \addlegendimage{no markers, dash dot, color={rgb,255:red,214;green,39;blue,40}, thick}
        \addlegendentry{Oracle}

        \addlegendimage{no markers, dash pattern=on 6pt off 2pt, color={rgb,255:red,148;green,103;blue,189}, thick}
        \addlegendentry{Non-adaptive method}

        \addlegendimage{no markers, dash pattern=on 4pt off 2pt on 1pt off 2pt, color={rgb,255:red,140;green,86;blue,75}, thick}
        \addlegendentry{SS.SE}
    \end{axis}
\end{tikzpicture}

    \caption{Performance comparison across metrics (rows: MSE, Variance, Bias, Selection Probability) under different parameter configurations (columns: $(\Delta, \tfrac{\sigma_2}{\sigma_1})$).}
    \label{fig:performance_compare}
\end{figure}

\subsection{Metric for comparison}

To evaluate performance, let \( \hat{\tau}_{\hat{w}_{\max}} \) denote the estimated treatment effect for the winner identified by each method, we adopt the following criteria for comparison.
\begin{itemize}
    \item {Proposed MSE objective}: \( \EE[(\hat{\tau}_{\hat{w}_{\max}} - \tau_{w_{\max}})^2] \).
    
    \item {Probability of correct selection~\eqref{eq:selection.error}}: \( \PP(\hat{w}_{\max} = w_{\max}) \).

    \item {Conditional Bias~\eqref{eq:conditional.unbiasedness}}: $\max_{w\in\{1,2\}}|\EE[\hat{\tau}_{\hat{w}_{\max}} - \tau_{\hat{w}_{\max}}\mid \hat{w}_{\max}=w]|$.
    
    \item {Variance}~\eqref{eq:variance}: \( \EE[\var(\hat{\tau}_{\hat{w}_{\max}} \mid \hat{w}_{\max})] \).
    \end{itemize}

\subsection{Results}

We summarize the results in Figure~\ref{fig:performance_compare}. In the top row (proposed MSE objective), our method remains comparable to the oracle and the gap vanishes as $T$ grows.
In contrast, sample-splitting baselines (SS.SE, SS.Hyper) incur larger MSEs because their selection and estimation do not use the full data.
We remark that 
(1) the non-adaptive design can be competitive when its equal allocation is near optimal (e.g., $\Delta>0$, $\sigma_2/\sigma_1=0.8$), but not uniformly so due to its lack of adaptivity (e.g., $\Delta>0$, $\sigma_2/\sigma_1=1.25$);
(2) the method without bias correction can attain an MSE even smaller than the bias-corrected oracle because it minimizes the proposed MSE without the conditional unbiasedness constraint~\eqref{eq:conditional.unbiasedness}. However, the MSE-gain comes at the undesirable cost of a significant bias as discussed below.

For the estimation accuracy, the third row plots the maximum of conditional biases $\max_{w\in\{1,2\}}|\EE[\hat{\tau}_{\hat{w}_{\max}} - \tau_{\hat{w}_{\max}}\mid \hat{w}_{\max}=w]|$. The method without bias correction exhibits significant selection bias, particularly for small $T$. 
Moreover, the bias would not vanish as $T$ increases.
As for variance, the second row plots $ \EE[\var(\hat{\tau}_{\hat{w}_{\max}} \mid \hat{w}_{\max})]$, where our method lies close to the oracle. 
Equal allocation is the second worst because it does not adapt and thus fails to devote more budget to the better treatment.
The no–bias-correction variant can even outperform the oracle on this criterion, for the same reason noted for the proposed MSE objective.

For selection accuracy, the bottom row reports the probability of identifying the best treatment $\PP(\hat{w}_{\max}=w_{\max})$. 
The oracle, our proposal, and the no–bias-correction variant perform comparably across settings, whereas the sample-splitting based procedures exhibit a higher mis-selection rate due to the inefficient use of data for selection. 
We remark that the non-adaptiv method attains the highest selection accuracy even compared to the oracle because it does not prioritize the better treatment and preserves exploration across both treatments, which comes at the cost of less accurate estimations discussed above.

\section{Real data}\label{sec:real.data}

We compare designs using the dataset from \textcite{Krueger1999star}, originating from the late-1980s Tennessee Student/Teacher Achievement Ratio (STAR) experiment conducted by the state Department of Education  (details in \Cref{sec:STAR.dataset}). 
We fix a total budget of $T=1000$ and apply the methods in \Cref{sec:simulation}, excluding the oracle due to the unavailability of oracle knowledge. 
To emulate an experiment, we sample without replacement from the original STAR dataset, and repeat the emulation multiple times to assess the variability.

In terms of comparison metrics, in \Cref{sec:real.data.strong.effect}, we omit the proposed MSE because the true treatment effects are unknown. 
Variance is assessed by repeating the emulated experiment and comparing the resulting estimates across repetitions. 
Bias again can not be computed without the ground truth, so we instead report estimated effects conditional on the selected winner.
For selection accuracy, we evaluate the consistency with the original study's conclusion.

As bias is problematic for the downstream decision-making based on the experiments, we design a placebo-type analysis to enable the bias assessment (\Cref{sec:real.data.mild.effect}). 
Specifically, we drop the small-class treatment and split the original control arm into two groups: one retained as control and the other used as a pseudo-treatment. 
By construction, the pseudo-treatment's effect is known to be \emph{zero}; thus, when it is selected, the associated estimate can be compared against the true treatment effect (zero) to compute the bias. 
In addition, we believe this crafted, near-null setting also reflects more typical practical regimes than the original dataset with a very strong treatment effect.

\subsection{STAR dataset}\label{sec:STAR.dataset}
The STAR datatset is collected from an experiment examined the impact of reduced class sizes on early-grade test scores, involving over $7,000$ students across $79$ schools. 
Students were randomly assigned to one of three intervention groups: small class, regular class, and regular-with-aide class.  
We treat the regular class as the control group and the small class and regular-with-aide class as treatment groups. 
The study followed students from kindergarten through third grade, generating four yearly subsets of data. We take the students' second-grade math scores on the Stanford Achievement Test as the primary outcome. 
In the original study, small classes outperform regular classes with an aide, which in turn outperform regular classes.

\subsection{Results from original STAR data}\label{sec:real.data.strong.effect}

As shown in Table~\ref{tab:results.1}, our method reduces variance by more than $10\%$ relative to the non-adaptive design and by about $43\%$ (or more) relative to the sample-splitting baselines. 
The variant without bias correction attains even smaller variance because it solves a less constrained optimization problem as discussed in \Cref{sec:simulation}.
Across estimated treatment effects conditional on each possible selection, all methods except SS.Hyper produce positive estimates, in line with the STAR study’s reported effects. 
For selection consistency with the original study, our proposal, the non-adaptive design, and the biased variant each achieve $\ge 98\%$, whereas the SS.Hyper and SS.SE are less consistent, reaching approximately $93\%$ and $96\%$, respectively.

\begin{table}[ht]
\centering
\small
\begin{tabular}{ccccc}
\toprule
Method & Var. & Consistency (\%) & TE (Reg.-with-aide) & TE (Small) \\
\midrule
Proposal & 9.481 & 98.3 & 3.183 & 8.446 \\
SS.Hyper & 23.042 & 92.9 & -0.085 & 8.462 \\
Non-adaptive & 10.796 & 99.4 & 3.981 & 8.524 \\
Proposal (w/o corr.) & 8.245 & 98.3 & 4.433 & 8.693 \\
SS.SE & 16.774 & 95.7 & 0.302 & 8.736 \\
\bottomrule
\end{tabular}
\caption{Comparison of methods on the original STAR dataset. We report variance (Var.), proportion of selection consistent with the original STAR study (Consistency), treatment effect estimation conditional on selecting the regular-with-aide (TE (Reg.-with-aide)), and treatment effect estimation conditional on selecting the the small class (TE (Small)).}
\label{tab:results.1}
\end{table}

\subsection{Results with pseudo-treatment}\label{sec:real.data.mild.effect}

To assess bias, we remove the small-class treatment (a very strong effect atypical in social-science experiments) and randomly split the regular-class group into a control and a pseudo-treatment. This construction facilitates the bias evaluation when the pseudo-treatment is selected, since the true effect is known to be zero. 
The corresponding results are summarized in Table~\ref{tab:results.2}. 

For variance and selection consistency, the patterns are similar to those in \Cref{sec:real.data.strong.effect} and our procedure performs strongly. For bias, when the pseudo-treatment is evaluated, the variant without bias correction exhibits a bias of about $1.625$, whereas our procedure reduces this by over $60\%$. 
Among all other methods reported in the table, the bias of our procedure most closely aligns with that of the sample-splitting estimation and selection method (SS.SE), which is unbiased by design.

\begin{table}[ht]
\centering
\small
\begin{tabular}{ccccc}
\toprule
Method & Var. & Consistency (\%) & Cond. Bias/TE (Pseudo) & TE (Reg.-with-aide) \\
\midrule
Proposal & 6.802 & 78.1 & 0.632 & 1.099 \\
SS.Hyper & 20.336 & 70.3 & -0.317 & 0.184 \\
Non-adaptive & 8.280 & 83.6 & 0.862 & 1.306 \\
Proposal (w/o corr.) & 5.523 & 78.1 & 1.625 & 1.996 \\
SS.SE & 12.006 & 73.8 & 0.354 & 0.892 \\
\bottomrule
\end{tabular}
\caption{Comparison of methods on the semi-synthetic STAR dataset with pseudo-treatment.
We report variance (Var.), proportion of selection consistent with the original STAR study  (Consistency), treatment effect estimation conditional on selecting the pseudo-treatment (TE (Pseudo)), which corresponds to the conditional bias, and treatment effect estimation conditional on selecting the regular-with-aide class (TE (Reg.-with-aide)).}
\label{tab:results.2}
\end{table}

\section{Discussion}\label{sec:discussion}

In multi-treatment experiments, the allocations that optimize winner selection and effect estimation typically differ. 
We introduce a unified mean-squared-error objective that trades off selection and estimation, and determine allocations by optimizing it.
By jointly optimizing selection and estimation, our design achieves favorable performance in both the risk of selecting a suboptimal treatment and the estimation accuracy for downstream inference and decision making.

We discuss several future directions.
\begin{itemize}
    \item Extension to $K>2$ treatments. 
    In this paper, we focus on the two-treatment case ($K=2$) to illustrate the selection–estimation trade-off in experiment design, and extending our proposal to arbitrary $K$ is of interest. 
    The challenge is that the optimization problem for a larger $K$ is associated with a higher-dimensional search space and requires more efficient computational shortcuts.

    \item Personalized experiment design.
    Beyond choosing a single winner based on the average treatment effect, one can consider personalized selection, assigning different treatments across subpopulations.    
    When subpopulations are prespecified, our procedure can be applied stratum by stratum, and it is worth investigating how to borrow strength across strata to improve efficiency.
    When subpopulations are data-adaptive, an additional layer of selection is introduced, and the bias correction for this extra selection step requires further research.

    \item Improving the non-asymptotic convergence rate of \Cref{prop:adp.alloc.convergence}. The current $O(T^{-1/14})$ rate is a result of uniform convergence over expanding parameter sets, which forces a delicate balance between the choice of $\alpha$ and $\beta$ and ultimately yields the slow rate. Refining the analysis with sharper empirical process techniques to better control the discrepancy between $\mathcal{R}_T$ and $\mathcal{R}$, or exploring multi-stage extensions, may yield faster convergence.

\end{itemize}

\printbibliography

\appendix

\section{Additional examples}\label{appe:sec:example}

\begin{example}\label{exam:counter.selected.winner}
   Consider $T=200$, $\sigma_w^2=1$ for $w\in\{0,1,2\}$, $\mathcal{I}=[-1,1]\setminus[-0.1,0.1]$, and
\[
   \begin{cases}
      n_1(\Delta) = n_2(\Delta) = 100, & \Delta < 0, \\
      n_1(\Delta) = 199, ~ n_2(\Delta) = 1, & \Delta \ge 0. 
   \end{cases}
\]
The selection rule~\eqref{eq:selected.winner} yields a worst-case error of $\Phi(-{0.1}/{\sqrt{1+1/199}})$.
However, the alternative rule based on $\hat{\Delta}$
\[
   \hat{w}_{\max} = 
   \begin{cases}
      2, & \hat{\Delta} > 0 \ \text{or}\ \hat{\Delta} < -1.1, \\
      1, & \text{otherwise},
   \end{cases}
\]
has a strictly smaller worst-case error $
   \Phi\!\left(-{0.1}/{\sqrt{1+1/199}}\right) - 
   \Phi\!\left(-{1.2}/{\sqrt{1+1/199}}\right)$.
\end{example}

\begin{example}\label{exam:optimize.full.sample}
Let $T=120$ with a pilot $T_0=30$ (10 per arm) and pilot estimates $\hat{\sigma}_w$ and $\hat{\Delta}$. 
The full-budget plug-in splits all $120$ units as $(36,24,60)$ based on the estimated parameters, topping up from $(10,10,10)$ assigns $(26,14,50)$ over the $90$ remaining units. 
In contrast, optimizing only the post-pilot stage allocates the rest $90$ units as $(27,18,45)$, yielding totals $(37,28,55)$ that deviate from the oracle plan of the full budget.
\end{example}

\section{Estimator derivations}\label{appe:derive:mse.A}

\subsection{Treatment effect estimator}\label{sec:debias}

\begin{figure}[t]
  \centering
  \begin{minipage}{0.4\textwidth}
    \centering
    \includegraphics[clip, trim = 0cm 0cm 0cm 0cm, width = 0.9\textwidth]{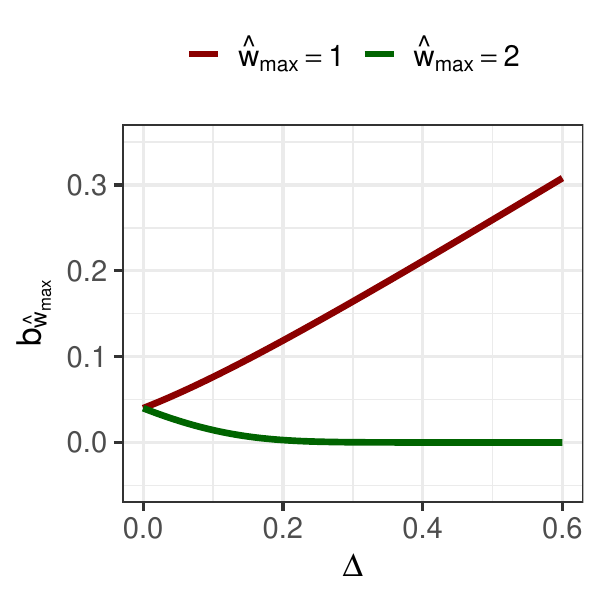}
    \caption*{\quad (a) $\sigma_1=\sigma_2=1$}
  \end{minipage}
    \begin{minipage}{0.4\textwidth}
    \centering
    \includegraphics[clip, trim = 0cm 0cm 0cm 0cm, width = 0.9\textwidth]{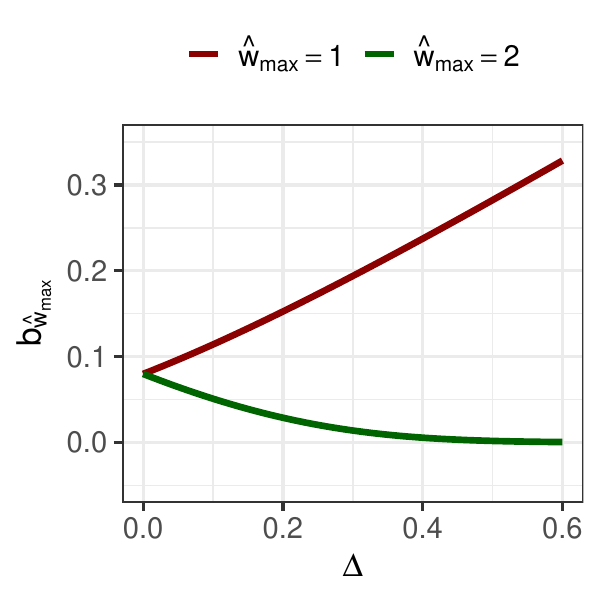}
    \caption*{\quad (b) $\sigma_1=\sigma_2=2$}
  \end{minipage}
    \caption{Conditional bias $b_{\hat{w}_{\max}}$ as a function of the effect difference $\Delta$ under equal allocation ($n_1=n_2=200$) and equal variances ($\sigma_1=\sigma_2\in\{1,2\}$). 
    Both conditional biases are positive: mis-selection ($\hat{w}_{\max}=1$, dark red) yields a more significant bias that grows with $\Delta$, whereas correct selection ($\hat{w}_{\max}=2$, dark green) has non-ignorable bias near $\Delta=0$. Higher outcome variance ($\sigma_1=\sigma_2=2$) induces larger bias.}

  \label{fig:conditional.bias}
\end{figure}

As illustrated in \Cref{fig:conditional.bias}, the selection bias $b_{\hat{w}_{\max}}$ is always positive regardless of which treatment is selected. 
When the winner is correctly identified, selection-induced bias is most severe for small effect differences $\Delta$, as estimates selected for being large tend to overstate the truth. 
On the other hand, when the winner is misidentified, the conditional bias upon selecting the first treatment is larger and increases with $\Delta$, since the inferior treatment is chosen only because of an overly positive realization.

Below we derive the bias.
We first state two useful lemma that's useful in deriving the bias correction term and MSE expression.
\begin{lemma}\label{lemma:gaussian.expectations}
For general 
\begin{align*}
        (X, Y) \sim \calN\left(( \mu_X, \mu_Y), \text{Diag}( \sigma^2_X, \sigma^2_Y)\right).
\end{align*}
with $\Phi(\cdot)$ the standard normal CDF and $\phi(\cdot)$ its PDF. we have 
\begin{align}\label{lemma:Gaussian.1.moment}
\EE\left[X\mid X>Y+a\right] = \mu_X + \frac{\sigma^2_X}{\sqrt{\sigma^2_X+\sigma^2_Y}}\frac{\phi\left(\frac{(\mu_X-\mu_Y)-a}{\sqrt{\sigma^2_X+\sigma^2_Y}}\right)}{\Phi\left(\frac{(\mu_X-\mu_Y)-a}{\sqrt{\sigma^2_X+\sigma^2_Y}}\right)}
\end{align}
and 
\begin{align}\label{lemma:Gaussian.2.moment}
\begin{split}
\EE\left[\left(X+b\right)^2\mid X>Y+a\right] 
=& (\mu_X+b)^2 + \sigma^2_X \\
&+ \frac{\sigma^2_X}{\sqrt{\sigma^2_X+\sigma^2_Y}}\frac{\phi\left(\frac{(\mu_X-\mu_Y)-a}{\sqrt{\sigma^2_X+\sigma^2_Y}}\right)}{\Phi\left(\frac{(\mu_X-\mu_Y)-a}{\sqrt{\sigma^2_X+\sigma^2_Y}}\right)}\left(\frac{\sigma^2_X}{\sqrt{\sigma^2_X+\sigma^2_Y}}\frac{a-(\mu_X-\mu_Y)}{\sqrt{\sigma^2_X+\sigma^2_Y}} + 2\mu_X+2b\right)
\end{split}
\end{align}
for any $a,b$.
\end{lemma} 
The proof of \Cref{lemma:gaussian.expectations} is deferred to the end of this section. We first work on the bias correction term \Cref{eq:tau.hat.debiased}.
We have
\begin{align*}
    (\hat{\mu}_1, \hat{\mu}_2) \sim \calN\left(( \mu_1, \mu_2), \text{Diag}\left( \frac{\sigma_1^2}{n_1}, \frac{\sigma_2^2}{n_2}\right)\right).
\end{align*}
Let $X =\hat{\mu}_1$ and $Y =\hat{\mu}_2$,
$\Delta = \mu_2 - \mu_1$ denote the effect size and $V = \frac{\sigma^2_1}{n_1}+\frac{\sigma^2_2}{n_2}$.
We have
\begin{align*}
\EE\left[\hat{\mu}_{\hat{w}_{\max}}-\mu_{\hat{w}_{\max}} \mid \hat{w}_{\max}=1\right] &= 
\EE\left[\hat{\mu}_{1}-\mu_{1} \mid \hat{w}_{\max}=1\right] \\
& = \EE\left[\hat{\mu}_{1}-\mu_{1} \mid \hat{\mu}_1 > \hat{\mu}_2\right] \\
& = \EE\left[X-\mu_{1} \mid X>Y\right]
\end{align*}
Applying \Cref{lemma:gaussian.expectations}, we get
\begin{align*}
\EE\left[\hat{\mu}_{\hat{w}_{\max}}-\mu_{\hat{w}_{\max}} \mid \hat{w}_{\max}=1\right] = \frac{\sigma^2_1}{n_1\sqrt{V}}\frac{\phi\left(\frac{\Delta}{\sqrt{V}}\right)}{\Phi\left(\frac{-\Delta}{\sqrt{V}}\right)}   
\end{align*}
Similarly
\begin{align*}
\EE\left[\hat{\mu}_{\hat{w}_{\max}}-\mu_{\hat{w}_{\max}} \mid \hat{w}_{\max}=2\right] &= 
\EE\left[\hat{\mu}_{2}-\mu_{2} \mid \hat{w}_{\max}=2\right] \\
&= \EE\left[\hat{\mu}_{2}-\mu_{2} \mid \hat{\mu}_1 < \hat{\mu}_2\right] \\
&= \EE\left[Y-\mu_{2} \mid X<Y\right] \\
&= \frac{\sigma^2_2}{n_2\sqrt{V}}\frac{\phi\left(\frac{\Delta}{\sqrt{V}}\right)}{\Phi\left(\frac{\Delta}{\sqrt{V}}\right)}
\end{align*}
Next we work on the derivation of the MSE expression 
\Cref{eq:n1.oracle}.
Since our estimator $\hat{\mu}^\debiased$ satisfies the conditional unbiased property, we have
\begin{align*}
\EE\left[\left(\hat{\mu}^\debiased_{\hat{w}_{\max}} - \mu_{w_{\max}}\right)^2\right] = \EE\left[(\hat{\mu}^\debiased_{\hat{w}_{\max}} - \mu_{\hat{w}_\text{max}})^2\right] + \EE\left[\left(\mu_{\hat{w}_{\max}} - \mu_{w_{\max}}\right)^2\right]
\end{align*}

We first consider the first component in the MSE decomposition.
\begin{align*}
    \EE\left[(\hat{\mu}^\debiased_{\hat{w}_{\max}} - \mu_{\hat{w}_\text{max}})^2\right] = \underbrace{\EE\left[(\hat{\mu}^\debiased_{2} - \mu_2)^2\1\left\{\hat{\mu}_1<\hat{\mu}_2\right\}\right]}_{:=(a)} + \underbrace{\EE\left[(\hat{\mu}^\debiased_{1} - \mu_1)^2\1\left\{\hat{\mu}_1>\hat{\mu}_2\right\}\right]}_{:=(b)}
\end{align*}
For the term $(a)$
\begin{align*}
\EE\left[(\hat{\mu}^\debiased_{2} - \mu_2)^2\1\left\{\hat{\mu}_1<\hat{\mu}_2\right\}\right] & = \EE\left[(\hat{\mu}_{2}- \frac{\sigma^2_2}{n_2\sqrt{V}}\frac{\phi\left(\frac{\Delta}{\sqrt{V}}\right)}{\Phi\left(\frac{\Delta}{\sqrt{V}}\right)}   - \mu_2)^2\1\left\{\hat{\mu}_1<\hat{\mu}_2\right\}\right]  \\
& = \EE\left[(Y- \frac{\sigma^2_2}{n_2\sqrt{V}}\frac{\phi\left(\frac{\Delta}{\sqrt{V}}\right)}{\Phi\left(\frac{\Delta}{\sqrt{V}}\right)}   - \mu_2)^2\1\left\{X<Y\right\}\right] \\
& = \EE\left[(Y- \frac{\sigma^2_2}{n_2\sqrt{V}}\frac{\phi\left(\frac{\Delta}{\sqrt{V}}\right)}{\Phi\left(\frac{\Delta}{\sqrt{V}}\right)}   - \mu_2)^2\mid X<Y\right]\cdot \PP\left(X<Y\right)
\end{align*}
Let $b = -\frac{\sigma^2_2}{n_2\sqrt{V}}\frac{\phi\left(\frac{\Delta}{\sqrt{V}}\right)}{\Phi\left(\frac{\Delta}{\sqrt{V}}\right)}   - \mu_2$ and $a=0$. Applying \Cref{lemma:gaussian.expectations}, we get
\begin{align*}
\EE\left[(\hat{\mu}^\debiased_{2} - \mu_2)^2\1\left\{\hat{\mu}_1<\hat{\mu}_2\right\}\right]   =  \frac{\sigma^2_2}{n_2}\left(\Phi(\frac{\Delta}{\sqrt{V}})-\frac{\sigma^2_2} {n_2\cdot V}\left[\frac{\Delta}{\sqrt{V}}\phi(\frac{\Delta}{\sqrt{V}})+\frac{\phi^2(\frac{\Delta}{\sqrt{V}})}{\Phi(\frac{\Delta}{\sqrt{V}})}\right]\right)
\end{align*}
Similarly,
\begin{align*}
\EE\left[(\hat{\mu}^\debiased_{1} - \mu_1)^2\1\left\{\hat{\mu}_1>\hat{\mu}_2\right\}\right]  &=
\EE\left[(X- \frac{\sigma^2_1}{n_1\sqrt{V}}\frac{\phi\left(\frac{\Delta}{\sqrt{V}}\right)}{\Phi\left(\frac{-\Delta}{\sqrt{V}}\right)}      - \mu_1)^2\1\left\{X>Y\right\}\right] \\
& = \EE\left[(X- \frac{\sigma^2_1}{n_1\sqrt{V}}\frac{\phi\left(\frac{\Delta}{\sqrt{V}}\right)}{\Phi\left(\frac{-\Delta}{\sqrt{V}}\right)}      - \mu_1)^2\mid X>Y\right]\cdot \PP\left(X>Y\right)
\end{align*}
Let $b = -\frac{\sigma^2_1}{n_1\sqrt{V}}\frac{\phi\left(\frac{\Delta}{\sqrt{V}}\right)}{\Phi\left(\frac{-\Delta}{\sqrt{V}}\right)}      - \mu_1$ and $a=0$. Applying \Cref{lemma:gaussian.expectations}, we get
\begin{align*}
\EE\left[(\hat{\mu}^\debiased_{1} - \mu_1)^2\1\left\{\hat{\mu}_1>\hat{\mu}_2\right\}\right]  &=
\frac{\sigma^2_1}{n_1}\left(\Phi(\frac{-\Delta}{\sqrt{V}})+\frac{\sigma^2_1}{n_1\cdot V}\left[\frac{\Delta}{\sqrt{V}}\phi(\frac{\Delta}{\sqrt{V}})-\frac{\phi^2(\frac{\Delta}{\sqrt{V}})}{\Phi(\frac{-\Delta}{\sqrt{V}})}\right]\right)
\end{align*}
Putting altogether, we arrive at
\begin{align*}
    \begin{split}   \EE\left[(\hat{\mu}^\debiased_{\hat{w}_{\max}} - \mu_{\hat{w}_\text{max}})^2\right]
    &= \frac{\sigma^2_2}{n_2}\Phi(\frac{\Delta}{\sqrt{V}})- \frac{\sigma^4_2}{n_2^2\cdot V}\phi(\frac{\Delta}{\sqrt{V}})\cdot\frac{\Delta}{\sqrt{V}} - \frac{\sigma^4_2}{n_2^2\cdot V}\frac{\phi^2(\frac{\Delta}{\sqrt{V}})}{\Phi(\frac{\Delta}{\sqrt{V}})} \\
    &\quad~+ \frac{\sigma^2_1}{n_1}\Phi(\frac{-\Delta}{\sqrt{V}})+ \frac{\sigma^4_1}{n_1^2\cdot V}\phi(\frac{\Delta}{\sqrt{V}})\cdot\frac{\Delta}{\sqrt{V}} - \frac{\sigma^4_1}{n_1^2\cdot V}\frac{\phi^2(\frac{\Delta}{\sqrt{V}})}{\Phi(\frac{-\Delta}{\sqrt{V}})}.
    \end{split}
\end{align*}
For the second component in the MSE expression, we first work on the case $\Delta>0$ and treatment 2 is the true optimal treatment arm.
\begin{align*}
\EE\left[\left(\mu_{\hat{w}_{\max}} - \mu_{w_{\max}}\right)^2\right] &= \EE\left[\left(\mu_{\hat{w}_{\max}} - \mu_{2}\right)^2\right] \\
& = \EE\left[\left(\mu_{1} - \mu_{2}\right)^2\mid \hat{w}_{\max}=1\right]\PP\left(\hat{w}_{\max}=1\right) + \EE\left[\left(\mu_{2} - \mu_{2}\right)^2\mid \hat{w}_{\max}=2\right]\PP\left(\hat{w}_{\max}=2\right) \\
& = \Delta^2 \cdot \PP\left(\hat{w}_{\max}=1\right) \\
& = \Delta^2 \cdot \PP\left(X>Y\right) \\
& = \Delta^2\Phi(\frac{-\Delta}{\sqrt{V}})
\end{align*}
Summing the first and second components in the MSE expression
\begin{align*}
    \begin{split}   \EE\left[\left(\hat{\mu}^\debiased_{\hat{w}_{\max}} - \mu_{w_{\max}}\right)^2\right]
    &= \frac{\sigma^2_2}{n_2}\Phi(\frac{\Delta}{\sqrt{V}})- \frac{\sigma^4_2}{n_2^2\cdot V}\phi(\frac{\Delta}{\sqrt{V}})\cdot\frac{\Delta}{\sqrt{V}} - \frac{\sigma^4_2}{n_2^2\cdot V}\frac{\phi^2(\frac{\Delta}{\sqrt{V}})}{\Phi(\frac{\Delta}{\sqrt{V}})} \\
    &\quad~+ (\Delta^2+\frac{\sigma^2_1}{n_1})\Phi(\frac{-\Delta}{\sqrt{V}})+ \frac{\sigma^4_1}{n_1^2\cdot V}\phi(\frac{\Delta}{\sqrt{V}})\cdot\frac{\Delta}{\sqrt{V}} - \frac{\sigma^4_1}{n_1^2\cdot V}\frac{\phi^2(\frac{\Delta}{\sqrt{V}})}{\Phi(\frac{-\Delta}{\sqrt{V}})}.
    \end{split}
\end{align*}
for $\Delta>0$. Similarly, we can obtain
\begin{align*}
    \begin{split}   \EE\left[\left(\hat{\mu}^\debiased_{\hat{w}_{\max}} - \mu_{w_{\max}}\right)^2\right]
    &= \frac{\sigma^2_1}{n_1}\Phi(\frac{-\Delta}{\sqrt{V}})+ \frac{\sigma^4_1}{n_1^2\cdot V}\phi(\frac{\Delta}{\sqrt{V}})\cdot\frac{\Delta}{\sqrt{V}} - \frac{\sigma^4_1}{n_1^2\cdot V}\frac{\phi^2(\frac{\Delta}{\sqrt{V}})}{\Phi(\frac{-\Delta}{\sqrt{V}})} \\
    &\quad~+ (\Delta^2+\frac{\sigma^2_2}{n_2})\Phi(\frac{\Delta}{\sqrt{V}})- \frac{\sigma^4_2}{n_2^2\cdot V}\phi(\frac{\Delta}{\sqrt{V}})\cdot\frac{\Delta}{\sqrt{V}} - \frac{\sigma^4_2}{n_2^2\cdot V}\frac{\phi^2(\frac{\Delta}{\sqrt{V}})}{\Phi(\frac{\Delta}{\sqrt{V}})}.
    \end{split}
\end{align*}
for $\Delta <0$.
Putting altogether we obtain
\begin{align*}
    \begin{split}   \EE\left[\left(\hat{\mu}^\debiased_{\hat{w}_{\max}} - \mu_{w_{\max}}\right)^2\right]
    &= \frac{\sigma^2_1}{n_1}\Phi(\frac{-\Delta}{\sqrt{V}})+ \frac{\sigma^4_1}{n_1^2\cdot V}\phi(\frac{\Delta}{\sqrt{V}})\cdot\frac{\Delta}{\sqrt{V}} - \frac{\sigma^4_1}{n_1^2\cdot V}\frac{\phi^2(\frac{\Delta}{\sqrt{V}})}{\Phi(\frac{-\Delta}{\sqrt{V}})} \\
    &\quad~+ \frac{\sigma^2_2}{n_2}\Phi(\frac{\Delta}{\sqrt{V}})- \frac{\sigma^4_2}{n_2^2\cdot V}\phi(\frac{\Delta}{\sqrt{V}})\cdot\frac{\Delta}{\sqrt{V}} - \frac{\sigma^4_2}{n_2^2\cdot V}\frac{\phi^2(\frac{\Delta}{\sqrt{V}})}{\Phi(\frac{\Delta}{\sqrt{V}})} \\
    &\quad~+ \1\left\{\Delta>0\right\}\cdot\Delta^2\Phi(\frac{-\Delta}{\sqrt{V}}) + \1\left\{\Delta<0\right\}\cdot\Delta^2\Phi(\frac{\Delta}{\sqrt{V}})
    \end{split} 
\end{align*}   
$\forall \Delta \in  \mathbb{R}$.
\begin{proof}[Proof of \Cref{lemma:gaussian.expectations}]
Since $X$ and $Y$ are independent:
$Z:=X-Y \sim \mathcal{N}(\mu_X - \mu_Y, \sigma_X^2 + \sigma_Y^2)$
Let $\sigma_Z^2 = \sigma_X^2 + \sigma_Y^2$ and $\mu_Z = \mu_X - \mu_Y$.
Since $X$ and $Z$ are joint normal, we have
\begin{align}\label{eq:joint.normal}
X = \mu_X + \frac{\mathrm{Cov}(X,Z)}{\sigma_Z^2}(Z - \mu_Z) + \epsilon
\end{align}
and \begin{align*}
\epsilon \sim \mathcal{N}\left(0, \sigma_X^2 - \frac{\mathrm{Cov}^2(X,Z)}{\sigma_Z^2}\right),    \quad \epsilon \perp Z
\end{align*}
We have
\[
\text{Cov}(Z, X) = \text{Cov}(X - Y, X) = \sigma_X^2 - 0 = \sigma_X^2
\]
Plugging it to the Eq.~\eqref{eq:joint.normal}, we have
\begin{align*}
X = \mu_X + \frac{\sigma_X^2}{\sigma_Z^2}(Z - \mu_Z) + \epsilon
\end{align*}
where 
\begin{align*}
\epsilon \sim \mathcal{N}\left(0, \sigma_X^2 - \frac{\sigma_X^4}{\sigma_Z^2}\right) = \mathcal{N}\left(0, \frac{\sigma_X^2\sigma_Y^2}{\sigma_X^2 + \sigma_Y^2}\right)
\end{align*}
then
\begin{align*}
\mathbb{E}[X \mid Z > a] &= \mu_X + \mathbb{E}[\frac{\sigma_X^2}{\sigma_Z^2}(Z - \mu_Z)\mid Z > a] \\
&= \mu_X + \frac{\sigma_X^2}{\sigma_Z^2}\cdot \sigma_Z \frac{\phi\left(\frac{a-\mu_Z}{\sigma_Z}\right)}{1 - \Phi\left(\frac{a-\mu_Z}{\sigma_Z}\right)} \\
&= \mu_X + \frac{\sigma_X^2}{\sigma_Z} \frac{\phi\left(\frac{a-\mu_Z}{\sigma_Z}\right)}{1 - \Phi\left(\frac{a-\mu_Z}{\sigma_Z}\right)} \\
& = \mu_X + \frac{\sigma_X^2}{\sqrt{\sigma_X^2 + \sigma_Y^2}} \cdot \frac{\phi\left(\frac{a - (\mu_X - \mu_Y)}{\sqrt{\sigma_X^2 + \sigma_Y^2}}\right)}{1 - \Phi\left(\frac{a - (\mu_X - \mu_Y)}{\sqrt{\sigma_X^2 + \sigma_Y^2}}\right)} \\
& = \mu_X + \frac{\sigma_X^2}{\sqrt{\sigma_X^2 + \sigma_Y^2}} \cdot \frac{\phi\left(\frac{(\mu_X - \mu_Y)-a}{\sqrt{\sigma_X^2 + \sigma_Y^2}}\right)}{\Phi\left(\frac{(\mu_X - \mu_Y)-a}{\sqrt{\sigma_X^2 + \sigma_Y^2}}\right)} 
\end{align*} \\
which gives the result of Eq.~\eqref{lemma:Gaussian.1.moment}

Squaring Eq.~\eqref{eq:joint.normal}:
\begin{align*}
X^2 &= \left(\mu_X + \frac{\sigma_X^2}{\sigma_Z^2}(Z - \mu_Z) + \epsilon\right)^2 \\
&= \mu_X^2 + \frac{\sigma_X^4}{\sigma_Z^4}(Z - \mu_Z)^2 + \epsilon^2 \\
&\quad + 2\mu_X\frac{\sigma_X^2}{\sigma_Z^2}(Z - \mu_Z) + 2\mu_X\epsilon + 2\frac{\sigma_X^2}{\sigma_Z^2}(Z - \mu_Z)\epsilon
\end{align*}
then
\begin{align*}
\mathbb{E}[X^2 \mid Z > a] &= \mu_X^2 + \frac{\sigma_X^4}{\sigma_Z^4}\mathbb{E}[(Z - \mu_Z)^2 \mid Z > a] + \mathbb{E}[\epsilon^2] \\
&\quad + 2\mu_X\frac{\sigma_X^2}{\sigma_Z^2}\mathbb{E}[Z - \mu_Z \mid Z > a]
\end{align*}
since $\mathbb{E}[\epsilon \mid Z > a] = 0$ and $\mathbb{E}[(Z - \mu_Z)\epsilon \mid Z > a] = 0$.
Let $\delta = \frac{(\mu_X-\mu_Y)-a}{\sigma_Z}$, using the property of the second moment of the truncated normal, we get
\begin{align}\label{eq:second.trunct}
\mathbb{E}[X^2 \mid Z > a]  &=  \mu_X^2 + \frac{\sigma_X^4}{\sigma_Z^2}\left(1-\delta\frac{\phi(\delta)}{\Phi(\delta)}\right) + \frac{\sigma^2_X\sigma^2_Y}{\sigma^2_X+\sigma^2_Y} + 2\mu_X\frac{\sigma_X^2}{\sigma_Z}\frac{\phi(\delta)}{\Phi(\delta)} 
\end{align}
The left hand side of Eq.~\eqref{lemma:Gaussian.2.moment} can be written as 
\begin{align*}
\mathbb{E}[(X+b)^2 \mid X >Y+ a] 
& = \mathbb{E}[(X+b)^2 \mid Z> a] \\
& = \mathbb{E}[X^2 \mid Z > a] + 2b\cdot\mathbb{E}[X \mid Z > a]+b^2 \\
\end{align*}
Combining Eq.~\eqref{lemma:Gaussian.1.moment},
    \eqref{eq:second.trunct}, we obtain 
\begin{align*}
\mathbb{E}[(X+b)^2 \mid X >Y+ a] 
& = \mu_X^2 + \frac{\sigma_X^4}{\sigma_Z^2}\left(1-\frac{(\mu_X-\mu_Y)-a}{\sigma_Z}\frac{\phi(\frac{(\mu_X-\mu_Y)-a}{\sigma_Z})}{\Phi(\frac{(\mu_X-\mu_Y)-a}{\sigma_Z})}\right) + \frac{\sigma^2_X\sigma^2_Y}{\sigma^2_X+\sigma^2_Y} \\
&\quad + 2\mu_X\frac{\sigma_X^2}{\sigma_Z}\frac{\phi(\frac{(\mu_X-\mu_Y)-a}{\sigma_Z})}{\Phi(\frac{(\mu_X-\mu_Y)-a}{\sigma_Z})}  + 2b\left(\mu_X + \frac{\sigma_X^2}{\sqrt{\sigma_X^2 + \sigma_Y^2}} \cdot \frac{\phi\left(\frac{(\mu_X - \mu_Y)-a}{\sqrt{\sigma_X^2 + \sigma_Y^2}}\right)}{\Phi\left(\frac{(\mu_X - \mu_Y)-a}{\sqrt{\sigma_X^2 + \sigma_Y^2}}\right)}\right)+b^2 \\
& = (\mu_X+b)^2 + \sigma^2_X + \frac{\sigma^2_X}{\sqrt{\sigma^2_X+\sigma^2_Y}}\frac{\phi\left(\frac{(\mu_X-\mu_Y)-a}{\sqrt{\sigma^2_X+\sigma^2_Y}}\right)}{\Phi\left(\frac{(\mu_X-\mu_Y)-a}{\sqrt{\sigma^2_X+\sigma^2_Y}}\right)}\left(\frac{\sigma^2_X}{\sqrt{\sigma^2_X+\sigma^2_Y}}\frac{(a-(\mu_X-\mu_Y)}{\sqrt{\sigma^2_X+\sigma^2_Y}} + 2\mu_X+2b\right)
\end{align*}
which is exactly the expression claimed in Eq.~\eqref{lemma:Gaussian.2.moment}.
\end{proof}

\section{Proof}

\subsection{Proof of \Cref{theo:optimal.rule}}\label{appe:proof:optimal.rule}

\begin{proof}[Proof of \Cref{theo:optimal.rule}]
We observed $\boldsymbol{Y}_1 := \left\{Y_i(1)\right\}_{i=1}^{n_1}$ and  $\boldsymbol{Y}_2 := \left\{Y_i(2)\right\}_{i=1}^{n_2}$ random samples. Let $\phi(\hat{\Delta}^N):\hat{\Delta}^N(\boldsymbol{Y}_1, \boldsymbol{Y}_2) \to \left\{0,1\right\}$ mapping experimental outcomes via the test statistic $\hat{\Delta}^N$ into a decision, regarding which treatment will be identified as the best. Specifically, if the action is 0, then treatment 1 will be identified as the best; if the action is 1, treatment 2 will be identified as the best. $\phi\left(\hat{\Delta}^N\right)$ is the action chosen when $\hat{\Delta}^N(\boldsymbol{Y}_1, \boldsymbol{Y}_2)$ is observed.
The set of all measurable decision is labeled as $\mathcal{D}$. Define the $\phi^*\left(\hat{\Delta}^N\right)$ to be such a decision that is made according to the sign of $\hat{\Delta}^N$ estimator, i.e.,
\begin{align}\label{eq:dm.rule}
\phi^*\left(\hat{\Delta}^N\right) = \II\left\{\hat{\Delta}^N>0\right\}
\end{align}
we want to show this decision rule maximizes the probability of selecting the true winner. Define the regret for a decision rule as
\begin{align}\label{eq:regret.rule}
R\left(\phi, \Delta\right) =     \begin{cases}
         1-\phi\left(\hat{\Delta}^N\right), & \Delta>0, \\
         \phi\left(\hat{\Delta}^N\right), & \Delta<0, \\
         0, & \Delta = 0
    \end{cases} 
\end{align}
and we have
\begin{align}
\EE_{\boldsymbol{Y}_1, \boldsymbol{Y}_2}\left[R\left(\phi\left(\hat{\Delta}^N\right), \Delta\right)\right] = \begin{cases}
         1-\EE\left[\phi\left(\hat{\Delta}^N\right)\right], & \Delta>0, \\
         \EE\left[\phi\left(\hat{\Delta}^N\right)\right], & \Delta<0, \\
         0, & \Delta = 0
    \end{cases} 
\end{align}
which is the probability of false selection. We have 
\begin{align}
      \hat{\Delta}^N \sim \calN\left(r(\Delta), 1\right),
\end{align}
$f\left(\hat{\Delta}^N=x; \Delta_2,\sigma^2_1,\sigma^2_2\right)/f\left(\hat{\Delta}^N=x;\Delta_1,\sigma^2_1,\sigma^2_2\right) \propto \exp\left\{(r(\Delta_2)-r(\Delta_1))\cdot x\right\}$ is an increasing function of $x$ for any $\Delta_2>\Delta_1$ under the assumption of $r(\Delta)$ is an increasing function of $\Delta$, and (may) unknown but fixed $\sigma^2_1$, $\sigma^2_2$.
Also, 
\begin{align}
R\left(0, \Delta\right)-R\left(1, \Delta\right) =     \begin{cases}
         1, & \Delta>0, \\
         -1, & \Delta<0, \\
         0, & \Delta = 0
    \end{cases} 
\end{align}
so the function $R\left(0, \Delta\right)-R\left(1, \Delta\right)$ has precisely one change point at 0, where it change from non-positive value to non-negative value. It follows from the results of Theorem 1 from \cite{Karlin1956decision}, for any feasible decision rule $\phi$ in $\mathcal{D}$, there exists a decision $\phi^0$ of the form $\phi^0:= \II\left\{\hat{\Delta}^N>t\right\}$ for some $t\in \mathbb{R}$ s.t.
\begin{align}
\EE_{\boldsymbol{Y}_1, \boldsymbol{Y}_2}\left[R\left(\phi\left(\hat{\Delta}^N\right), \Delta\right)\right] \geq \EE_{\boldsymbol{Y}_1, \boldsymbol{Y}_2}\left[R\left(\phi^0\left(\hat{\Delta}^N\right), \Delta\right)\right], \quad \forall \Delta \in \mathbb{R}
\end{align}
It means for any decision rule, there always exists a $\phi^0:= \II\left\{\hat{\Delta}^N>t\right\}$ for some $t\in \mathbb{R}$ s.t. it has less probability of false selection(larger probability of correct selection). Thus, a smaller class of threshold  decision rules 
\begin{align}
\phi \in \left\{\II\left(\hat{\Delta}^N>t\right): t\in \mathbb{R}\right\}
\end{align}
is essentially complete.
Now our goal is to find the decision rule in this class such that it has the lowest regret.
Notice that 
\begin{align*}
\EE_{\boldsymbol{Y}_1, \boldsymbol{Y}_2}\left[R\left(\phi^0\left(\hat{\Delta}^N\right), \Delta\right)\right] &=\PP\left(\text{false selection}\right)\\ &= \II\left\{\Delta>0\right\}\cdot\PP\left(\phi^0\left(\hat{\Delta}^N\right)=0 \right)+ \II\left\{\Delta<0\right\}\cdot\PP\left(\phi^0\left(\hat{\Delta}^N\right)=1 \right) \\
& = \II\left\{\Delta>0\right\}\cdot\PP\left(\hat{\Delta}^N<t\right)+ \II\left\{\Delta<0\right\}\cdot\PP\left(\hat{\Delta}^N>t\right)\\
& = \II\left\{\Delta>0\right\}\cdot\Phi\left(t-r(\Delta)\right)+\II\left\{\Delta<0\right\}\cdot\Phi\left(r(\Delta) - t\right)
\end{align*}
There does not exists a $t \in \mathbb{R}$ s.t. the above term is minimized across all $\Delta \in \calI$. 

Next we show 
\[    \phi^* \in \text{arg}\min_{\phi \in \mathcal{D}}\max_{\Delta \in \calI}\EE_{\boldsymbol{Y}_1, \boldsymbol{Y}_2}\left[R\left(\phi\left(\hat{\Delta}^N\right), \Delta\right)\right]\]
From the analysis above, it sufficient to consider 
\[\min_{t \in \mathbb{R}}\max_{\Delta \in \calI}\EE_{\boldsymbol{Y}_1, \boldsymbol{Y}_2}\left[R\left(\II\left\{\hat{\Delta}^N>t\right\}, \Delta\right)\right]\] \\
when $t=0$,
\begin{align}
\EE_{\boldsymbol{Y}_1, \boldsymbol{Y}_2}\left[R\left(\phi^0\left(\hat{\Delta}^N\right), \Delta\right)\right] =\II\left\{\Delta \neq0\right\}\cdot\Phi\left(-r(|\Delta|)\right)= \II\left\{\Delta \neq0\right\}\cdot\Phi\left(\frac{-|\Delta|}{\sqrt{V(\Delta)}}\right)
\end{align}
Suppose by contradiction that t = 0 is not minimax-regret. If the minimax-regret choice is $t>0$, then by choosing a $\Delta>0$ from $\calI$, the adversary achieves
\begin{align}
\EE_{\boldsymbol{Y}_1, \boldsymbol{Y}_2}\left[R\left(\phi^0\left(\hat{\Delta}^N\right), \Delta\right)\right] =  \Phi\left(t-r(|\Delta|)\right) > \Phi\left(-r(|\Delta|)\right)
\end{align}
which implies 
\begin{align}
\max_{\Delta \in \calI}\EE_{\boldsymbol{Y}_1, \boldsymbol{Y}_2}\left[R\left(\II\left\{\hat{\Delta}^N>t\right\}, \Delta\right)\right] \geq \max_{\Delta \in \calI}\EE_{\boldsymbol{Y}_1, \boldsymbol{Y}_2}\left[R\left(\II\left\{\hat{\Delta}^N>0\right\}, \Delta\right)\right] ,\quad \forall t >0
\end{align}
so we can not choose $t>0$. Similarly, If the minimax-regret choice is $t<0$, then by choosing a $\Delta<0$ from $\calI$, the adversary achieves
\begin{align}
\EE_{\boldsymbol{Y}_1, \boldsymbol{Y}_2}\left[R\left(\phi^0\left(\hat{\Delta}^N\right), \Delta\right)\right] =  \Phi\left(-r(|\Delta|)-t\right) > \Phi\left(-r(|\Delta|)\right)
\end{align}
which implies 
\begin{align}
\max_{\Delta \in \calI}\EE_{\boldsymbol{Y}_1, \boldsymbol{Y}_2}\left[R\left(\II\left\{\hat{\Delta}^N>t\right\}, \Delta\right)\right] \geq \max_{\Delta \in \calI}\EE_{\boldsymbol{Y}_1, \boldsymbol{Y}_2}\left[R\left(\II\left\{\hat{\Delta}^N>0\right\}, \Delta\right)\right], \quad \forall t <0
\end{align}
so we can not choose $t<0$.
Thus, $t=0$ is minimax-regret.

\end{proof} 

\subsection{Proof of \Cref{prop:adp.alloc.convergence}}\label{sec:proof:nonasymptotic}

\begin{proof}[Proof of \Cref{prop:adp.alloc.convergence}]

We multiply the objective function \eqref{eq:MSE.estimate} by $T_1$ and denote it as $\mathcal{R}_{T}$ to emphasis its dependence on $T$. For any fixed $T$, the optimal allocation proportions that minimize objective function~\eqref{eq:MSE.estimate}  is same as the one minimizes $\mathcal{R}_{T}$. Hence it's sufficient to work on the function $\mathcal{R}_{T}$. $\mathcal{R}_{T}$ can be written as a function of $p_{0,T}, p_{1,T},p_{2,T}$. For abbreviation, we drop the subscript $T$ in $p$. 

\begin{align}
\begin{split}
 \mathcal{R}_{T}(p_0, p_1, p_2) &= \frac{\hat{\sigma}^2_2}{p_2+m_2/T_1}\Phi\left(\hat{\Delta}\sqrt{\frac{T_1}{v(p)}}\right)- \frac{\hat{\sigma}^4_2}{(p_2+m_2/T_1)^2\cdot v(p)}\phi\left(\hat{\Delta}\sqrt{\frac{T_1}{v(p)}}\right)\cdot\hat{\Delta}\sqrt{\frac{T_1}{v(p)}}\\ &\quad~- \frac{\hat{\sigma}^4_2}{(p_2+m_2/T_1)^2\cdot v(p)}\frac{\phi^2\left(\hat{\Delta}\sqrt{\frac{T_1}{v(p)}}\right)}{\Phi\left(\hat{\Delta}\sqrt{\frac{T_1}{v(p)}}\right)} + (T_1 \cdot \hat{\Delta}^2+\frac{\hat{\sigma}^2_1}{p_1+m_1/T_1})\Phi\left(-\hat{\Delta}\sqrt{\frac{T_1}{v(p)}}\right) \\
  &\quad~+ \frac{\hat{\sigma}^4_1}{(p_1+m_1/T_1)^2\cdot v(p)}\phi\left(\hat{\Delta}\sqrt{\frac{T_1}{v(p)}}\right)\cdot \hat{\Delta}\sqrt{\frac{T_1}{v(p)}}- \frac{\hat{\sigma}^4_1}{(p_1+m_1/T_1)^2\cdot v(p)}\frac{\phi^2\left(\hat{\Delta}\sqrt{\frac{T_1}{v(p)}}\right)}{\Phi\left(-\hat{\Delta}\sqrt{\frac{T_1}{v(p)}}\right)} \\
 &\quad~+ \frac{\hat{\sigma}^2_0}{p_0+m_0/T_1}
\end{split}
\end{align}
where 
\begin{align}
v(p) = \frac{\hat{\sigma}^2_2}{p_2 + m_2/T_1}+\frac{\hat{\sigma}^2_1}{p_1 + m_1/T_1}
\end{align}
Without loss of generality, we assume $\Delta>0$ and the second treatment has largest mean. Define $\mathcal{P}_{T} = \{(p_0, p_1, p_2) \in [\delta_T, 1-2\delta_T]^3 : p_0 + p_1 + p_2 = 1\}$. Let  $\hat{p}_T$ as the optimal solution solving \eqref{eq:MSE.estimate} on $\mathcal{P}_T$. \\
To characterize the limiting behavior of the optimal solution, we first find a limit function for the original objective $
\mathcal{R}_T$ and determine its optimal solution. Then, we argue that the optimal solution for  $\mathcal{R}_T$ converges to the optimal solution of the limit function by leveraging the uniform convergence of
$\mathcal{R}_T$   on $\mathcal{P}_{T} $ and the fact that the minimizer of the limit function is a well-separated point. \\
 Define 
\begin{align}
\mathcal{R}(p_0, p_1, p_2) = \frac{\sigma^2_2}{p_2}+ \frac{\sigma^2_0}{p_0} 
\end{align}
We now establish the uniform consistency of $\mathcal{R}_T(p)$ to $\mathcal{R}(p)$ over the compact set $\mathcal{P}_T$ 
\begin{align*}
\sup_{p \in \mathcal{P}_{T}} \mid \mathcal{R}_{T}(p) - \mathcal{R}(p)\mid 
\end{align*}
To that end, we present the concentration results for the sample variance.
\begin{lemma}[High-probability bound for the sample variance under a finite fourth moment]\label{lemma:variance.hp}
Let \(X_1,\dots,X_n\) be i.i.d. real random variables with mean \(\mu=\mathbb{E}[X]\), variance \(\sigma^2=\var(X)\), and finite fourth moment
\[
\kappa_4:=\mathbb{E}\big[(X-\mu)^4\big]<\infty.
\]
Let
\[
\bar X_n=\frac{1}{n}\sum_{i=1}^n X_i,\qquad \widehat\sigma_n^2 := \frac{1}{n-1}\sum_{i=1}^n (X_i-\bar X_n)^2,
\]
Then for any \(\varepsilon\in(0,1)\), with probability at least \(1-\varepsilon\),
\begin{align*}
|\widehat\sigma_n^2-\sigma^2\big|
\le \frac{n}{n-1}\,\sqrt{\frac{2\kappa_4}{n\varepsilon}}
\;+\;\frac{2\sigma^2}{(n-1)\varepsilon}
\;+\;\frac{\sigma^2}{n-1}.
\end{align*}
Equivalently,
\[
\big|\widehat\sigma_n^2-\sigma^2\big|
\le \frac{n}{n-1}\,\sqrt{\frac{2\kappa_4}{n\varepsilon}}
\;+\;\frac{\sigma^2}{\,n-1\,}\Big(1+\frac{2}{\varepsilon}\Big).
\]
\end{lemma}

\begin{corollary}
\label{cor:app:variance-sample-size}
Under the assumptions of Lemma~\ref{lemma:variance.hp}, fix confidence $\varepsilon\in(0,1)$. 
Define
\[
n_0 := \max\!\left\{\,2,\;\frac{128\,\kappa_4}{\sigma^4\,\varepsilon},\; 8\Big(1+\frac{2}{\varepsilon}\Big) \right\}.
\]
Then for every integer $n\ge n_0$ it holds with probability at least $1-\varepsilon$ that
\[
\big|\widehat\sigma_n^2-\sigma^2\big| < \frac{\sigma^2}{2}.
\]
\end{corollary}
Define the following events.
\begin{align}
&\mathcal{E}(\Delta) := \left\{\big|\widehat\Delta-\Delta\big|
\le \sqrt{\sigma^2_1 +\sigma^2_2}\sqrt{\frac{12}{T_0\varepsilon}}\right\} \\
&\mathcal{E}(\sigma^2_0) := \left\{\big|\widehat\sigma_0^2-\sigma^2_0\big|
\le \frac{T_0}{T_0-3}\,\sqrt{\frac{24\kappa_0}{T_0\varepsilon}}
\;+\;\frac{3\sigma^2_0}{\,T_0-3\,}\Big(1+\frac{8}{\varepsilon}\Big)\right\} \\
&\mathcal{E}(\sigma^2_1) := \left\{\big|\widehat\sigma_1^2-\sigma^2_1\big|
\le \frac{T_0}{T_0-3}\,\sqrt{\frac{24\kappa_1}{T_0\varepsilon}}
\;+\;\frac{3\sigma^2_1}{\,T_0-3\,}\Big(1+\frac{8}{\varepsilon}\Big)\right\}\\
&\mathcal{E}(\sigma^2_2) := \left\{\big|\widehat\sigma_2^2-\sigma^2_2\big|
\le \frac{T_0}{T_0-3}\,\sqrt{\frac{24\kappa_2}{T_0\varepsilon}}
\;+\;\frac{3\sigma^2_2}{\,T_0-3\,}\Big(1+\frac{8}{\varepsilon}\Big)\right\}
\end{align}
Let \(n:=T_0/3\). \(\widehat\Delta\) is the sample mean of \(n\) i.i.d.\ observations with mean \(\Delta\) and variance \(\sigma_1^2+\sigma_2^2\). Then
\[
\var(\widehat\Delta)=\frac{\sigma_1^2+\sigma_2^2}{n}
=\frac{\sigma_1^2+\sigma_2^2}{T_0/3}
=\frac{3(\sigma_1^2+\sigma_2^2)}{T_0}.
\]
Applying Chebyshev's inequality with threshold
\[
t \;:=\; \sqrt{\sigma_1^2+\sigma_2^2}\,\sqrt{\frac{12}{T_0\varepsilon}},
\]
we obtain
\begin{align*}
\PP\big(\big|\widehat\Delta-\Delta\big|>t\big)
&\le \frac{\var(\widehat\Delta)}{t^2}
= \frac{\dfrac{3(\sigma_1^2+\sigma_2^2)}{T_0}}
{\big(\sigma_1^2+\sigma_2^2\big)\dfrac{12}{T_0\varepsilon}}\\[6pt]
&= \frac{3}{12/\varepsilon}
= \frac{\varepsilon}{4}.
\end{align*}
Therefore
\[
\PP\big(\mathcal{E}(\Delta)\big)
= 1 - \PP\big(\big|\widehat\Delta-\Delta\big|>t\big)
\ge 1 - \frac{\varepsilon}{4}.
\]
Applying Lemma~\ref{lemma:variance.hp} to the unit receiving treatment 0, 1, 2 in  stage 0 respectively, we have
\begin{align*}
\PP(\mathcal{E}(\sigma^2_0)) \geq 1-\frac{\varepsilon}{4} \\
\PP(\mathcal{E}(\sigma^2_1)) \geq 1-\frac{\varepsilon}{4}\\
\PP(\mathcal{E}(\sigma^2_2)) \geq 1-\frac{\varepsilon}{4}
\end{align*}
Denote the intersect of all above events as $\mathcal{E}$, i.e.,
\[= \mathcal{E}(\Delta) \cap\mathcal{E}(\sigma^2_0)\cap\mathcal{E}(\sigma^2_1)\cap\mathcal{E}(\sigma^2_2)\]
we have
\begin{align*}
\PP(\mathcal{E}) \geq 1-\PP(\overline{\mathcal{E}(\Delta)})-\PP(\overline{\mathcal{E}(\sigma^2_0)}) - \PP(\overline{\mathcal{E}(\sigma^2_1)})- \PP(\overline{\mathcal{E}(\sigma^2_2)}) \geq 1- \varepsilon
\end{align*}
For the reminder of the proof, we will assume this event hold. Furthermore, for $T_0 > \frac{48(\sigma^2_1+\sigma^2_2)}{\Delta^2\varepsilon}$, we have $\sqrt{\sigma^2_2 +\sigma^2_1}\sqrt{\frac{12}{T_0\varepsilon}} < \frac{\Delta}{2}$, namely
\[\big|\widehat\Delta-\Delta\big| \leq \
\Delta/2\]
Applying Corollary~\ref{cor:app:variance-sample-size} to the sample variance for treatment 0,1,2 respectively with $n=T_0/3$ and confidence $\varepsilon/4$ and taking the maximum over all the bound, we have for all $T_0 \geq  \max\!\left\{\,6,\;\frac{1536\,\kappa_0}{\sigma^4_0\,\varepsilon},\frac{1536\,\kappa_1}{\sigma^4_1\,\varepsilon},\frac{1536\,\kappa_2}{\sigma^4_2\,\varepsilon},\; 24\Big(1+\frac{8}{\varepsilon}\Big) \right\}$ 
\begin{align*}
\big|\widehat\sigma_0^2-\sigma^2_0\big| \leq \sigma^2_0/2 \\
\big|\widehat\sigma_1^2-\sigma^2_1\big| \leq \sigma^2_1/2 \\
\big|\widehat\sigma_2^2-\sigma^2_2\big| \leq \sigma^2_2/2
\end{align*}
Also, for sufficient large $T_0$, the $T_0^{-1/2}$ term in the error of sample variance will dominate  the term involving $T_0^{-1}$, leading to 
\begin{align*}
\big|\widehat\sigma_0^2-\sigma^2_0\big|
\le 3\,\sqrt{\frac{24\kappa_0}{T_0\varepsilon}} \\
\big|\widehat\sigma_1^2-\sigma^2_1\big|
\le 3\,\sqrt{\frac{24\kappa_1}{T_0\varepsilon}} \\
\big|\widehat\sigma_2^2-\sigma^2_2\big|
\le 3\,\sqrt{\frac{24\kappa_2}{T_0\varepsilon}} 
\end{align*}
for $T_0 \geq  \max\!\left\{\frac{9\sigma^2_0(1+8/\varepsilon)^2\varepsilon+72\kappa_0}{24\kappa_0} , \frac{9\sigma^2_1(1+8/\varepsilon)^2\varepsilon+72\kappa_1}{24\kappa_1}, \frac{9\sigma^2_2(1+8/\varepsilon)^2\varepsilon+72\kappa_2}{24\kappa_2}\right\}$.
In the following proof, we will work for the 
\begin{equation*}
T_0 \geq \max \left\{
\begin{multlined}
\frac{48(\sigma^2_1+\sigma^2_2)}{\Delta^2\varepsilon},\; 6,\;
\frac{1536\kappa_0}{\sigma^4_0\varepsilon},\;
\frac{1536\kappa_1}{\sigma^4_1\varepsilon},\;
\frac{1536\kappa_2}{\sigma^4_2\varepsilon},\; \\
24\left(1+\frac{8}{\varepsilon}\right),\;
\frac{9\sigma^2_0(1+8/\varepsilon)^2\varepsilon+72\kappa_0}{24\kappa_0}, \\
\frac{9\sigma^2_1(1+8/\varepsilon)^2\varepsilon+72\kappa_1}{24\kappa_1},\;
\frac{9\sigma^2_2(1+8/\varepsilon)^2\varepsilon+72\kappa_2}{24\kappa_2}
\end{multlined}
\right\}
\end{equation*}
The parameter space $\mathcal{P}_T$ is a compact set where each allocation proportion $p_w$, $w \in\left\{0,1,2\right\}$ is bounded below by $\delta_T > 0$. This compactness allows us to uniformly control the function $v(p)$ for each fixed $T$. Specifically, for all $p \in \mathcal{P}_T$, we have 
\begin{align*}
v(p) &= \frac{\hat{\sigma}^2_2}{p_2 + m_2/T_1}+\frac{\hat{\sigma}^2_1}{p_1 + m_1/T_1} \\
& \quad~\leq    \frac{\hat{\sigma}^2_2}{p_2}+\frac{\hat{\sigma}^2_1}{p_1} \\
& \quad~\leq \frac{ \hat{\sigma}^2_1+\hat{\sigma}^2_2}{\delta_T} := v_{\text{max}} \quad \text{a.s.}
\end{align*}
and similarly,
\begin{align*}
v(p) \geq \frac{ \hat{\sigma}^2_1+\hat{\sigma}^2_2}{2} := v_{\text{min}}
\end{align*}
These bounds allow us to control the deviation 
\[\sup_{p \in \mathcal{P}_{T}} \mid \mathcal{R}_{T}(p) - \mathcal{R}(p)\mid \]
Specifically, we obtain
\begin{align}\label{proof:eq:consistency}
\begin{split}
\sup_{p \in \mathcal{P}_{T}} \mid \mathcal{R}_{T}(p) - \mathcal{R}(p)\mid 
&\overset{(a)}{\leq} \sup_{p \in \mathcal{P}_{T}} \mid \frac{\hat{\sigma}^2_2}{p_2+m_2/T_1}\Phi\left(\hat{\Delta}\sqrt{\frac{T_1}{v(p)}}\right)- \frac{\sigma^2_2}{p_2+m_2/T_1}\mid \\
&\quad~\overset{(b)}{+}\sup_{p \in \mathcal{P}_{T}}\mid \frac{\sigma^2_2}{p_2+m_2/T_1} - \frac{\sigma^2_2}{p_2} \mid\\
&\quad~\overset{(c)}{+} \sup_{p \in \mathcal{P}_{T}}\mid\frac{\hat{\sigma}^2_0}{p_0+m_0/T_1}- \frac{\sigma^2_0}{p_0+m_0/T_1}\mid \\
&\quad~\overset{(d)}{+} \sup_{p \in \mathcal{P}_{T}}\mid\frac{\sigma^2_0}{p_0+m_0/T_1}- \frac{\sigma^2_0}{p_0}\mid \\
&\quad~\overset{(e)}{+} \frac{\hat{\sigma}^4_2}{\delta^2_T \cdot v(p_{\text{min}})}\phi\left(\hat{\Delta}\sqrt{\frac{T_1}{v(p_{\text{max}})}}\right)\cdot\hat{\Delta}\sqrt{\frac{T_1}{v(p_{\text{min}})}} \\
&\quad~\overset{(f)}{+} \frac{\hat{\sigma}^4_2}{\delta^2_T \cdot v(p_{\text{min}})}\frac{\phi^2\left(\hat{\Delta}\sqrt{\frac{T_1}{v(p_{\text{max}})}}\right)}{\Phi\left(\hat{\Delta}\sqrt{\frac{T_1}{v(p_{\text{max}})}}\right)} \\
&\quad~\overset{(g)}{+} (T_1 \cdot \hat{\Delta}^2+\frac{\hat{\sigma}^2_1}{\delta_T})\Phi\left(-\hat{\Delta}\sqrt{\frac{T_1}{v(p_{\text{max}})}}\right) \\
&\quad~\overset{(h)}{+} \frac{\hat{\sigma}^4_1}{\delta^2_T\cdot v(p_{\text{min}})}\phi\left(\hat{\Delta}\sqrt{\frac{T_1}{v(p_{\text{max}})}}\right)\cdot\hat{\Delta}\sqrt{\frac{T_1}{v(p_{\text{min}})}} \\
&\quad~\overset{(i)}{+} \frac{\hat{\sigma}^4_1}{\delta^2_T\cdot v(p_{\text{min}})}\frac{\phi^2\left(\hat{\Delta}\sqrt{\frac{T_1}{v(p_{\text{max}})}}\right)}{\Phi\left(-\hat{\Delta}\sqrt{\frac{T_1}{v(p_{\text{min}})}}\right)}
\end{split}  
\quad \text{a.s. }
\end{align}
For $(a)$, we have
\begin{align*}
&\sup_{p \in \mathcal{P}_{T}} \mid \frac{\hat{\sigma}^2_2}{p_2+m_2/T_1}\Phi\left(\hat{\Delta}\sqrt{\frac{T_1}{v(p)}}\right)- \frac{\sigma^2_2}{p_2+m_2/T_1}\mid   \leq \sup_{p \in \mathcal{P}_{T}} \mid \frac{\hat{\sigma}^2_2}{\delta_T}\Phi\left(\hat{\Delta}\sqrt{\frac{T_1}{v(p)}}\right)- \frac{\hat{\sigma}^2_2}{\delta_T}\mid +  \sup_{p \in \mathcal{P}_{T}} \mid \frac{\hat{\sigma}^2_2}{\delta_T} - \frac{\sigma^2_2}{\delta_T}\mid \\
&= \frac{\hat{\sigma}^2_2}{\delta_T}\cdot \sup_{p \in \mathcal{P}_{T}} \left(1-\Phi\left(\hat{\Delta}\sqrt{\frac{T_1}{v(p)}}\right)\right)  + \frac{1}{\delta_T}\cdot \mid \hat{\sigma}^2_2 - \sigma^2_2\mid \\
&= \frac{\hat{\sigma}^2_2}{\delta_T}\cdot \sup_{p \in \mathcal{P}_{T}} \Phi\left(-\hat{\Delta}\sqrt{\frac{T_1}{v(p)}}\right) + 2T_0^{\alpha}\cdot3\,\sqrt{\frac{24\kappa_2}{T_0\varepsilon}}  \\
& \leq 3\sigma^2_2 \cdot T^{\alpha}_0 \cdot  \Phi\left(-\Delta/2 \sqrt{\frac{T_1}{v_\text{max}}}\right) + 6\cdot\,\sqrt{\frac{24\kappa_2}{\varepsilon}}T_0^{-(1/2-\alpha)} \\
& = 3\sigma^2_2 \cdot T^{\alpha}_0\cdot  \Phi\left(-\Delta/2 \sqrt{\frac{T_1/2 \cdot T^{-\alpha}_0}{\hat{\sigma}^2_1+\hat{\sigma}^2_2 }}\right) + 6\cdot\,\sqrt{\frac{24\kappa_2}{\varepsilon}}T_0^{-(1/2-\alpha)} \\
& \leq 3\sigma^2_2 \cdot T^{\alpha}_0\cdot  \Phi\left(-\Delta/2 \sqrt{\frac{T_1 T^{-\alpha}_0}{3(\sigma^2_1+\sigma^2_2 )}}\right) + 6\cdot\,\sqrt{\frac{24\kappa_2}{\varepsilon}}T_0^{-(1/2-\alpha)}
\end{align*}
By applying the Gaussian tail bound 
\[
\Phi(-x) < \frac{1}{x\sqrt{2\pi}} e^{-x^2/2}
\]
we obtain the following results
\begin{align*}
&\sup_{p \in \mathcal{P}_{T}} \mid \frac{\hat{\sigma}^2_2}{p_2+m_2/T_1}\Phi\left(\hat{\Delta}\sqrt{\frac{T_1}{v(p)}}\right)- \frac{\sigma^2_2}{p_2+m_2/T_1}\mid   
 \leq 3\sigma^2_2 \cdot T^{\alpha}_0\cdot  \sqrt{\frac{12\cdot  T^{\alpha}_0 (\sigma^2_1+\sigma^2_2) }{T_1  \cdot \Delta^2}} \frac{1}{\sqrt{2 \pi}} e^{-\frac{\Delta^2}{8} \cdot \frac{T_1  T_0^{-\alpha}}{3(\sigma^2_1+\sigma^2_2 )}} + 6\cdot\,\sqrt{\frac{24\kappa_2}{\varepsilon}}T_0^{-(1/2-\alpha)}\\
& \leq 3\sigma^2_2 \cdot T^{3/2}_1 \cdot  \sqrt{\frac{12\cdot  (\sigma^2_1+\sigma^2_2) }{  \Delta^2}} \frac{1}{\sqrt{2 \pi}} e^{-\frac{\Delta^2}{8} \cdot \frac{T_1^{1/2}}{3(\sigma^2_1+\sigma^2_2 )}} + 6\cdot\,\sqrt{\frac{24\kappa_2}{\varepsilon}}T_0^{-(1/2-\alpha)} \\
\end{align*} 
For $T > 2^{\tfrac{1}{1-\beta}}$, we have $T_0 < T_1$. 
Let 
\[
a = \frac{\Delta^2}{24(\sigma_1^2 + \sigma_2^2)}.
\]
If 
\[
T_1 > \left(\frac{12}{a}\ln\!\left(\frac{12}{a}\right)\right)^2,
\]
then 
\[
\exp(-aT_1)\, T_1^{3/2} \;\leq\; T_1^{-(1/2-\alpha)} \;\leq\; T_0^{-(1/2-\alpha)} \quad \text{when } T_0 < T_1.
\]
Thus, it suffices to require
\[
T_0 = T^{\beta} > \left(\frac{12}{a}\ln\!\left(\frac{12}{a}\right)\right)^2.
\]
Consequently, for
\[
T > \max \left\{ 2^{\tfrac{1}{1-\beta}}, \left(\frac{12}{a}\ln\!\left(\frac{12}{a}\right)\right)^{2/\beta} \right\},
\]
we have
\begin{align*}
\sup_{p \in \mathcal{P}_{T}} \mid \frac{\hat{\sigma}^2_2}{p_2+m_2/T_1}\Phi\left(\hat{\Delta}\sqrt{\frac{T_1}{v(p)}}\right)- \frac{\sigma^2_2}{p_2+m_2/T_1}\mid \leq 3\sigma^2_2 \cdot  \sqrt{\frac{1}{2a}} \frac{1}{\sqrt{2 \pi}} T_0^{-(1/2-\alpha)} + 6\cdot\,\sqrt{\frac{24\kappa_2}{\varepsilon}}T_0^{-(1/2-\alpha)}
\end{align*}
For $(b)$
\begin{align*}
\sup_{p \in \mathcal{P}_{T}}\mid \frac{\sigma^2_2}{p_2+m2/T_1} - \frac{\sigma^2_2}{p_2} \mid &=   \sup_{p \in \mathcal{P}_{T}}\mid \frac{\sigma^2_2\cdot m_2/T_1} {(p_2+m_2/T_1)\cdot p_2}\mid\\
& \leq  \frac{\sigma^2_2\cdot m_2/T_1} {\delta_T^2}\\ 
& = \frac{4\sigma^2_2\cdot T^{\beta}\cdot T^{2\alpha\beta}} {3(T-T^{\beta})} \\
& = \frac{4\sigma^2_2\cdot T^{(2\alpha+1)\beta}} {3(T-T^{\beta})} \\
& \leq \frac{8\sigma^2_2\cdot T^{(2\alpha+1)\beta}} {3T} \\
& = \frac{8\sigma^2_2\cdot T^{(2\alpha+1)\beta-1}} {3}
\end{align*}
Similarly, for $(d)$
\begin{align*}
\sup_{p \in \mathcal{P}_{T}}\mid\frac{\sigma^2_0}{p_0+m_0/T_1}- \frac{\sigma^2_0}{p_0}\mid &\leq \frac{4\sigma^2_0\cdot T^{(2\alpha+1)\beta}} {3(T-T^{\beta})} \\
& \leq \frac{8\sigma^2_0\cdot T^{(2\alpha+1)\beta-1}} {3}
\end{align*}
For $(c)$ 
\begin{align*}
\sup_{p \in \mathcal{P}_{T}}\mid\frac{\hat{\sigma}^2_0}{p_0+m_0/T_1}- \frac{\sigma^2_0}{p_0+m_0/T_1}\mid &\leq \frac{1}{\delta_T}\cdot \mid \hat{\sigma}^2_0 - \sigma^2_0\mid \\
& = 2T_0^{\alpha}\cdot3\,\sqrt{\frac{24\kappa_0}{T_0\varepsilon}} \\
& = 6\cdot\,\sqrt{\frac{24\kappa_0}{\varepsilon}}T_0^{-(1/2-\alpha)}
\end{align*}
For $(e)$
\begin{align*}
\frac{\hat{\sigma}^4_2}{\delta^2_T \cdot v(p_{\text{min}})}\phi\left(\hat{\Delta}\sqrt{\frac{T_1}{v(p_{\text{max}})}}\right)\cdot\hat{\Delta}\sqrt{\frac{T_1}{v(p_{\text{min}})}}
& = \frac{2\hat{\sigma}^4_2}{ (\hat{\sigma}^2_1+\hat{\sigma}^2_2 )\cdot \delta^2_T} \cdot \phi\left(\hat{\Delta}\sqrt{\frac{T_1\delta_T}{\hat{\sigma}^2_1+\hat{\sigma}^2_2 }}\right)\cdot \hat{\Delta}\sqrt{\frac{2\cdot T_1\delta_T}{\hat{\sigma}^2_1+\hat{\sigma}^2_2 }} \\
& \leq \frac{36\sigma^4_2 T^{2\alpha}_0}{ (\sigma^2_1+\sigma^2_2 )} \cdot \phi\left(\Delta/2 \sqrt{\frac{T_1 \cdot T^{-\alpha}_0}{3({\sigma}^2_1+\sigma^2_2)}}\right)\cdot 3\Delta /2 \sqrt{\frac{2 \cdot T_1 \cdot T^{-\alpha}_0}{\sigma^2_1+\sigma^2_2 }} \\
& \leq  \frac{36\sigma^4_2 T^{2\alpha}_0}{ (\sigma^2_1+\sigma^2_2 )}\cdot\frac{1}{\sqrt{2 \pi}} e^{-\frac{\Delta^2}{8} \cdot \frac{T_1^{1/2}}{3(\sigma^2_1+\sigma^2_2 )}}\cdot 3\Delta /2 \sqrt{\frac{2 \cdot T_1 \cdot T^{-\alpha}_0}{\sigma^2_1+\sigma^2_2 }}\\
& \leq \frac{36\sigma^4_2 T^{2\alpha+1/2}_1}{ (\sigma^2_1+\sigma^2_2 )}\cdot\frac{1}{\sqrt{2 \pi}} e^{-\frac{\Delta^2}{8} \cdot \frac{T_1^{1/2}}{3(\sigma^2_1+\sigma^2_2 )}}\cdot 3\Delta /2 \sqrt{\frac{2   }{\sigma^2_1+\sigma^2_2 }}
\end{align*}
Following the same analysis as in $(a)$, we arrive at for $T > \max \left\{ 2^{\frac{1}{1-\beta}}, (\frac{12}{a}\ln(\frac{12}{a}))^{2/\beta}\right\}$
\begin{align*}
\frac{\hat{\sigma}^4_2}{\delta^2_T \cdot v(p_{\text{min}})}\phi\left(\hat{\Delta}\sqrt{\frac{T_1}{v(p_{\text{max}})}}\right)\cdot\hat{\Delta}\sqrt{\frac{T_1}{v(p_{\text{min}})}} \leq \frac{54\sigma^4_2 }{ \sigma^2_1+\sigma^2_2 }\cdot\frac{1}{\sqrt{2 \pi}} \cdot \Delta  \sqrt{\frac{2   }{\sigma^2_1+\sigma^2_2 }}T^{-(1/2 -\alpha)}_0
\end{align*}
The same analysis holds for $(h)$. For $T > \max \left\{ 2^{\frac{1}{1-\beta}}, (\frac{12}{a}\ln(\frac{12}{a}))^{2/\beta}\right\}$,
\begin{align*}
\frac{\hat{\sigma}^4_1}{\delta^2_T\cdot v(p_{\text{min}})}\phi\left(\hat{\Delta}\sqrt{\frac{T_1}{v(p_{\text{max}})}}\right)\cdot\hat{\Delta}\sqrt{\frac{T_1}{v(p_{\text{min}})}} \leq \frac{54\sigma^4_1 }{ \sigma^2_1+\sigma^2_2 }\cdot\frac{1}{\sqrt{2 \pi}} \cdot \Delta  \sqrt{\frac{2   }{\sigma^2_1+\sigma^2_2 }}T^{-(1/2 -\alpha)}_0
\end{align*}
For $(f)$, 
\begin{align*}
\frac{\hat{\sigma}^4_2}{\delta^2_T \cdot v(p_{\text{min}})}\frac{\phi^2\left(\hat{\Delta}\sqrt{\frac{T_1}{v(p_{\text{max}})}}\right)}{\Phi\left(\hat{\Delta}\sqrt{\frac{T_1}{v(p_{\text{max}})}}\right)} 
& \leq \frac{2\hat{\sigma}^4_2}{ (\hat{\sigma}^2_1+\hat{\sigma}^2_2 )\cdot \delta^2_T}\frac{\phi^2\left(\Delta/2\sqrt{\frac{T_1}{v(p_{\text{max}})}}\right)}{\Phi\left(\Delta/2\sqrt{\frac{T_1}{v(p_{\text{max}})}}\right)} \\
& \leq \frac{36\sigma^4_2 T^{2\alpha}_0}{ (\sigma^2_1+\sigma^2_2 )}\frac{\phi^2\left(\Delta/2\sqrt{\frac{T_1}{v(p_{\text{max}})}}\right)}{\Phi\left(-\Delta/2\sqrt{\frac{T_1}{v(p_{\text{max}})}}\right)} \\
& \leq\frac{36\sigma^4_2 T^{2\alpha}_0}{ (\sigma^2_1+\sigma^2_2 )} \frac{\frac{\Delta^2 T_1  T_0^{-\alpha}}{12(\sigma^2_1+\sigma^2_2 )}+1}{\sqrt{\frac{\Delta^2 T_1  T_0^{-\alpha}}{12(\sigma^2_1+\sigma^2_2 )}}}\frac{1}{\sqrt{2 \pi}} e^{-\frac{\Delta^2}{8} \cdot \frac{T_1  T_0^{-\alpha}}{3(\sigma^2_1+\sigma^2_2 )}} \\
\end{align*}
 For $ T > (\frac{12(\sigma^2_1+\sigma^2_2)}{\Delta^2})^{\frac{2}{\beta}}$, we have $\frac{\Delta^2 T_1  T_0^{-\alpha}}{12(\sigma^2_1+\sigma^2_2 )} >1$, which implies
 \[\frac{\frac{\Delta^2 T_1  T_0^{-\alpha}}{12(\sigma^2_1+\sigma^2_2 )}+1}{\sqrt{\frac{\Delta^2 T_1  T_0^{-\alpha}}{12(\sigma^2_1+\sigma^2_2 )}}} < 2\sqrt{\frac{\Delta^2 T_1  T_0^{-\alpha}}{12(\sigma^2_1+\sigma^2_2 )}}\]
 Hence  for $T > \max \left\{(\frac{12(\sigma^2_1+\sigma^2_2)}{\Delta^2})^{\frac{2}{\beta}}, 2^{\frac{1}{1-\beta}}, (\frac{12}{a}\ln(\frac{12}{a}))^{2/\beta}\right\}$
\begin{align*}
\frac{\hat{\sigma}^4_2}{\delta^2_T \cdot v(p_{\text{min}})}\frac{\phi^2\left(\hat{\Delta}\sqrt{\frac{T_1}{v(p_{\text{max}})}}\right)}{\Phi\left(\hat{\Delta}\sqrt{\frac{T_1}{v(p_{\text{max}})}}\right)} & \leq \frac{72\sigma^4_2 T^{2\alpha}_0}{ (\sigma^2_1+\sigma^2_2 )}\sqrt{\frac{\Delta^2 T_1  T_0^{-\alpha}}{12(\sigma^2_1+\sigma^2_2 )}}\frac{1}{\sqrt{2 \pi}}e^{-\frac{\Delta^2}{8} \cdot \frac{T_1  T_0^{-\alpha}}{3(\sigma^2_1+\sigma^2_2 )}} \\
& \leq \frac{72\sigma^4_2 T^{2\alpha+1}}{ (\sigma^2_1+\sigma^2_2 )}\sqrt{\frac{\Delta^2 }{12(\sigma^2_1+\sigma^2_2 )}}\frac{1}{\sqrt{2 \pi}}e^{-\frac{\Delta^2}{8} \cdot \frac{T_1  T_0^{-\alpha}}{3(\sigma^2_1+\sigma^2_2 )}} \\
& \leq \frac{72\sigma^4_2 }{ (\sigma^2_1+\sigma^2_2 )}\sqrt{\frac{\Delta^2 }{12(\sigma^2_1+\sigma^2_2 )}}\frac{1}{\sqrt{2 \pi}}T^{-(1/2 -\alpha)}_0
\end{align*}
Similarly, for $(i)$, 
\begin{align*}
\frac{\hat{\sigma}^4_1}{\delta^2_T\cdot v(p_{\text{min}})}\frac{\phi^2\left(\hat{\Delta}\sqrt{\frac{T_1}{v(p_{\text{max}})}}\right)}{\Phi\left(-\hat{\Delta}\sqrt{\frac{T_1}{v(p_{\text{min}})}}\right)} \leq \frac{72\sigma^4_1 }{ (\sigma^2_1+\sigma^2_2 )}\sqrt{\frac{3\Delta^2 }{4(\sigma^2_1+\sigma^2_2 )}}\frac{1}{\sqrt{2 \pi}}T^{-(1/2 -\alpha)}_0
\end{align*}
for $T > \max \left\{(\frac{12(\sigma^2_1+\sigma^2_2)}{\Delta^2})^{\frac{2}{\beta}}, 2^{\frac{1}{1-\beta}}, (\frac{12}{a}\ln(\frac{12}{a}))^{2/\beta}\right\}$.
Lastly, for $(g)$,
\begin{align*}
(T_1 \cdot \hat{\Delta}^2+\frac{\hat{\sigma}^2_1}{\delta_T})\Phi\left(-\hat{\Delta}\sqrt{\frac{T_1}{v(p_{\text{max}})}}\right) & \leq \left(T_1 \cdot \frac{9\Delta^2}{4}+3\sigma^2_1 T_0^{\alpha}\right)\sqrt{\frac{T_0^{\alpha} \cdot12\cdot  (\sigma^2_1+\sigma^2_2) }{  \Delta^2}} \frac{1}{\sqrt{2 \pi}} e^{-\frac{\Delta^2}{8} \cdot \frac{T_1^{1/2}}{3(\sigma^2_1+\sigma^2_2 )}} \\
& \leq \left(T_1^{1+\alpha/2} \cdot \frac{9\Delta^2}{4}+3\sigma^2_1 T_0^{3\alpha/2}\right)\sqrt{\frac{12\cdot  (\sigma^2_1+\sigma^2_2) }{  \Delta^2}} \frac{1}{\sqrt{2 \pi}} e^{-\frac{\Delta^2}{8} \cdot \frac{T_1^{1/2}}{3(\sigma^2_1+\sigma^2_2 )}} \\
& \leq  \left(\frac{9\Delta^2}{4}+3\sigma^2_1 \right)\sqrt{\frac{12\cdot  (\sigma^2_1+\sigma^2_2) }{  \Delta^2}} \frac{1}{\sqrt{2 \pi}} T_0^{-(1/2-\alpha)} \\
& = \left(\frac{9\Delta^2}{4}+3\sigma^2_1 \right)\sqrt{\frac{1}{2  a}} \frac{1}{\sqrt{2 \pi}} T_0^{-(1/2-\alpha)}
\end{align*}
for $T > \max \left\{ 2^{\frac{1}{1-\beta}}, (\frac{12}{a}\ln(\frac{12}{a}))^{2/\beta}\right\}$. \\
Putting all together, we arrive at
\begin{align}\label{eq:unif.consis}
\begin{split}
\sup_{p \in \mathcal{P}_{T}} \mid \mathcal{R}_{T}(p) - \mathcal{R}(p)\mid &\leq 3(\sigma^2_2+\sigma^2_1) \cdot  \sqrt{\frac{1}{2a}} \frac{1}{\sqrt{2 \pi}} T_0^{-(1/2-\alpha)} + 12\cdot\,\left(\sqrt{\frac{6\kappa_2}{\varepsilon}}+\sqrt{\frac{6\kappa_2}{\varepsilon}}\right)T_0^{-(1/2-\alpha)} + \frac{9\Delta^2}{4}\sqrt{\frac{1}{  2a}} \frac{1}{\sqrt{2 \pi}} T_0^{-(1/2-\alpha)} \\
&\quad~+ \frac{8(\sigma^2_2+\sigma^2_0)\cdot T^{(2\alpha+1)\beta-1}} {3} + \frac{54(\sigma^4_1+\sigma^4_2) }{ \sigma^2_1+\sigma^2_2 }\cdot\frac{1}{\sqrt{2 \pi}} \cdot \Delta  \sqrt{\frac{2   }{\sigma^2_1+\sigma^2_2 }}T^{-(1/2 -\alpha)}_0 \\
&\quad~+ \frac{216(\sigma^4_1+\sigma^4_2) }{ (\sigma^2_1+\sigma^2_2 )}\sqrt{3a}\frac{1}{\sqrt{2 \pi}}T^{-(1/2 -\alpha)}_0
\end{split}
\end{align}
Let $p^*_T$ be the optimal solution for $\mathcal{R}(p)$ solved on $\mathcal{P}_T$, we have
\begin{align}\label{proof:eq:R.optimal}
p^*_T = \left(\frac{\sigma_0}{\sigma_2+\sigma_0}\cdot(1-\delta_T), \delta_T, \frac{\sigma_2}{\sigma_2+\sigma_0}\cdot(1-\delta_T)\right)
\end{align}
Define $p^*$ as the limit point of $p^*_T$ as $T \to \infty$.
\begin{align}\label{proof:eq:R.limit.optimal}
p^*_T = \left(\frac{\sigma_0}{\sigma_2+\sigma_0}, 0, \frac{\sigma_2}{\sigma_2+\sigma_0}\right)
\end{align}
we have
\begin{align*}
\big|\mathcal{R}(p^*_T) - \mathcal{R}(p^*)\big| &= {}\frac{(\sigma_2+\sigma_1)^2}{1-\delta_T} - (\sigma_2+\sigma_1)^2 \\
&  = \frac{\delta_T}{1-\delta_T}(\sigma_2+\sigma_1)^2 \\
& = \frac{\frac{1}{2}T_0^{-\alpha}}{1-\frac{1}{2}T_0^{-\alpha}}(\sigma_2+\sigma_1)^2
\end{align*}
For $T > (\frac{2}{3})^{\frac{1}{\alpha \beta}}$, we have $\frac{\frac{1}{2}T_0^{-\alpha}}{1-\frac{1}{2}T_0^{-\alpha}} < 2T_0^{-\alpha}$, hence
\begin{align}\label{proof:eq:R.convergence}
\big|\mathcal{R}(p^*_T) - \mathcal{R}(p^*)\big| \leq 2T_0^{-\alpha}(\sigma_2+\sigma_1)^2
\end{align}
Since $\hat{p}_T$ is  the minimizer of $\mathcal{R}_T$ on $\mathcal{P}_T$ and $p^*_T \in \mathcal{P}_T$, we have
\begin{align*}
\mathcal{R}_T(\hat{p}_T) \leq \mathcal{R}_T(p^*_T)
\end{align*}
Hence
\begin{align}\label{proof:eq:R.convergence.2}
\begin{split}
\mathcal{R}(\hat{p}_T) - \mathcal{R}(p^*) &= \mathcal{R}(\hat{p}_T)  - \mathcal{R}_T(p^*_T) +  \mathcal{R}_T(p^*_T) - R(p^*)\\
&  =\mathcal{R}(\hat{p}_T)  - \mathcal{R}_T(p^*_T) +  \mathcal{R}_T(p^*_T) - R(p^*_T)+ R(p^*_T)- R(p^*) \\
& \leq  \mathcal{R}(\hat{p}_T)  - \mathcal{R}_T(\hat{p}_T) +  \mathcal{R}_T(p^*_T) - R(p^*_T)+ R(p^*_T)- R(p^*)\\
& \leq \sup_{p \in \mathcal{P}_{T}} \mid \mathcal{R}_{T}(p) - \mathcal{R}(p)\mid + \sup_{p \in \mathcal{P}_{T}} \mid \mathcal{R}_{T}(p) - \mathcal{R}(p)\mid + \big|\mathcal{R}(p^*_T) - \mathcal{R}(p^*)\big| \\
\end{split}
\end{align}
which implies 
\begin{align*}
\begin{split}
\mathcal{R}(\hat{p}_T) - \mathcal{R}(p^*) &\leq 6(\sigma^2_2+\sigma^2_1) \cdot  \sqrt{\frac{1}{2a}} \frac{1}{\sqrt{2 \pi}} T_0^{-(1/2-\alpha)} + 24\cdot\,\left(\sqrt{\frac{6\kappa_2}{\varepsilon}}+\sqrt{\frac{6\kappa_2}{\varepsilon}}\right)T_0^{-(1/2-\alpha)} + \frac{9\Delta^2}{2}\sqrt{\frac{1}{  2a}} \frac{1}{\sqrt{2 \pi}} T_0^{-(1/2-\alpha)} \\
&\quad~+ \frac{16(\sigma^2_2+\sigma^2_0)\cdot T^{(2\alpha+1)\beta-1}} {3} + \frac{108(\sigma^4_1+\sigma^4_2) }{ \sigma^2_1+\sigma^2_2 }\cdot\frac{1}{\sqrt{2 \pi}} \cdot \Delta  \sqrt{\frac{2   }{\sigma^2_1+\sigma^2_2 }}T^{-(1/2 -\alpha)}_0 \\
&\quad~+ \frac{432(\sigma^4_1+\sigma^4_2) }{ (\sigma^2_1+\sigma^2_2 )}\sqrt{3a}\frac{1}{\sqrt{2 \pi}}T^{-(1/2 -\alpha)}_0 + 2T_0^{-\alpha}(\sigma_2+\sigma_1)^2 \\
& = O(T^{\min \left\{-(1/2-\alpha)\beta, -\alpha\beta, (2\alpha+1)\beta-1\right\}})
\end{split}
\end{align*}
for  
\begin{equation*}
T \geq \max \left\{
\begin{multlined}
\left(\frac{48(\sigma^2_1+\sigma^2_2)}{\Delta^2\varepsilon}\right)^{\frac{1}{\beta}},\;
6^{\frac{1}{\beta}},\;
\left(\frac{1536\kappa_0}{\sigma^4_0\varepsilon}\right)^{\frac{1}{\beta}},\;
\left(\frac{1536\kappa_1}{\sigma^4_1\varepsilon}\right)^{\frac{1}{\beta}},\;
\left(\frac{1536\kappa_2}{\sigma^4_2\varepsilon}\right)^{\frac{1}{\beta}},\; \\
\left(24\left(1+\frac{8}{\varepsilon}\right)\right)^{\frac{1}{\beta}},\;
\left(\frac{9\sigma^2_0(1+8/\varepsilon)^2\varepsilon+72\kappa_0}{24\kappa_0}\right)^{\frac{1}{\beta}},\;
\left(\frac{9\sigma^2_1(1+8/\varepsilon)^2\varepsilon+72\kappa_1}{24\kappa_1}\right)^{\frac{1}{\beta}},\; \\
\left(\frac{9\sigma^2_2(1+8/\varepsilon)^2\varepsilon+72\kappa_2}{24\kappa_2}\right)^{\frac{1}{\beta}},\;
\left(\frac{12(\sigma^2_1+\sigma^2_2)}{\Delta^2}\right)^{\frac{2}{\beta}},\;
2^{\frac{1}{1-\beta}},\;
\left(\frac{12}{a}\ln\left(\frac{12}{a}\right)\right)^{2/\beta},\;
\left(\frac{2}{3}\right)^{\frac{1}{\alpha \beta}}
\end{multlined}
\right\}
\end{equation*}
 where  $a = \frac{\Delta^2}{24(\sigma^2_1+\sigma^2_2 )}$.
By setting $\alpha = \frac{1}{4}$ and $\beta = \frac{4}{7}$, we have $(2\alpha+1)\beta-1 = -(1/2-\alpha)\beta= -\alpha\beta = -1/7$. \\
Now we have bounded $\mathcal{R}(\hat{p}_T) - \mathcal{R}(p^*)$. The function $\mathcal{R}(p_0, p_1, p_2)$ only depends on variable $p_0, p_2$. It's strictly convex on the domain  $\Bar{\mathcal{P}}:=\{(p_0,p_1, p_2) \in (0,1)\times[0,1]\times(0,1) : p_0+p_1 + p_2 = 1\}$ and admits a global minimizer $p_0 = \frac{\sigma_0}{\sigma_0 + \sigma_2},p_1=0,  p_2=\frac{\sigma_2}{\sigma_0 + \sigma_2}$. The following proposition allow us to further bounds the distance between $\hat{p}_T$ and $p^*$.

\begin{proposition}\label{prop.neyman.delta}
Let $\sigma_1,\sigma_2>0$. For $p=(p_0,p_1,p_2)$  on $\mathcal{P} :=\{(p_0,p_1, p_2) \in[0,1] \times(0,1)\times(0,1) : p_0+p_1 + p_2 = 1\}$, define the Neyman-type risk
\[
R(p)=\frac{\sigma_2^2}{p_2}+\frac{\sigma_1^2}{p_1}.
\]
The Neyman minimizer is
\[
p_0^*=0,\qquad p_1^*=\frac{\sigma_1}{\sigma_1+\sigma_2},\qquad
p_2^*=\frac{\sigma_2}{\sigma_1+\sigma_2},
\]
with minimal risk $R^*=(\sigma_1+\sigma_2)^2$. 

For any $1>\delta>0$, if $R(p)\le R^*+\delta$, then:

\begin{enumerate}
    \item 
    \[
    p_0 \le \frac{\delta}{R^*+\delta}.
    \]
    \item  $q:=p_1+p_2$ and $s:=p_1/q$, and $s^*=\sigma_1/(\sigma_1+\sigma_2)$,
    \[
    |s-s^*|\le \sqrt{\frac{\delta}{\sigma_1^2+\sigma_2^2}}.
    \]
    \item 
    \[
    |p_i-p_i^*| \le \frac{\delta}{R^*+\delta} \;+\; \sqrt{\frac{\delta}{\sigma_1^2+\sigma_2^2}}
    \qquad (i=1,2).
    \]
\end{enumerate}
\end{proposition}
By Proposition~\ref{prop.neyman.delta}, for any $\delta \in (0,1)$ such that 
$\mathcal{R}(\hat{p}_T) - \mathcal{R}(p^*) < \delta$, we have $\|\hat{p}_T - p^* \|_2 = O(\sqrt{\delta})$. Hence we have
\begin{align*}
\|\hat{p}_T - p^* \|_2 = O(T^{-1/14})
\end{align*}

\end{proof}

\begin{proof}[Proof of \Cref{lemma:variance.hp}]
We first record the algebraic relation between the two variance estimators. Define the ``population-centered'' variance estimator
\[
V_n:=\frac{1}{n}\sum_{i=1}^n (X_i-\mu)^2.
\]
and 
\[S_n^2=\frac{1}{n}\sum_{i=1}^n (X_i-\bar X_n)^2\]
By the usual identity
\[
S_n^2 = V_n - (\bar X_n-\mu)^2,
\]
and since \(\widehat\sigma_n^2 = \dfrac{n}{n-1}S_n^2\), we obtain
\begin{align}\label{proof:eq:hp.decomp}
\widehat\sigma_n^2-\sigma^2
&= \frac{n}{n-1}(S_n^2-\sigma^2) + \Big(\frac{n}{n-1}-1\Big)\sigma^2 \nonumber\\
&= \frac{n}{n-1}(V_n-\sigma^2) - \frac{n}{n-1}(\bar X_n-\mu)^2 + \frac{\sigma^2}{\,n-1\,}. 
\end{align}
Thus, by the triangle inequality,
\begin{equation}\label{proof:eq:hp.bound-decomp}
\big|\widehat\sigma_n^2-\sigma^2\big|
\le \frac{n}{n-1}\,\big|V_n-\sigma^2\big| \;+\; \frac{n}{n-1}\,(\bar X_n-\mu)^2 \;+\; \frac{\sigma^2}{\,n-1\,}.
\end{equation}
We now control the two random terms \(|V_n-\sigma^2|\) and \((\bar X_n-\mu)^2\) using  Chebyshev's inequality.
We first control the term $V_n-\sigma^2$.  Let \(Y:=(X-\mu)^2\). Then \(V_n=n^{-1}\sum_{i=1}^n Y_i\) with \(Y_i\) i.i.d. and
\[
\var(V_n)=\frac{1}{n}\var(Y)=\frac{1}{n}\var\big((X-\mu)^2\big)
\le \frac{1}{n}\mathbb{E}\big[(X-\mu)^4\big]=\frac{\kappa_4}{n}.
\]
By Chebyshev's inequality, for any \(t>0\),
\begin{align}\label{proof:eq:hp.bound.1}
\PP\big(|V_n-\sigma^2|\ge t\big)\le \frac{\var(V_n)}{t^2}\le \frac{\kappa_4}{n t^2}.
\end{align}
The sample mean satisfies \(\var(\bar X_n)=\sigma^2/n\). Applying Markov inequality to the nonnegative random variable \((\bar X_n-\mu)^2\) yields, for any \(u>0\),
\begin{align}\label{proof:eq:hp.bound.2}
\PP\big((\bar X_n-\mu)^2\ge u\big)
\le \frac{\mathbb{E}[(\bar X_n-\mu)^2]}{u}
= \frac{\sigma^2}{n u}.
\end{align}
To produce a total failure probability at most \(\varepsilon\), split \(\varepsilon\) into two equal parts and require each of the tail bounds ~\eqref{proof:eq:hp.bound.1} and ~\eqref{proof:eq:hp.bound.2} to be at most \(\varepsilon/2\). Concretely, choose
\[
t := \sqrt{\frac{2\kappa_4}{n\varepsilon}}
\qquad\text{and}\qquad
u := \frac{2\sigma^2}{n\varepsilon}.
\]
With these choices, ~\eqref{proof:eq:hp.bound.1} gives \(\PP(|V_n-\sigma^2|\ge t)\le \varepsilon/2\), and ~\eqref{proof:eq:hp.bound.2}  gives \(\PP((\bar X_n-\mu)^2\ge u)\le \varepsilon/2\). By the union bound,
\[
\PP\Big(|V_n-\sigma^2|<t\ \text{ and }\ (\bar X_n-\mu)^2<u\Big)\ge 1-\varepsilon.
\]
On the event \(\{|V_n-\sigma^2|<t\}\cap\{(\bar X_n-\mu)^2<u\}\) we plug the thresholds \(t,u\) into \eqref{proof:eq:hp.bound-decomp} to get
\[
\big|\widehat\sigma_n^2-\sigma^2\big|
\le \frac{n}{n-1}\,t \;+\; \frac{n}{n-1}\,u \;+\; \frac{\sigma^2}{\,n-1\,}.
\]
Substituting the definitions of \(t\) and \(u\) yields
\[
\big|\widehat\sigma_n^2-\sigma^2\big|
\le \frac{n}{n-1}\,\sqrt{\frac{2\kappa_4}{n\varepsilon}}
\;+\;\frac{2\sigma^2}{(n-1)\varepsilon}
\;+\;\frac{\sigma^2}{\,n-1\,},
\]
and this event has probability at least \(1-\varepsilon\). This proves the lemma.
\end{proof}

\begin{proof}[Proof of \Cref{cor:app:variance-sample-size}]
Lemma~\ref{lemma:variance.hp} gives, for any $n\ge2$, with probability at least $1-\varepsilon$,
\[
\big|\widehat\sigma_n^2-\sigma^2\big|
\le \frac{n}{n-1}\sqrt{\frac{2\kappa_4}{n\varepsilon}}
\;+\;\frac{2\sigma^2}{(n-1)\varepsilon}
\;+\;\frac{\sigma^2}{\,n-1\,}.
\]
To obtain a sufficient condition ensuring the above is less than $<\sigma^2/2$, we use the fact that $\dfrac{n}{n-1}\le 2$ and $\dfrac{1}{n-1}\le \dfrac{2}{n}$ (valid for $n\ge2$) to obtain the inequality
\begin{align}\label{proof:eq:hp.size.1}
\big|\widehat\sigma_n^2-\sigma^2\big|
\le 2\sqrt{\frac{2\kappa_4}{n\varepsilon}}
\;+\;\frac{2\sigma^2}{n}\Big(1+\frac{2}{\varepsilon}\Big).
\end{align}
It suffices to force each term on the right-hand side of ~\eqref{proof:eq:hp.size.1} to be at most $\sigma^2/4$. Solving these two scalar inequalities yields the two lower bounds
\[
n \ge \frac{128\,\kappa_4}{\sigma^4\,\varepsilon}
\qquad\text{and}\qquad
n \ge 8\Big(1+\frac{2}{\varepsilon}\Big).
\]
Including the trivial requirement $n\ge2$ and taking the maximum of the three bounds produces the stated $n_0$. Therefore for every integer $n\ge n_0$ the bound \(|\widehat\sigma_n^2-\sigma^2|<\sigma^2/2\) holds with probability at least $1-\varepsilon$, as claimed.
\end{proof}

\begin{proof}[Proof of \Cref{prop.neyman.delta}]
Set $a:=\sigma_2^2$ and $b:=\sigma_1^2$. Define the one-dimensional function
\[
\phi(s)=\frac{a}{1-s}+\frac{b}{s},\qquad s\in(0,1).
\]
With $q=p_1+p_2$ and $s=p_1/q$ we can rewrite
\[
R(p)=\frac{1}{q}\,\phi(s).
\]
The Neyman minimizer corresponds to $q=1$, $s=s^*$, and
$\phi(s^*)=R^*=(\sigma_1+\sigma_2)^2$.

We decompose the excess risk as
\[
R(p)-R^*
=\frac{1}{q}\phi(s)-\phi(s^*)
=\Big(\frac{1}{q}-1\Big)\phi(s^*)+\frac{1}{q}\big(\phi(s)-\phi(s^*)\big).
\]
Both terms on the right-hand side are nonnegative because $q\le1$ (so $\tfrac1q-1\ge0$) and $\phi(s)\ge\phi(s^*)$ (by optimality of $s^*$). Thus if $R(p)\le R^*+\delta$ then
\[
\Big(\frac{1}{q}-1\Big)\phi(s^*)\le \delta
\qquad\text{and}\qquad
\phi(s)-\phi(s^*)\le \delta.
\]
The first inequality rearranges to
\[
\frac{1-q}{q}\,\phi(s^*)\le\delta
\]
namely\[
p_0=1-q\le \frac{\delta}{\phi(s^*)+\delta}=\frac{\delta}{R^*+\delta},
\]
which proves the first statement. \\
To bound $|s-s^*|$ we use strong convexity of $\phi$ on $(0,1)$. Compute the second derivative
\[
\phi''(s)=\frac{2a}{(1-s)^3}+\frac{2b}{s^3}.
\]
Since $(1-s)^3\le1$ and $s^3\le1$ for $s\in(0,1)$, we have
\[
\phi''(s)\ge 2a+2b = 2(\sigma_1^2+\sigma_2^2).
\]
Thus $\phi$ is $\mu$-strongly convex with $\mu:=2(\sigma_1^2+\sigma_2^2)$. Strong convexity gives
\[
\phi(s)-\phi(s^*)\ge \frac{\mu}{2}(s-s^*)^2 = (\sigma_1^2+\sigma_2^2)(s-s^*)^2.
\]
Using $\phi(s)-\phi(s^*)\le\delta$ we obtain
\[
|s-s^*|\le \sqrt{\frac{\delta}{\sigma_1^2+\sigma_2^2}},
\]
which proves the second statement.
For the componentwise deviation note $p_1=qs$ and $p_1^*=s^*$. Thus
\[
|p_1-p_1^*| = |qs-s^*|
\le |q-1|\,s^* + |s-s^*|
\le p_0 + |s-s^*|.
\]
Combining the bound on $p_0$   and the bound on $|s-s^*|$ from  yields the thid statement. The same argument gives the identical bound for $p_2$.
\end{proof}

\subsection{Proof of  \Cref{prop:adp.clt}}\label{sec:proof:asymptotic}
\begin{proof}[Proof of \Cref{prop:adp.clt}]
We first state the asymptomatic normality results for the sample mean estimator in each batch and corresponding difference in mean estimator.
\begin{theorem}\label{theorem:adp.clt.1} For any fixed $T,T_0,T_1,m_0,m_1,m_2$ in the two batch design, let $\hat{p}_{T} := \left(\hat{p}_{0,T}, \hat{p}_{1,T}, \hat{p}_{2,T}\right)$ be the optimal allocation proportions for each arm in the second batch that minimize objective function~\eqref{eq:MSE.estimate} on the domain $\mathcal{P}_{T} = \{(p_0, p_1, p_2) \in [\delta_T, 1-2\cdot\delta_T]^3 : p_0 + p_1 + p_2 = 1\}$ for adaptive two-stage algorithm, where $\delta_T= \frac{1}{2}t^{-\alpha}$ and $\alpha \in (0, \frac{1}{2})$. Let $n_0,n_1,n_2$ be the corresponding numbers of allocation for each treatment in the second batch. Then as $T \to \infty$, \begin{align}\label{prop:eq:adp.clt.1} \begin{bmatrix} m_0^{1/2} (\hat{\mu}_0^{(0)}-\mu_0)\\ n_0^{1/2} (\hat{\mu}_0^{(1)}-\mu_0) \\ m_1^{1/2} (\hat{\mu}_1^{(0)}-\mu_1)\\ n_1^{1/2} (\hat{\mu}_1^{(1)}-\mu_1)\\ m_2^{1/2} (\hat{\mu}_2^{(0)}-\mu_2)\\ n_2^{1/2} (\hat{\mu}_2^{(1)}-\mu_2) \end{bmatrix} \xrightarrow{d} \mathcal{N}\left(0, \text{diag}\left(\sigma^2_0,\sigma^2_0,\sigma^2_1,\sigma^2_1,\sigma^2_2,\sigma^2_2\right) \right) \end{align} and \begin{align} \begin{pmatrix}\label{prop:eq:adp.clt.2} \sqrt{\frac{n_w n_{w'}}{n_w\sigma^2_{w'}+n_{w'}\sigma^2_{w}}}\left(\hat{\mu}_w^{(1)}-\hat{\mu}_{w'}^{(1)}-\mu_w +\mu_{w'}\right) \\ \sqrt{\frac{m_w m_{w'}}{m_w\sigma^2_{w'}+m_{w'}\sigma^2_{w}}}\left(\hat{\mu}_w^{(0)}-\hat{\mu}_{w'}^{(0)}-\mu_w + \mu_{w'}\right) \end{pmatrix} \xrightarrow{d} \mathcal{N}\left(0,\bm{I}_2\right) \end{align} for any $w,w' \in \left\{0,1,2\right\}$.
\end{theorem}
The proof of \Cref{theorem:adp.clt.1} is deferred to the end of this section. We come back to proving  \Cref{prop:adp.clt}.
\Cref{theorem:adp.clt.1} implies
\begin{align}\label{proof:eq:second.batch.clt}
\begin{bmatrix}
 n_0^{1/2} (\hat{\mu}_0^{(1)}-\mu_0) \\
 n_1^{1/2} (\hat{\mu}_1^{(1)}-\mu_1)\\
 n_2^{1/2} (\hat{\mu}_2^{(1)}-\mu_2)
\end{bmatrix} 
 \xrightarrow{d} \mathcal{N}\left(0, \text{diag}\left(\sigma^2_0,\sigma^2_1,\sigma^2_2\right) \right)
\end{align}
We define the aggregate sample mean estimator as 
\[
\hat{\mu}_w
:= \frac{\sum_{j \in \text{pilot}, W_j = w} Y_j + \sum_{i \in \text{post-pilot}, W_i = w} Y_i}{m_w + n_w}
= \omega \hat{\mu}^{(0)}_w + (1 - \omega) \hat{\mu}^{(1)}_w, \quad \omega = \frac{m_w}{m_w + n_w}, ~w \in \{0,1, 2\}.
\]
Since we have $\hat{p}_{w,T} \geq \delta_T$ and $\delta_T = w(\frac{1}{T^{1/2}})$,  $n_w = T_1 \cdot \hat{p}_{w,T} = w(T^{1/2})$. Together with the assumption that $T_0 = O(\sqrt{T})$, we have $m_w = o(n_w)$, which implies the following results
\begin{align*}
\lVert \sqrt{n_w}\left(\hat{\mu}_w - \hat{\mu}_w^{(1)}\right)\rVert & = \lVert\sqrt{n_w}\left(\frac{m_w\cdot\hat{\mu}_w^{(0)}+ n_w\cdot\hat{\mu}_w^{(1)}}{m_w +n_w} - \frac{(m_w + n_w)\cdot\hat{\mu}_w^{(1)}}{m_w + n_w}\right)\rVert \\
& = \lVert \sqrt{n_w}\left(\frac{m_w\cdot\hat{\mu}_w^{(0)}- m_w\cdot\hat{\mu}_w^{(1)}}{m_w +n_w}\right)\rVert \\
& \leq \frac{m_w\cdot \sqrt{n_w}}{m_w + n_w}\lVert \hat{\mu}_w^{(0)}-\hat{\mu}_w^{(1)} \rVert \\
& = \frac{m_w\cdot \sqrt{n_w}}{m_w + n_w}\cdot  O_p(\frac{1}{\sqrt{m_w}}+ \frac{1}{\sqrt{n_w}}) \\
& = o_p(1)
\end{align*}
for any $w, \in \left\{0,1,2\right\}$.\\
Combine with ~\eqref{proof:eq:second.batch.clt}, we arrive at
\begin{align}\label{proof:eq:pooled.clt}
    \begin{bmatrix}
 n_0^{1/2} (\hat{\mu}_0-\mu_0) \\
 n_1^{1/2} (\hat{\mu}_1-\mu_1)\\
 n_2^{1/2} (\hat{\mu}_2-\mu_2)
\end{bmatrix} 
 \xrightarrow{d} \mathcal{N}\left(0, \text{diag}\left(\sigma^2_0,\sigma^2_1,\sigma^2_2\right) \right)
\end{align}
Namely, the pooled sample mean estimator $\hat{\mu}_w$, $w \in \left\{0,1,2\right\}$ are asymptotically normal.
Further, 
\begin{align*}
&\lVert\sqrt{\frac{n_w n_{w'}}{n_w\sigma^2_{w'}+n_{w'}\sigma^2_{w}}}\left(\hat{\mu}_w-\hat{\mu}_w^{(1)}-\hat{\mu}_{w'}+\hat{\mu}_{w'}^{(1)}\right)\lVert \\
& \leq \lVert\sqrt{\frac{n_w n_{w'}}{n_w\sigma^2_{w'}+n_{w'}\sigma^2_{w}}}\lVert \cdot\left(\|\hat{\mu}_w-\hat{\mu}_w^{(1)}\|+\|\hat{\mu}_{w'}-\hat{\mu}_{w'}^{(1)}\|\right) \\
& \leq \lVert\sqrt{\frac{n_w n_{w'}}{n_w\sigma^2_{w'}+n_{w'}\sigma^2_{w}}}\lVert \cdot \left(\frac{m_w}{m_w+n_w}\cdot O_P(\frac{1}{\sqrt{m_w}}+ \frac{1}{\sqrt{n_w}})+\frac{m_w'}{m_w'+n_w'} \cdot O_P(\frac{1}{\sqrt{m_w'}}+ \frac{1}{\sqrt{n_w'}})\right) \\
& = o_p(1) + o_p(1)  \\
\end{align*}
Combine with ~\eqref{prop:eq:adp.clt.2}, we arrive at
\begin{align}\label{proof:eq:pooled.clt.2}
\sqrt{\frac{n_w n_{w'}}{n_w\sigma^2_{w'}+n_{w'}\sigma^2_{w}}}\left(\hat{\mu}_w-\hat{\mu}_{w'}-\mu_w + \mu_{w'}\right) \xrightarrow{d} \mathcal{N}\left(0,1\right)
\end{align}
for any $w,w' \in \left\{0,1,2\right\}$

Next we show the asymptotic normality of 
$\hat{\tau}_{\hat{w}_{\max}}$. Without loss of generality, we assume $\Delta>0$ and the second treatment has largest mean.
Consider the normalized form $\sqrt{T}\left(\hat{\tau}_{\hat{w}_{\max}}-\tau_{w_{\max}}\right)$
\begin{align*}
\sqrt{T}\left(\hat{\tau}_{\hat{w}_{\max}}-\tau_{w_{\max}}\right) 
& = \sqrt{T}\left(\hat{\mu}_2 - \hat{b}_2\right) \1\left\{\hat{\mu}_2 > \hat{\mu}_1 \right\} + \sqrt{T}\left(\hat{\mu}_1 - \hat{b}_1\right) \1\left\{\hat{\mu}_2 < \hat{\mu}_1 \right\} - \sqrt{T}\hat{\mu}_0-\sqrt{T}\tau_{w_{\max}} \\
& = \underbrace{\sqrt{T}\left(\hat{\mu}_2-\hat{\mu}_0-\tau_{w_{\max}}\right)}_{:=(a)} +\underbrace{\sqrt{T}\hat{\mu}_1 \1\left\{\hat{\mu}_2 < \hat{\mu}_1 \right\}}_{:=(b)} + \underbrace{\sqrt{T}\hat{\mu}_2\cdot\left(\1\left\{\hat{\mu}_2 > \hat{\mu}_1 \right\}-1\right)}_{:=(c)}\\
&\quad~- \underbrace{\sqrt{T}\cdot\hat{b}_2\1\left\{\hat{\mu}_2 > \hat{\mu}_1 \right\}}_{:=(d)}   - \underbrace{\sqrt{T}\cdot\hat{b}_1\1\left\{\hat{\mu}_2 < \hat{\mu}_1 \right\}}_{:=(e)}
\end{align*}
We analyze each term separately.
For $(a)$, by ~\eqref{proof:eq:pooled.clt.2},
\begin{align}\label{proof:eq:tau.max.clt.a}
\begin{split}
    \sqrt{T}\left(\hat{\mu}_2-\hat{\mu}_0-\tau_{w_{\max}}\right) & = \sqrt{T}\cdot\sqrt{(\frac{\sigma^2_2}{n_2}+ \frac{\sigma^2_0}{n_0})}\cdot\left(\sqrt{\frac{n_2n_0}{n_2\sigma^2_0+n_0\sigma^2_2}}\right)\left(\hat{\mu}_2-\hat{\mu}_0-\mu_2 + \mu_0\right)\\
& \xrightarrow{d}\sqrt{(\frac{\sigma^2_2}{p^*_2}+ \frac{\sigma^2_0}{p^*_0})} \cdot \mathcal{N}\left(0, 1\right)\\
& = \mathcal{N}\left(0, (\sigma_0+\sigma_2)^2\right)
\end{split}
\end{align}
where we have $p^*_0 = \frac{\sigma_0}{\sigma_0 +\sigma_2}$ , $p^*_2 = \frac{\sigma_2}{\sigma_0 +\sigma_2}$ and use the fact that $\|\hat{p}_T - p^*\|_2 \xrightarrow{p} 0$ from \Cref{prop:adp.alloc.convergence}. \\
Next consider the term $1\left\{\hat{\mu}_2 < \hat{\mu}_1 \right\}$, which indicates the happening of false selection. Define $Y_T = \sqrt{\frac{n_2n_1}{n_2\sigma^2_1+n_1\sigma^2_2}}\cdot (\hat{\mu}_2-\hat{\mu}_1 - \mu_2 + \mu_1)$, we have shown $Y_T \xrightarrow{d} \mathcal{N}\left(0,1\right)$ as $T \to \infty$ in ~\eqref{proof:eq:pooled.clt.2}. By Prohorov's theorem, for any $\epsilon>0$, there exists a constant $M$ s.t. for any $T$
\begin{align}
\PP \left(\mid Y_T\mid > M\right) < \epsilon
\end{align}
We have $n_w = \delta_T\cdot T_1 = \omega(T^{1/2})$ and $n_w = O(T)$. which further implies
\begin{align*}
\sqrt{\frac{n_2n_1}{n_2\sigma^2_1+n_1\sigma^2_2}} & \geq \sqrt{\frac{n_2n_1}{(n_2+n_1)\cdot(\sigma^2_1+\sigma^2_2)}}  \\
& \geq \sqrt{\frac{\min (n_1,n_2)^2}{2\cdot \max (n_1,n_2)\cdot(\sigma^2_1+\sigma^2_2)}} \\
&= \omega(1)
\end{align*}
Thus, for any $c>0$, there exists a constant $T_0$ s.t. $\sqrt{\frac{n_2n_1}{n_2\sigma^2_1+n_1\sigma^2_2}} > c$ for all $T > T_0$. 
Pick $T$ large enough such that $\sqrt{\frac{n_2n_1}{n_2\sigma^2_1+n_1\sigma^2_2}}\cdot (\mu_2 - \mu_1) > M>0$. Hence
\begin{align*}
\PP\left(\hat{\mu}_2 - \hat{\mu}_1 <0 \right) &= \PP\left(Y_T  <   -\sqrt{\frac{n_2n_1}{n_2\sigma^2_1+n_1\sigma^2_2}} \cdot (\mu_2 - \mu_1)\right)\\
& \leq \PP\left(Y_T < -M\right) \\
& \leq \PP \left(\mid Y_T\mid > M\right) < \epsilon
\end{align*}
Namely, for every $\epsilon>0$, there exists a $T_\epsilon$ s.t. for every $T>T_\epsilon$
\begin{align}\label{proof:eq:correct.convergence}
\PP\left(\hat{\mu}_2 - \hat{\mu}_1 <0 \right) < \epsilon
\end{align}
Thus  $\PP\left(\hat{\mu}_2 - \hat{\mu}_1 <0 \right) \to 0 $ as $T \to \infty$.\\
For term $(b)$, consider the term $\sqrt{T}\cdot\1\left\{\hat{\mu}_2 < \hat{\mu}_1 \right\}$. It takes value $\sqrt{T}$ w.p. $\PP\left(\hat{\mu}_2 - \hat{\mu}_1 <0 \right)$ and 0 o.w. By ~\eqref{proof:eq:correct.convergence}, $\sqrt{T}\cdot\1\left\{\hat{\mu}_2 < \hat{\mu}_1 \right\}  \xrightarrow{p} 0 $.
Hence
\begin{align}\label{proof:eq:tau.max.clt.b}
\big|\sqrt{T}\cdot\hat{\mu}_1 \1\left\{\hat{\mu}_2 < \hat{\mu}_1 \right\}\big| \leq \big|\underbrace{\sqrt{T}\1\left\{\hat{\mu}_2 < \hat{\mu}_1 \right\}}_{o_p(1)} \cdot \underbrace{n_1^{1/2}\hat{\mu}_1}_{O_p(1)} \big| = o_p(1)
\end{align}
For term $(c)$, we have $1\left\{\hat{\mu}_2 > \hat{\mu}_1 \right\}-1 \xrightarrow{p} 0$ by ~\eqref{proof:eq:correct.convergence}, which implies
\begin{align}\label{proof:eq:tau.max.clt.c}
\sqrt{T}\hat{\mu}_2\cdot\left(\1\left\{\hat{\mu}_2 > \hat{\mu}_1 \right\}-1\right) &= \underbrace{\sqrt{\frac{T}{n_2}}}_{O_P(1)}\cdot \underbrace{n_2^{1/2}\hat{\mu}_2}_{O_p(1)}\cdot\underbrace{\left(\1\left\{\hat{\mu}_2 > \hat{\mu}_1 \right\}-1\right)}_{o_p(1)} = o_p(1)
\end{align}
For term $(d)$, it can be bounded above
\begin{align*}
\sqrt{T}\cdot\hat{b}_2\cdot\1\left\{\hat{\mu}_2 > \hat{\mu}_1 \right\} &=\sqrt{T}\cdot\frac{\hat{\sigma}^2_2}{n_2\sqrt{\hat{V}}}\frac{\phi\left(\frac{\hat{\Delta}}{\sqrt{\hat{V}}}\right)}{\Phi\left(\frac{\hat{\Delta}}{\sqrt{\hat{V}}}\right)}\1\left\{\hat{\mu}_2 > \hat{\mu}_1 \right\} \\
& \leq \sqrt{T}\cdot\hat{\sigma}^2_2\frac{\phi\left(\frac{\hat{\Delta}}{\sqrt{\hat{V}}}\right)}{\Phi\left(\frac{\hat{\Delta}}{\sqrt{\hat{V}}}\right)}\cdot \sqrt{\frac{n_2n_1}{n_2\hat{\sigma}^2_1 + n_1\hat{\sigma}^2_2}} \\
& \lesssim \sqrt{T}\cdot\hat{\sigma}^2_2\cdot\phi\left(\frac{\hat{\Delta}}{\sqrt{\hat{V}}}\right)\cdot\frac{1}{\Phi\left(\frac{\hat{\Delta}}{\sqrt{\hat{V}}}\right)}(\hat{\sigma}^2_1+\hat{\sigma}^2_2)^{-1/2} \cdot T^{3/4} \\
& \lesssim T^{5/4}\cdot\frac{\hat{\sigma}^2_2}{\sqrt{\hat{\sigma}^2_1+\hat{\sigma}^2_2}}\cdot e^{-1/2\cdot \left(\frac{\hat{\Delta}^2}{\hat{V}}\right)}\cdot\frac{1}{\Phi\left(\frac{\hat{\Delta}}{\sqrt{\hat{V}}}\right)}
\end{align*}
Consider the term  $\frac{1}{\Phi\left(\frac{\hat{\Delta}}{\sqrt{\hat{V}}}\right)}$, where 
\begin{align*}
\frac{1}{\Phi\left(\frac{\hat{\Delta}}{\sqrt{\hat{V}}}\right)} = \frac{1}{\Phi\left(\hat{\Delta}\cdot\sqrt{\frac{n_2n_1}{n_2\hat{\sigma}^2_1 + n_1\hat{\sigma}^2_2 }}\right)} 
\end{align*}
Notice that 
\begin{align*}
\hat{\Delta}\cdot \sqrt{\frac{n_2n_1}{n_2\hat{\sigma}^2_1+n_1\hat{\sigma}^2_2}} 
& \geq \underbrace{\hat{\Delta}\cdot(\hat{\sigma}^2_1+\hat{\sigma}^2_2)^{-1/2}}_{O_p(1)} \cdot \underbrace{\sqrt{\frac{\min (n_1,n_2)^2}{2\cdot \max (n_1,n_2)}}}_{\omega(1)} \\
\end{align*}
Which implies, for any $M>0$
\begin{align*}
\PP\left(\hat{\Delta}\cdot \sqrt{\frac{n_2n_1}{n_2\hat{\sigma}^2_1+n_1\hat{\sigma}^2_2}} > M\right) \to 1
\end{align*}
whence
\begin{align*}
\Phi\left(\hat{\Delta}\cdot\sqrt{\frac{n_2n_1}{n_1\hat{\sigma}^2_2 + n_1\hat{\sigma}^2_2 }}\right)\xrightarrow{p} 1, \quad \text{and}\quad  \frac{1}{\Phi\left(\hat{\Delta}\cdot\sqrt{\frac{n_2n_1}{n_1\hat{\sigma}^2_2 + n_1\hat{\sigma}^2_2 }}\right)} \xrightarrow{p} 1
\end{align*}
On the other hand
\begin{align*}
 T^{5/4}\cdot e^{-1/2\cdot \left(\frac{\hat{\Delta}^2}{\hat{V}}\right)} &=T^{5/4}\cdot \exp\left\{ -1/2\cdot (\hat{\Delta}^2 \frac{n_2n_1}{n_1\hat{\sigma}^2_2 + n_2\hat{\sigma}^2_1 })\right\} \\
 & =  \exp\left\{ -1/2\cdot (\hat{\Delta}^2\cdot \frac{n_2n_1}{n_1\hat{\sigma}^2_2 + n_2\hat{\sigma}^2_1 }-\frac{5}{2}\log(T))\right\}
\end{align*}
Since  $\hat{\Delta}\xrightarrow{p}\Delta$, $\hat{\sigma}^2_1\xrightarrow{p}\sigma^2_1$, $\hat{\sigma}^2_2\xrightarrow{p}\sigma^2_2$ and $n_w = \omega(T^{1/2})$, $\frac{n_2}{T} \xrightarrow{p} \frac{\sigma_2}{\sigma_0+\sigma_2} \in (0,1)$, we have for any $M>0$
\begin{align*}
\PP\left(\hat{\Delta}^2\cdot \frac{n_2n_1}{n_1\hat{\sigma}^2_2 + n_2\hat{\sigma}^2_1 }-\frac{5}{2}\log(T) >M\right) \to 1
\end{align*}
which implies
\begin{align*}
 T^{5/4}\cdot e^{-1/2\cdot \left(\frac{\hat{\Delta}^2}{\hat{V}}\right)} =    \exp\left\{ -1/2\cdot (\hat{\Delta}^2\cdot \frac{n_2n_1}{n_1\hat{\sigma}^2_2 + n_2\hat{\sigma}^2_1 }-\frac{5}{2}\log(T))\right\} \xrightarrow{p}0
\end{align*}
whence
\begin{align}\label{proof:eq:tau.max.clt.d}
\sqrt{T}\cdot\hat{b}_2\cdot\1\left\{\hat{\mu}_2 > \hat{\mu}_1 \right\} \leq \underbrace{T^{5/4}\cdot e^{-1/2\cdot \left(\frac{\hat{\Delta}^2}{\hat{V}}\right)}}_{o_p(1)}\cdot\underbrace{\Phi\left(\frac{\hat{\Delta}}{\sqrt{\hat{V}}}\right)^{-1}\cdot\frac{\hat{\sigma}^2_2}{\sqrt{\hat{\sigma}^2_1+\hat{\sigma}^2_2}}}_{O_p(1)} = o_p(1)
\end{align}
For the last term $(e)$, It takes value $\sqrt{T}\cdot \hat{b}_1$ w.p. $\PP\left(\hat{\mu}_2 - \hat{\mu}_1 <0 \right)$ and 0 o.w. By ~\eqref{proof:eq:correct.convergence}, we arrive at
\begin{align}\label{proof:eq:tau.max.clt.e}
  \sqrt{T}\cdot \hat{b}_1\cdot\1\left\{\hat{\mu}_2 < \hat{\mu}_1 \right\}  \xrightarrow{p} 0   
\end{align}
Combine ~\eqref{proof:eq:tau.max.clt.a},
\eqref{proof:eq:tau.max.clt.b}, \eqref{proof:eq:tau.max.clt.c},
\eqref{proof:eq:tau.max.clt.d},
\eqref{proof:eq:tau.max.clt.e}
we arrive it 
\begin{align*}
\sqrt{T}\left(\hat{\tau}_{\hat{w}_{\max}}-\tau_2\right) \xrightarrow{d}\mathcal{N}\left(0, (\sigma_0+\sigma_2)^2\right)
\end{align*}
When $\Delta>0$ and second treatment has larger mean. Similar analysis holds for the case 
$\Delta<0$ and the first treatment has larger mean. Overall, we have
\begin{align*}
\begin{split}
\sqrt{T}\left(\hat{\tau}_{\hat{w}_{\max}} - \tau_{w_{\max}}\right) \xrightarrow{d} \mathcal{N}\left(0, (\sigma_0 +\sigma_{w_{\text{max}}})^2 \right)
\end{split}
\end{align*}
which completes the proof.
\end{proof}

\begin{proof}[Proof of \Cref{theorem:adp.clt.1}]
We aim to prove the following results
\begin{align}
\begin{bmatrix}
 m_0^{1/2} (\hat{\mu}_0^{(0)}-\mu_0)\\
 n_0^{1/2} (\hat{\mu}_0^{(1)}-\mu_0) \\
 m_1^{1/2} (\hat{\mu}_1^{(0)}-\mu_1)\\
 n_1^{1/2} (\hat{\mu}_1^{(1)}-\mu_1)\\
 m_2^{1/2} (\hat{\mu}_2^{(0)}-\mu_2)\\
 n_2^{1/2} (\hat{\mu}_2^{(1)}-\mu_2)
\end{bmatrix} =
\begin{bmatrix}
 m_0^{-1/2} (\sum_{i=1}^{T_0}\1\left\{W_{0,i}=0\right\}(Y_{0,i}- \mu_0))\\
 n_0^{-1/2} (\sum_{i=1}^{T_1}\1\left\{W_{1,i}=0\right\}(Y_{1,i}- \mu_0)) \\
  m_1^{-1/2} (\sum_{i=1}^{T_0}\1\left\{W_{0,i}=1\right\}(Y_{0,i}- \mu_1))\\
 n_1^{-1/2} (\sum_{i=1}^{T_1}\1\left\{W_{1,i}=1\right\}(Y_{1,i}- \mu_1))\\
  m_2^{-1/2} (\sum_{i=1}^{T_0}\1\left\{W_{0,i}=2\right\}(Y_{0,i}- \mu_1))\\
 n_2^{-1/2} (\sum_{i=1}^{T_1}\1\left\{W_{1,i}=2\right\}(Y_{1,i}- \mu_2))
\end{bmatrix}
 \xrightarrow{d} \mathcal{N}\left(0, \text{diag}\left(\sigma^2_0,\sigma^2_0,\sigma^2_1,\sigma^2_1,\sigma^2_2,\sigma^2_2\right) \right)
\end{align}
By Cramer-Wold device, it's sufficient to show for any fixed vector $\bm{a} \in \mathbb{R}^{6}$
\begin{align*}
\bm{a}^\top\begin{bmatrix}
 m_0^{-1/2} (\sum_{i=1}^{T_0}\1\left\{W_{0,i}=0\right\}(Y_{0,i}- \mu_0))\\
 n_0^{-1/2} (\sum_{i=1}^{T_1}\1\left\{W_{1,i}=0\right\}(Y_{1,i}- \mu_0)) \\
  m_1^{-1/2} (\sum_{i=1}^{T_0}\1\left\{W_{0,i}=1\right\}(Y_{0,i}- \mu_1))\\
 n_1^{-1/2} (\sum_{i=1}^{T_1}\1\left\{W_{1,i}=1\right\}(Y_{1,i}- \mu_1))\\
  m_2^{-1/2} (\sum_{i=1}^{T_0}\1\left\{W_{0,i}=2\right\}(Y_{0,i}- \mu_2))\\
 n_2^{-1/2} (\sum_{i=1}^{T_1}\1\left\{W_{1,i}=2\right\}(Y_{1,i}- \mu_2))
\end{bmatrix}
 \xrightarrow{d} \mathcal{N}\left(0, \bm{a}^\top\text{diag}\left(\sigma^2_0,\sigma^2_0,\sigma^2_1,\sigma^2_1,\sigma^2_2,\sigma^2_2\right)\bm{a} \right)
\end{align*}
We can write the vector $\bm{a} \in \mathbb{R}^{6}$  as $\bm{a}^\top = [\bm{a}_0^\top,\bm{a}_1^\top]$ where $\bm{a}_t \in \mathbb{R}^{3}$ and $\bm{a}_t^\top = [a_{t,0}^\top,a_{t,1}^\top,a_{t,2}^\top]$, $t \in \left\{0,1\right\}$. To express the expression more compactly, let the treatment allocation $m_0, m_1, m_2, n_0,n_1,n_2$ be written as $n^{(0)}_{0},n^{(0)}_{1},n^{(0)}_{2},n^{(1)}_{0},n^{(1)}_{1},n^{(1)}_{2}$ Then the above expression can be written us
\begin{align*}
\sum_{t=0}^1 \bm{a}_t^\top \begin{bmatrix}
n^{(t)}_{0} &0  &  0\\
0 & n^{(t)}_{1} & 0 \\
0 & 0 &  n^{(t)}_{2}\\
\end{bmatrix}^{-1/2}
\sum_{i=1}^{T_t}\begin{bmatrix}
\1\left\{W_{t,i}=0\right\}(Y_{t,i}- \mu_0) \\
\1\left\{W_{t,i}=1\right\}(Y_{t,i}- \mu_1) \\
\1\left\{W_{t,i}=2\right\}(Y_{t,i}- \mu_2)
\end{bmatrix}
\end{align*}
Define the random variable $Z_{t,i} :=\bm{a}_t^\top \begin{bmatrix}
n^{(t)}_{0} &0  &  0\\
0 & n^{(t)}_{1} & 0 \\
0 & 0 &  n^{(t)}_{2}\\
\end{bmatrix}^{-1/2}
\begin{bmatrix}
\1\left\{W_{t,i}=0\right\}(Y_{t,i}- \mu_0) \\
\1\left\{W_{t,i}=1\right\}(Y_{t,i}- \mu_1) \\
\1\left\{W_{t,i}=2\right\}(Y_{t,i}- \mu_2)
\end{bmatrix}$ for  $(t,i) \in \left\{0,1\right\}\times [1:T_t]$. Let $\mathcal{H}_{t-1} = \left\{W_{t-1,i}, Y_{t-1,i}\right\}_{i=1}^{T_t}, t\in \left\{0,1\right\}$.\\
We have
\begin{align*}
\EE \left[Z_{t,i}\mid \mathcal{H}_{t-1}\right] & = \bm{a}_t^\top \begin{bmatrix}
n^{(t)}_{0} &0  &  0\\
0 & n^{(t)}_{1} & 0 \\
0 & 0 &  n^{(t)}_{2}\\
\end{bmatrix}^{-1/2}\cdot \EE \left[\begin{bmatrix}
\1\left\{W_{t,i}=0\right\}(Y_{t,i}- \mu_0) \\
\1\left\{W_{t,i}=1\right\}(Y_{t,i}- \mu_1) \\
\1\left\{W_{t,i}=2\right\}(Y_{t,i}- \mu_2)
\end{bmatrix}\mid \mathcal{H}_{t-1}\right]\\
& = \bm{a}_t^\top \begin{bmatrix}
n^{(t)}_{0} &0  &  0\\
0 & n^{(t)}_{1} & 0 \\
0 & 0 &  n^{(t)}_{2}\\
\end{bmatrix}^{-1/2}\cdot \EE \begin{bmatrix}
\PP\left(W_{t,i} =0 \mid \mathcal{H}_{t-1}\right)\cdot\EE \left[(Y_{t,i}- \mu_0)\mid \mathcal{H}_{t-1}, W_{t,i} =0\right] \\
\PP\left(W_{t,i} =1 \mid \mathcal{H}_{t-1}\right)\cdot\EE \left[(Y_{t,i}- \mu_1)\mid \mathcal{H}_{t-1}, W_{t,i} =1\right]\\
\PP\left(W_{t,i} =2 \mid \mathcal{H}_{t-1}\right)\cdot\EE \left[(Y_{t,i}- \mu_2)\mid \mathcal{H}_{t-1}, W_{t,i} =2\right]
\end{bmatrix} \\
& = \bm{a}_t^\top \begin{bmatrix}
n^{(t)}_{0} &0  &  0\\
0 & n^{(t)}_{1} & 0 \\
0 & 0 &  n^{(t)}_{2}\\
\end{bmatrix}^{-1/2}\cdot \EE \begin{bmatrix}
\PP\left(W_{t,i} =0 \mid \mathcal{H}_{t-1}\right)\cdot\EE \left[(Y_{t,i}- \mu_0)\mid W_{t,i} =0\right] \\
\PP\left(W_{t,i} =1 \mid \mathcal{H}_{t-1}\right)\cdot\EE \left[(Y_{t,i}- \mu_1)\mid W_{t,i} =1\right]\\
\PP\left(W_{t,i} =2 \mid \mathcal{H}_{t-1}\right)\cdot\EE \left[(Y_{t,i}- \mu_2)\mid W_{t,i} =2\right]
\end{bmatrix} \\
& = 0
\end{align*}
for  $(t,i) \in \left\{0,1\right\}\times [1:T_t]$. \\
Thus $\left\{Z_{t,i}\right\}$, where $(t,i) \in \left\{0,1\right\}\times [1:T_t]$ form a difference sequence w.r.t $\left\{\mathcal{H}_t\right\}_{t=0}^1$.
whence we can use the central limit theorem for martingale difference sequence to provide the asymptotic normally. We use the central limit theorem as stated in  \parencite{dvoretzky1972asymptotic, zhang2020inference, cook2023semiparametric}. In our two batch setting, the theorem states $\sum_{t=0}^1\sum_{i=1}^{T_t} z_i \xrightarrow{d}\mathcal{N}\left(0,\sigma^2\right)$ if the following Linderberg-type condition holds:
\begin{enumerate}[label=(\alph*)]
    \item $\sum_{t=0}^1\sum_{i=1}^{T_t}  \mathbb{E} [z_{t,i}^2 | \mathcal{H}_{t-1}] \xrightarrow{P} \sigma^2$
    \label{item:a}
    
    \item for any $\epsilon > 0$, $\sum_{t=0}^1\sum_{i=1}^{T_t}  \mathbb{E} [z_{t,i}^2 \1 [|z_{t,i}| > \epsilon | \mathcal{H}_{t-1}] \xrightarrow{P} 0$.
    \label{item:b}
\end{enumerate}

\noindent\textbf{Conditional Variance} \\
\begin{flalign*}
&\sum_{t=0}^1\sum_{i=1}^{T_t}  \mathbb{E} [z_{t,i}^2 | \mathcal{H}_{t-1}] = 
\sum_{t=0}^1\sum_{i=1}^{T_t}  \mathbb{E} [\left(\bm{a}_t^\top \begin{bmatrix}
n^{(t)}_{0} &0  &  0\\
0 & n^{(t)}_{1} & 0 \\
0 & 0 &  n^{(t)}_{2}\\
\end{bmatrix}^{-1/2}
\begin{bmatrix}
\1\left\{W_{t,i}=0\right\}(Y_{t,i}- \mu_0) \\
\1\left\{W_{t,i}=1\right\}(Y_{t,i}- \mu_1) \\
\1\left\{W_{t,i}=2\right\}(Y_{t,i}- \mu_2)
\end{bmatrix}\right)^2 | \mathcal{H}_{t-1}] \\
& = \sum_{t=0}^1\sum_{i=1}^{T_t}  \mathbb{E} [\bm{a}_t^\top \begin{bmatrix}
n^{(t)}_{0} &0  &  0\\
0 & n^{(t)}_{1} & 0 \\
0 & 0 &  n^{(t)}_{2}\\
\end{bmatrix}^{-1/2} 
\diag\left(\1\left\{W_{t,i}=w\right\}(Y_{t,i}- \mu_w)^2\right)_{3\times3}
 \begin{bmatrix}
n^{(t)}_{0} &0  &  0\\
0 & n^{(t)}_{1} & 0 \\
0 & 0 &  n^{(t)}_{2}\\
\end{bmatrix}^{-1/2}\bm{a}\mid\mathcal{H}_{t-1}] \\
& = \sum_{t=0}^1\sum_{i=1}^{T_t}  \bm{a}_t^\top \begin{bmatrix}
n^{(t)}_{0} &0  &  0\\
0 & n^{(t)}_{1} & 0 \\
0 & 0 &  n^{(t)}_{2}\\
\end{bmatrix}^{-1/2} \mathbb{E} [
\diag\left(\1\left\{W_{t,i}=w\right\}(Y_{t,i}- \mu_w)^2\right)_{3\times3}
| \mathcal{H}_{t-1}]
\begin{bmatrix}
n^{(t)}_{0} &0  &  0\\
0 & n^{(t)}_{1} & 0 \\
0 & 0 &  n^{(t)}_{2}\\
\end{bmatrix}^{-1/2} \bm{a} \\
& = \sum_{t=0}^1\sum_{i=1}^{T_t}  \bm{a}_t^\top \begin{bmatrix}
n^{(t)}_{0} &0  &  0\\
0 & n^{(t)}_{1} & 0 \\
0 & 0 &  n^{(t)}_{2}\\
\end{bmatrix}^{-1/2} 
\diag\left(\PP\left(W_{t,i}=w \mid \mathcal{H}_{t-1}\right)\cdot\sigma^2_w\right)_{3\times3}
\begin{bmatrix}
n^{(t)}_{0} &0  &  0\\
0 & n^{(t)}_{1} & 0 \\
0 & 0 &  n^{(t)}_{2}\\
\end{bmatrix}^{-1/2} \bm{a} \\
& = \sum_{t=0}^1\sum_{i=1}^{T_t}  \bm{a}_t^\top 
\diag\left(\frac{1}{n^{(t)}_{w}}\cdot\PP\left(W_{t,i}=w \mid \mathcal{H}_{t-1}\right)\cdot\sigma^2_w\right)_{3\times3}
 \bm{a} \\
 & = \sum_{t=0}^1\sum_{i=1}^{T_t}  \sum_{w=0}^2 a^2_{t,w} 
\frac{1}{n^{(t)}_{w}}\cdot\PP\left(W_{t,i}=w \mid \mathcal{H}_{t-1}\right)\cdot\sigma^2_w \\
& = \sum_{t=0}^1 \sum_{w=0}^2 a^2_{t,w} \cdot\sigma^2_w
\frac{1}{n^{(t)}_{w}}\sum_{i=1}^{T_t} \PP\left(W_{t,i}=w \mid \mathcal{H}_{t-1}\right) \\
& =  \sum_{t=0}^1 \sum_{w=0}^2 a^2_{t,w} \cdot\sigma^2_w
\frac{1}{n^{(t)}_{w}}\cdot n^{(t)}_{w} \\
& = \sum_{t=0}^1 \sum_{w=0}^2 a^2_{t,w} \cdot\sigma^2_w \\
& = \bm{a}^\top\text{diag}\left(\sigma^2_0,\sigma^2_0,\sigma^2_1,\sigma^2_1,\sigma^2_2,\sigma^2_2\right)\bm{a}
\end{flalign*}

\noindent\textbf{Linderberg Condition} \\
\begin{align*}
&\sum_{t=0}^1\sum_{i=1}^{T_t}  \mathbb{E} [z_{t,i}^2 \1 [|z_{t,i}| > \epsilon | \mathcal{H}_{t-1}] = \sum_{t=0}^1\sum_{i=1}^{T_t} \sum_{w=0}^2 \mathbb{E} [ a^2_{t,w} \frac{1}{n^{(t)}_{w}}\cdot\1\left\{W_{t,i}=w\right\}(Y_{t,i}- \mu_w)^2 \1 \left(|z_{t,i}| > \epsilon \right) | \mathcal{H}_{t-1}] \\
&=\sum_{t=0}^1\sum_{i=1}^{T_t} \sum_{w=0}^2 \mathbb{E} [ a^2_{t,w} \frac{1}{n^{(t)}_{w}}\cdot\1\left\{W_{t,i}=w\right\}(Y_{t,i}- \mu_w)^2 \1 \left(|\sum_{w=0}^2 a^2_{t,w} \frac{1}{n^{(t)}_{w}}\cdot\1\left\{W_{t,i}=w\right\}(Y_{t,i}- \mu_w)^2| > \epsilon^2 \right) | \mathcal{H}_{t-1}] \\
& =\ \sum_{t=0}^1\sum_{i=1}^{T_t} \sum_{w=0}^2 \mathbb{E} [ a^2_{t,w} \frac{1}{n^{(t)}_{w}}\cdot(Y_{t,i}- \mu_w)^2 \1 \left(| a^2_{t,w} \frac{1}{n^{(t)}_{w}}\cdot(Y_{t,i}(w)- \mu_w)^2| > \epsilon^2 \right) | \mathcal{H}_{t-1}, W_{t,i}=w]\PP\left(W_{t,i}=w \mid \mathcal{H}_{t-1}\right) \\
& = \sum_{t=0}^1\sum_{i=1}^{T_t} \sum_{w=0}^2 \frac{1}{T_t} \cdot \mathbb{E} [ a^2_{t,w} \cdot(Y_{t,i}(w)- \mu_w)^2 \1\left(| a^2_{t,w} \frac{1}{n^{(t)}_{w}}\cdot(Y_{t,i}(w)- \mu_w)^2| > \epsilon^2\right) | \mathcal{H}_{t-1}, W_{t,i}=w] \\
& \leq \sum_{t=0}^1\sum_{i=1}^{T_t} \sum_{w=0}^2 \frac{1}{T_t} \cdot \mathbb{E} [ a^2_{t,w} \cdot(Y_{t,i}(w)- \mu_w)^2 \1\left(| a^2_{t,w} \cdot(Y_{t,i}(w)- \mu_w)^2| > T_t\delta_T\cdot\epsilon^2\right) | \mathcal{H}_{t-1}, W_{t,i}=w]
\end{align*}
Consider the term $ \1\left( a^2_{t,w} \cdot(Y_{t,i}(w)- \mu_w)^2 > T_t\delta_T\cdot\epsilon^2\right)$
\begin{align*}
\PP\left( a^2_{t,w} \cdot(Y_{t,i}(w)- \mu_w)^2 > T_t\delta_T\cdot\epsilon^2\right) &\leq \frac{a^2_{t,w}\EE[(Y_{t,i}(w)- \mu_w)^2]}{T_t\delta_T\cdot\epsilon^2} \\
& = \frac{a^2_{t,w}\cdot\sigma^2_w}{T_t\delta_T\cdot\epsilon^2} \to 0 
\end{align*}
where the convergence results followed from the fact that $T_t\delta_T = \omega(1)$. Hence the term $ a^2_{t,w} \cdot(Y_{t,i}(w)- \mu_w)^2 \1\left(| a^2_{t,w} \cdot(Y_{t,i}(w)- \mu_w)^2| > T_t\delta_T\cdot\epsilon^2\right) \xrightarrow{p} 0$. Moreover, we have the uniform boundedness
\begin{align*}
\EE[\mid a^2_{t,w} \cdot(Y_{t,i}(w)- \mu_w)^2 \1\left(| a^2_{t,w} \cdot(Y_{t,i}(w)- \mu_w)^2| > T_t\delta_T\cdot\epsilon^2\right)\mid] 
& \leq \EE[ a^2_{t,w} \cdot(Y_{t,i}(w)- \mu_w)^2 ]\\
& \leq \lVert \bm{a} \rVert^2_{\infty} \max\left\{\sigma^2_0, \sigma^2_1, \sigma^2_2\right\} 
\end{align*}
for  $(t,i) \in \left\{0,1\right\}\times [1:T_t]$. By \textcite{loeve1977}, we have $\EE[a^2_{t,w} \cdot(Y_{t,i}(w)- \mu_w)^2 \1\left(| a^2_{t,w} \cdot(Y_{t,i}(w)- \mu_0)^2| > T_t\delta_T\cdot\epsilon^2\right)] \to 0$. \\
Then, for any $\epsilon>0$
\begin{align*}
& \PP\left(\sum_{t=0}^1\sum_{i=1}^{T_t} \sum_{w=0}^2 \frac{1}{T_t} \cdot \mathbb{E} [ a^2_{t,w} \cdot(Y_{t,i}(w)- \mu_w)^2 \1\left(| a^2_{t,w} \cdot(Y_{t,i}(w)- \mu_w)^2| > T_t\delta_T\cdot\epsilon^2\right) | \mathcal{H}_{t-1}, W_{t,i}=w] > \epsilon\right)  \\
& \leq \sum_{t=0}^1 \sum_{w=0}^2 \PP\left(\sum_{i=1}^{T_t} \frac{1}{T_t} \cdot \mathbb{E} [ a^2_{t,w} \cdot(Y_{t,i}(w)- \mu_w)^2 \1\left(| a^2_{t,w} \cdot(Y_{t,i}(w)- \mu_w)^2| > T_t\delta_T\cdot\epsilon^2\right) | \mathcal{H}_{t-1}, W_{t,i}=w] > \frac{\epsilon}{6} \right) \\
& \leq \sum_{t=0}^1 \sum_{w=0}^2\frac{6}{\epsilon \cdot T_t} \sum_{i=1}^{T_t}\mathbb{E} [ a^2_{t,w} \cdot(Y_{t,i}(w)- \mu_w)^2 \1\left(| a^2_{t,w} \cdot(Y_{t,i}(w)- \mu_w)^2| > T_t\delta_T\cdot\epsilon^2\right)] \to 0
\end{align*}
so we have 
\begin{align*}
\sum_{t=0}^1\sum_{i=1}^{T_t}  \mathbb{E} [z_{t,i}^2 \1 [|z_{t,i}| > \epsilon | \mathcal{H}_{t-1}] \xrightarrow{p} 0.
\end{align*}
which verifies the condition $(b)$.
Similarly, we can show 
\begin{align*}
 \begin{pmatrix}
\sqrt{\frac{n_w n_{w'}}{n_w\sigma^2_{w'}+n_{w'}\sigma^2_{w}}}\left(\hat{\mu}_w^{(1)}-\hat{\mu}_{w'}^{(1)}-\mu_w +\mu_{w'}\right) \\
\sqrt{\frac{m_w m_{w'}}{m_w\sigma^2_{w'}+m_{w'}\sigma^2_{w}}}\left(\hat{\mu}_w^{(0)}-\hat{\mu}_{w'}^{(0)}-\mu_w + \mu_{w'}\right)
\end{pmatrix}
 \xrightarrow{d} \mathcal{N}\left(0,\bm{I}_2\right)
\end{align*}
 for any $w,w' \in \left\{0,1,2\right\}$
\end{proof}

\section{Additional simulations}\label{appe:sec:simulation}

\begin{figure}[tbp]
\centering
\includegraphics[width=0.5\textwidth]{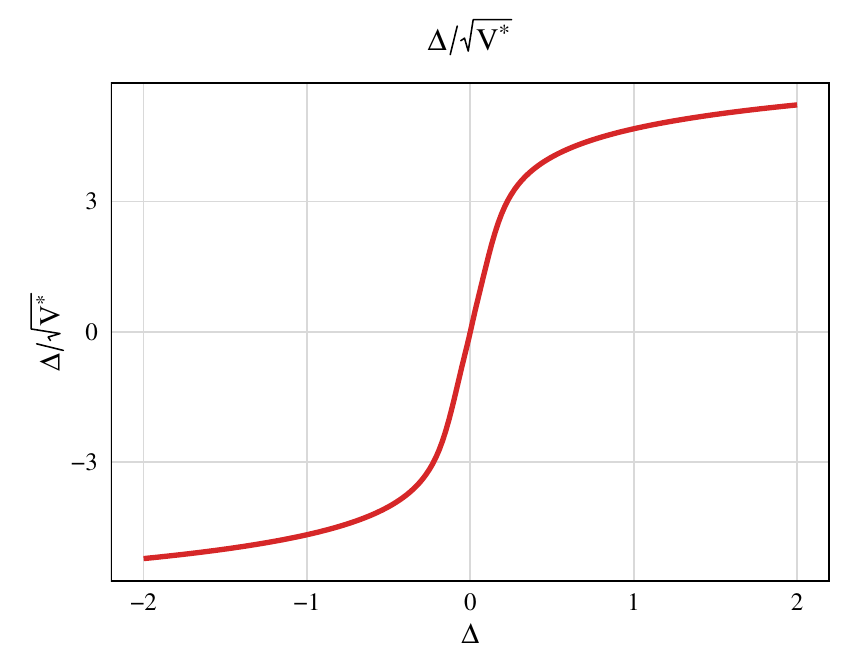}  %
\captionsetup{justification=centering}  
\caption{Monotonicity of $r(\Delta) = \frac{\Delta}{\sqrt{V(\Delta)}}$}
\label{fig:deri0}
\end{figure}

\begin{figure}[htbp]
    \centering
    \begin{subfigure}[t]{0.48\textwidth}
        \centering
        \includegraphics[width=\textwidth, trim=10 10 10 10, clip]{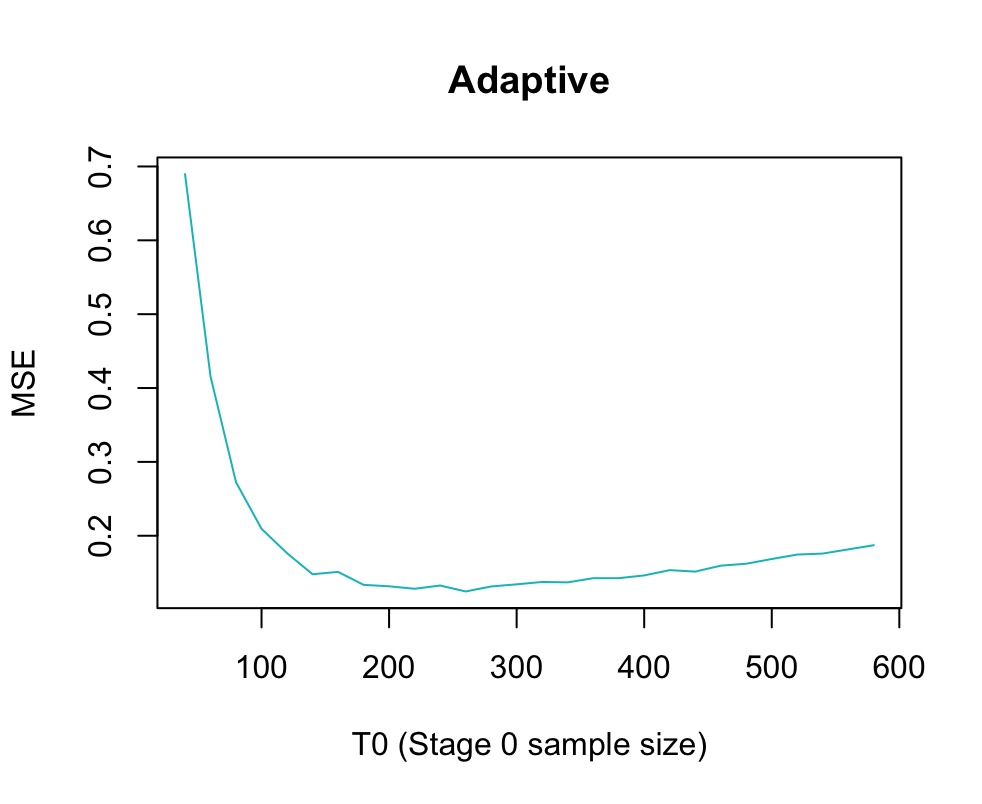}
        \caption{Adaptive method that leverages data from both stages(Proposal).}
        \label{fig:adap_T0}
    \end{subfigure}
    \hfill
    \begin{subfigure}[t]{0.48\textwidth}
        \centering
        \includegraphics[width=\textwidth, trim=10 10 10 10, clip]{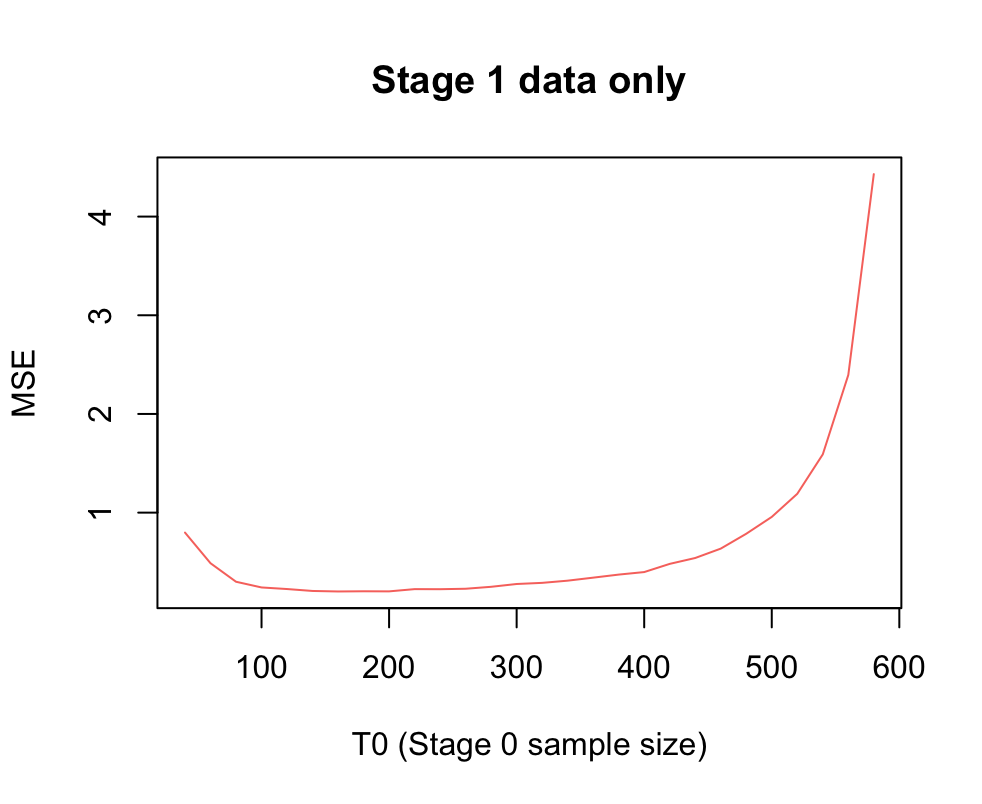}
        \caption{Method using only stage-1 data(SS.Hyper).}
        \label{fig:Stage1_T0}
    \end{subfigure}
    \caption{MSE as a function of $T_0$ with total budget $T$ fixed.}
    \label{fig:T_0.choose}
\end{figure}

\begin{figure}[tbp]
\centering
\includegraphics[width=0.7\textwidth]{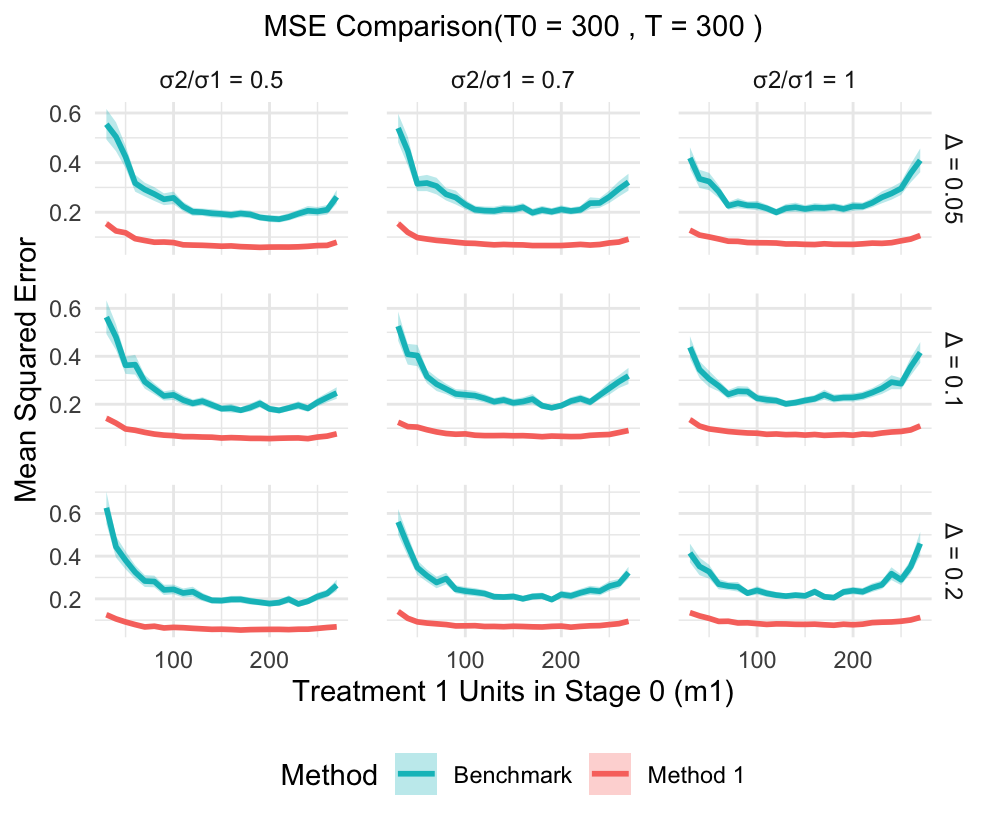}  %
\captionsetup{justification=centering}  
\caption{MSE comparison between two stage adaptive algorithm(Method 1) and SS.SE(benchmark) for $T_0$ = 300 and $T = 300$.}
\label{fig:two_stage_300}
\end{figure}

\begin{figure}[tbp]
\centering
\includegraphics[width=0.7\textwidth]{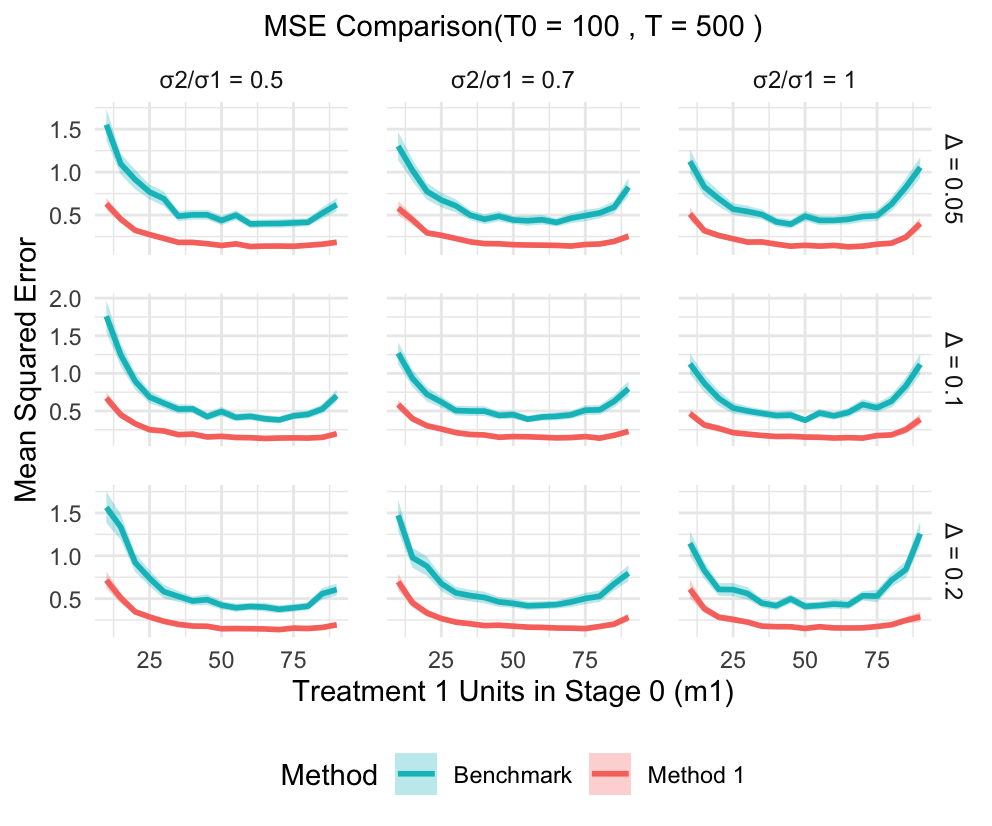}  %
\captionsetup{justification=centering}  
\caption{MSE comparison between two stage adaptive algorithm(Method 1) and SS.SE(benchmark) for $T_0$ = 100 and $T = 500$.}
\label{fig:two_stage_100}
\end{figure}

\begin{figure}[tbp]
\centering
\includegraphics[width=0.7\textwidth]{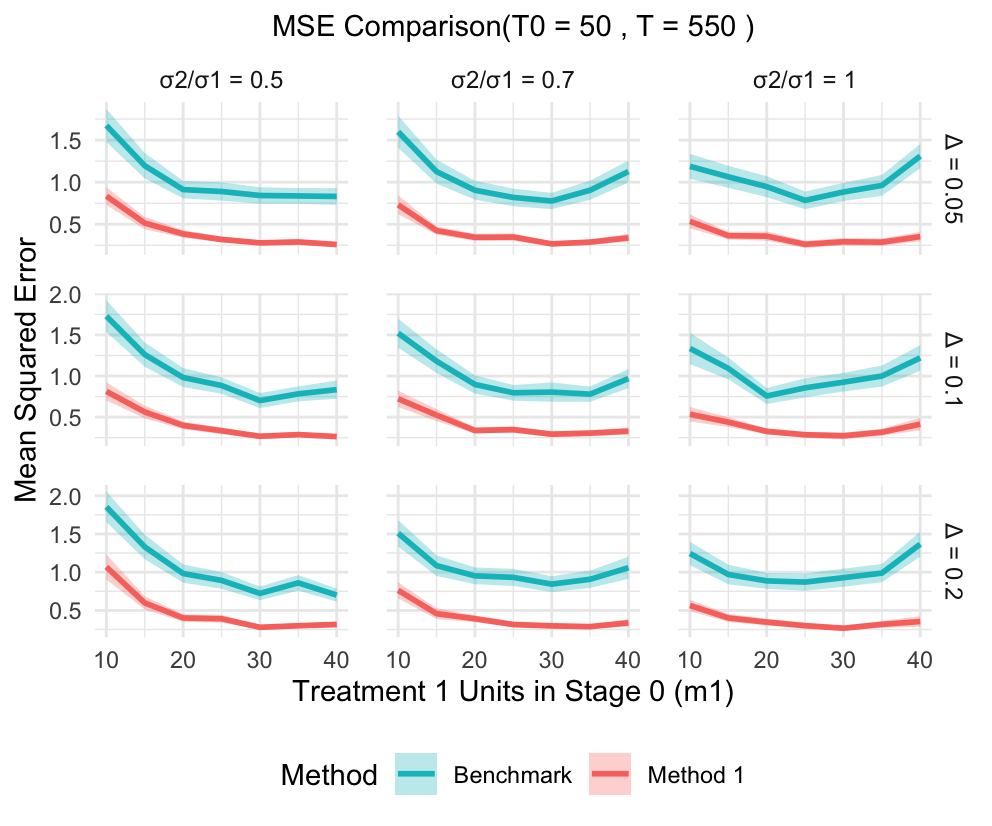}  %
\captionsetup{justification=centering}  
\caption{MSE comparison between two stage adaptive algorithm(Method 1) and SS.SE(benchmark) for $T_0 = 50$  and $T = 550$.}
\label{fig:two_stage_50}
\end{figure}

\end{document}